\documentclass[11pt]{article}
\usepackage{fullpage}
\usepackage[T1]{fontenc}
\usepackage{amsmath,amssymb,amsthm}
\usepackage{physics,braket}
\usepackage{relsize,dsfont,cancel,slashed,xfrac}
\usepackage{graphicx,wrapfig,float}
\usepackage{tikz-cd}
\usetikzlibrary{calc}
\usepackage[font=small]{caption}
\usepackage{subcaption}
\usepackage{enumitem,multirow,booktabs,tabularx,longtable,makecell}
\usepackage{etoolbox,import,listings,authblk}
\usepackage[
    backend=biber,
    style=phys,
    biblabel=brackets,
    articletitle=false,
    eprint=true
]{biblatex}
\usepackage[titles]{tocloft}
\usepackage[pdfpagelabels=false]{hyperref}
\hypersetup{
    colorlinks=true,
    linkcolor=blue,
    citecolor=purple,
    filecolor=blue,
    urlcolor=purple
}
\usepackage{orcidlink}

\allowdisplaybreaks
\numberwithin{equation}{section}

\newcommand{\id}{\mathds{1}}
\newcommand{\Acal}{{\mathcal{A}}}
\newcommand{\Fcal}{{\mathcal{F}}}
\newcommand{\Gcal}{{\mathcal{G}}}
\newcommand{\Hcal}{{\mathcal{H}}}

\newcommand{\mFP}{m_{\mathrm{FP}}}
\newcommand{\mpl}{m_{\mathrm{Pl}}}

\usepackage[compat=1.1.0]{tikz-feynman}

\newcommand{\SpinTwoLeg}[4]{%
    \ifstrequal{#1}{G}
        {\draw[/tikzfeynman/boson] (#2) -- (#3);}
        {\draw (#2) -- (#3);}
    \node[#4] at (#3) {$#1$};
}

\newcommand{\ThreeVertex}[3]{%
    \begin{tikzpicture}[scale=0.85,baseline={(current bounding box.center)}]
        \coordinate (v) at (0,0);
        \coordinate (a) at (150:1.5);
        \coordinate (b) at (210:1.5);
        \coordinate (c) at (0:1.7);

        \fill (v) circle (2.5pt);

        \SpinTwoLeg{#1}{v}{a}{above left}
        \SpinTwoLeg{#2}{v}{b}{below left}
        \SpinTwoLeg{#3}{v}{c}{right}
    \end{tikzpicture}%
}

\newcommand{\FourVertex}[4]{%
    \begin{tikzpicture}[scale=0.85,baseline={(current bounding box.center)}]
        \coordinate (v) at (0,0);
        \coordinate (a) at (145:1.5);
        \coordinate (b) at (215:1.5);
        \coordinate (c) at (35:1.5);
        \coordinate (d) at (-35:1.5);

        \fill (v) circle (2.5pt);

        \SpinTwoLeg{#1}{v}{a}{above left}
        \SpinTwoLeg{#2}{v}{b}{below left}
        \SpinTwoLeg{#3}{v}{c}{above right}
        \SpinTwoLeg{#4}{v}{d}{below right}
    \end{tikzpicture}%
}

\newcommand{\SpinTwoInternal}[3]{%
    \ifstrequal{#1}{G}
        {\draw[/tikzfeynman/boson] (#2) -- (#3);}
        {\draw (#2) -- (#3);}
}

\newcommand{\SChannel}[5]{%
    \begin{tikzpicture}[scale=0.75,baseline={(current bounding box.center)}]
        \path[use as bounding box] (-2.0,-1.15) rectangle (2.0,1.15);

        \coordinate (vL) at (-0.55,0);
        \coordinate (vR) at (0.55,0);

        \coordinate (a) at (-1.6,0.8);
        \coordinate (b) at (-1.6,-0.8);
        \coordinate (c) at (1.6,0.8);
        \coordinate (d) at (1.6,-0.8);

        \fill (vL) circle (2.5pt);
        \fill (vR) circle (2.5pt);

        \SpinTwoLeg{#1}{vL}{a}{above left}
        \SpinTwoLeg{#2}{vL}{b}{below left}
        \SpinTwoInternal{#5}{vL}{vR}
        \SpinTwoLeg{#3}{vR}{c}{above right}
        \SpinTwoLeg{#4}{vR}{d}{below right}
    \end{tikzpicture}%
}

\newcommand{\TChannel}[5]{%
    \begin{tikzpicture}[scale=0.75,baseline={(current bounding box.center)}]
        \path[use as bounding box] (-2.0,-1.15) rectangle (2.0,1.15);

        \coordinate (vT) at (0,0.4);
        \coordinate (vB) at (0,-0.4);

        \coordinate (a) at (-1.6,0.8);
        \coordinate (b) at (-1.6,-0.8);
        \coordinate (c) at (1.6,0.8);
        \coordinate (d) at (1.6,-0.8);

        \fill (vT) circle (2.5pt);
        \fill (vB) circle (2.5pt);

        \SpinTwoLeg{#1}{vT}{a}{above left}
        \SpinTwoLeg{#3}{vT}{c}{above right}
        \SpinTwoLeg{#2}{vB}{b}{below left}
        \SpinTwoLeg{#4}{vB}{d}{below right}
        \SpinTwoInternal{#5}{vT}{vB}
    \end{tikzpicture}%
}

\newcommand{\UChannel}[5]{%
    \begin{tikzpicture}[scale=0.75,baseline={(current bounding box.center)}]
        \path[use as bounding box] (-2.0,-1.15) rectangle (2.0,1.15);

        \coordinate (vT) at (0,0.4);
        \coordinate (vB) at (0,-0.4);

        \coordinate (a) at (-1.6,0.8);
        \coordinate (b) at (-1.6,-0.8);
        \coordinate (c) at (1.6,0.8);
        \coordinate (d) at (1.6,-0.8);

        \fill (vT) circle (2.5pt);
        \fill (vB) circle (2.5pt);

        \SpinTwoLeg{#1}{vT}{a}{above left}
        \SpinTwoLeg{#4}{vT}{d}{below right}
        \SpinTwoLeg{#2}{vB}{b}{below left}
        \SpinTwoLeg{#3}{vB}{c}{above right}
        \SpinTwoInternal{#5}{vT}{vB}
    \end{tikzpicture}%
}

\newcommand{\ContactDiagram}[4]{%
    \begin{tikzpicture}[scale=0.75,baseline={(current bounding box.center)}]
        \path[use as bounding box] (-2.0,-1.15) rectangle (2.0,1.15);

        \coordinate (v) at (0,0);

        \coordinate (a) at (-1.6,0.8);
        \coordinate (b) at (-1.6,-0.8);
        \coordinate (c) at (1.6,0.8);
        \coordinate (d) at (1.6,-0.8);

        \fill (v) circle (2.5pt);

        \SpinTwoLeg{#1}{v}{a}{above left}
        \SpinTwoLeg{#2}{v}{b}{below left}
        \SpinTwoLeg{#3}{v}{c}{above right}
        \SpinTwoLeg{#4}{v}{d}{below right}
    \end{tikzpicture}%
}

\title{Complete set of tree-level $2\to 2$ scattering amplitudes of\\ ghost-free bimetric theory}

\author[1]{F. Herber\thanks{\href{mailto:felix.herber@fysik.su.se}{felix.herber@fysik.su.se}}\,\orcidlink{0009-0008-1038-0410}}
\author[1,2]{J. Flinckman\thanks{\href{mailto:joakim.flinckman@fysik.su.se}{joakim.flinckman@fysik.su.se}}\,\orcidlink{0009-0004-4545-3123}}
\author[1,2]{S.F. Hassan\thanks{\href{mailto:fawad@fysik.su.se}{fawad@fysik.su.se}}\,\orcidlink{0000-0003-3910-431X}}

\affil[1]{Department of Physics, Stockholm University
}
\affil[2]{The Oskar Klein Centre, Stockholm University
}

\date{}

\begin{document}

\maketitle

\begin{abstract}
    Bimetric theory is an extension of general relativity that perturbatively describes one massless and one massive graviton. It admits observationally viable cosmologies, passes local tests of gravity, and provides candidates for dynamical dark energy and spin-2 dark matter. Scattering amplitudes provide a complementary probe of its consistency, e.g. through analyticity, unitarity, and causality. While the tree-level $2\to2$ amplitudes of the closely related theory of massive gravity have previously been computed and analysed, the richer amplitude structure of bimetric theory has mostly been studied for selected processes and helicity sectors. We present the complete set of tree-level $2\to2$ scattering amplitudes of ghost-free bimetric theory around proportional Minkowski backgrounds. We expand the action through quartic order in the mass eigenstates and use a computer-algebra workflow to compute all of the $199$ symmetry-inequivalent helicity amplitudes. We find a clear hierarchy: amplitudes with a single massive external state vanish, those with exactly two are independent of the nonlinear bimetric parameters, and nonlinear ghost-free parameter dependence first appears with three massive external states. Taking the massive-gravity limit yields the complete set of tree-level $2\to2$ massive gravity amplitudes, which agree exactly with previous results. Finally, all amplitudes grow with energy at most as $E^6$, corresponding to the characteristic $\Lambda_3$ strong-coupling scale, and this maximal growth cannot be eliminated by any nontrivial choice of the theory parameters.
\end{abstract}

\newpage
\tableofcontents
\newpage

\section{Introduction}

General relativity can be formulated perturbatively as a nonlinear theory of a self-interacting massless spin-2 field. At the fully nonlinear and background-independent level, however, it is naturally formulated geometrically in terms of a spacetime metric. Einstein's theory has proven remarkably successful across gravitational phenomena ranging from precision tests in the solar system to black holes, gravitational waves, and the large-scale evolution of the universe. Nevertheless, several major open questions remain closely connected to gravity, including the nature of dark matter \cite{Rubin:1980zd,Clowe:2006eq,Bertone:2016nfn}, the observed accelerated expansion of the universe \cite{SupernovaSearchTeam:1998fmf,SupernovaCosmologyProject:1998vns}, and the tension between early- and late-universe determinations of the Hubble constant \cite{Planck:2018vyg,Riess:2021jrx}. These questions motivate the exploration of extensions and modifications of general relativity.

A natural extension of the spin-2 formulation of general relativity is to endow the graviton with a mass, thereby giving the gravitational interaction a finite range. A consistent linear theory of a massive spin-2 field was formulated by Fierz and Pauli \cite{Pauli:1939xp,Fierz:1939ix}. However, its massless limit does not continuously reproduce the matter couplings of general relativity: the helicity-0 mode retains a finite coupling to the trace of the energy-momentum tensor, giving rise to the van Dam--Veltman--Zakharov (vDVZ) discontinuity \cite{vanDam:1970vg,Zakharov:1970cc}. This leads, for example, to a prediction that differs from that of general relativity for light deflection and is incompatible with solar-system observations.

A crucial observation was subsequently made by Vainshtein, who showed that the weak-field and small-mass limits do not commute \cite{Vainshtein:1972sx}. Nonlinear interactions can therefore become important near sufficiently dense sources, screening the additional helicity states and allowing general relativity to be recovered locally. The vDVZ discontinuity therefore need not rule out a nonlinear theory of a massive graviton. However, Boulware and Deser showed that generic nonlinear completions of the Fierz--Pauli theory propagate an additional scalar degree of freedom, now known as the Boulware--Deser ghost, which renders the theory unstable \cite{Boulware:1972yco,Boulware:1972zf,Creminelli:2005qk}.

It was later shown that the Boulware--Deser argument admits a loophole in the form of a special nonlinear interaction potential \cite{deRham:2010ik,deRham:2010kj}. The resulting theory, known as massive gravity (sometimes dRGT gravity after de Rham, Gabadadze and Tolley), was subsequently shown to be free of the Boulware--Deser ghost \cite{Hassan:2011hr,Hassan:2011tf,Hassan:2011ea}. Despite being ghost-free, massive gravity faces serious difficulties as a replacement for general relativity. For example, homogeneous and isotropic cosmological solutions in massive gravity are generically affected by instabilities or strong coupling \cite{DAmico:2011eto,Gumrukcuoglu:2011ew,DeFelice:2013bxa}.

Some of these difficulties can be alleviated by extending this ghost-free construction to a theory with two interacting dynamical metrics. The resulting bimetric theory retains the massive-gravity interaction potential and perturbatively describes interacting massless and massive spin-2 fields. Although generic nonlinear interactions between two metrics would reintroduce the Boulware--Deser ghost, this particular construction remains ghost-free and therefore provides a fully nonlinear theory of their interactions \cite{Hassan:2011zd,Hassan:2011ea,Hassan:2018mbl}.

Classical solutions in massive gravity and bimetric theory have been studied extensively \cite{Koyama:2011xz,Comelli:2011wq,Babichev:2014tfa,Gervalle:2020mfr}, particularly in cosmology \cite{DAmico:2011eto,Gumrukcuoglu:2011ew,DeFelice:2013bxa,Volkov:2011an,vonStrauss:2011mq,Koennig:2014ods}. In bimetric theory, observationally viable cosmologies can provide dynamical dark energy \cite{DeFelice:2014nja,Hogas:2021fmr,Hogas:2021lns,Akrami:2015qga}, the massive spin-2 mode can serve as dark matter \cite{Aoki:2016zgp,Babichev:2016bxi,Marzola:2017lbt}, and recent fits indicate that the Hubble tension may be partially alleviated \cite{Hogas:2025ahb}. Scattering amplitudes provide a complementary probe of the interactions between the physical spin-2 degrees of freedom. Already at the leading nonlinear orders, they identify possible decay channels and describe elastic scattering, pair production, and annihilation. When matter couplings are included, the same framework extends to massive-spin-2 decays into matter and to matter production or annihilation through spin-2 exchange. At a more structural level, the high-energy behaviour and decoupling limit reveal the strong-coupling scale of the effective theory, while the amplitudes also probe analyticity, unitarity, and causality through asymptotic superluminality.

Tree-level graviton scattering in general relativity is well understood \cite{DeWitt:1967uc,Berends:1974gk,Grisaru:1975bx,Sannan:1986tz,Benincasa:2007qj}, and explicit four-graviton amplitudes are also known at one and two loops \cite{Dunbar:1994bn,Donoghue:1999qh,Abreu:2020lyk}. In contrast, amplitude studies of massive gravity, ghost-free bimetric theory, and related interacting spin-2 effective theories have mostly concentrated on selected processes or helicity sectors \cite{Aubert:2003je,Falkowski:2020mjq}, special kinematic regimes relevant to positivity and asymptotic superluminality \cite{deRham:2018qqo,Alberte:2019xfh,Hinterbichler:2017qyt,Bonifacio:2017nnt}, or the leading high-energy behaviour and associated strong-coupling bounds \cite{Arkani-Hamed:2002bjr,Schwartz:2003vj,Fasiello:2013woa,Bonifacio:2014rba,Scargill:2015wxs,Bonifacio:2018vzv,Bonifacio:2018aon,Bonifacio:2019mgk}. Cheung and Remmen \cite{Cheung:2016yqr} computed the general tree-level massive-gravity amplitude, although the corresponding set of diagrams is considerably smaller than in bimetric theory.\footnote{Reference~\cite{Christensen:2014wra} previously computed all $5^4=625$ helicity amplitudes for a generic interacting massive spin-2 theory, though not for ghost-free massive gravity.} Our aim is to compute the complete set of tree-level $2\to2$ helicity amplitudes of ghost-free bimetric theory around proportional Minkowski backgrounds. Such a complete calculation also provides the means to compare the resulting amplitudes with corresponding results obtained using massive spinor-helicity methods \cite{Arkani-Hamed:2017jhn,Chung:2018kqs} and massive double-copy constructions \cite{Momeni:2020vvr}, and to compare the underlying cubic interactions with those obtained from string-theoretic calculations \cite{Lust:2021jps,Lust:2023sfk}, enabling independent cross-checks.

The calculation is carried out using a dedicated symbolic workflow in \textsc{Mathematica}, based on \textsc{xAct}, \textsc{FeynRules}, \textsc{FeynArts}, and \textsc{FeynCalc} \cite{xact,Alloul:2013bka,Hahn:2000kx,Shtabovenko:2020gxv}. Starting from the perturbative bimetric action, the workflow generates the interaction vertices, constructs the corresponding tree-level diagrams, and evaluates the resulting helicity amplitudes.

\subsection{Summary of results}
\label{sec:summary}

We obtain the complete set of tree-level $2\to2$ helicity amplitudes of ghost-free bimetric theory around proportional Minkowski backgrounds, encoded by $199$ symmetry-inequivalent representatives. The complete symbolic expressions are provided in an accompanying repository \cite{FelixHerber_TreeLevelBimetricCalculations_2025}, together with the code used to compute them and a simple interface for extracting any desired symbolic helicity amplitude. Beyond the individual expressions, the results reveal a clear hierarchy: processes containing precisely one massive external state are absent, while processes containing exactly two massive particles probe only the universal gravitational coupling of a Fierz--Pauli field and are independent of the detailed nonlinear bimetric potential. The ghost-free bimetric structure is first probed by amplitudes containing at least three massive external states. Although the amplitudes are generically lengthy, several classes admit compact expressions in the centre-of-mass frame. In the massive-gravity limit, we also verify that our results reproduce the previously computed tree-level massive-gravity amplitudes \cite{Cheung:2016yqr}.

At high energies, all amplitudes grow at most as $E^6$, corresponding to $\Lambda_3$ strong-coupling behaviour, and this maximal growth cannot be removed from every channel by tuning of the theory parameters. The amplitudes reproduce the known general relativity and massive-gravity results in the appropriate limits and agree with the subset of bimetric and massive-gravity amplitudes previously available in the literature.

\subsection{Outline}

The paper is organised as follows. In Section~\ref{sec:bim}, we begin with a brief review of ghost-free bimetric theory and its proportional Minkowski solutions, emphasising the parameter redundancies and symmetries that will be useful in the subsequent analysis. In Section~\ref{sec:bim_expansion}, we expand the bimetric action perturbatively around the proportional Minkowski background, identify the canonically normalised massless and massive spin-2 eigenstates and their linear equations of motion, and discuss the relevant limits of the theory. We then determine how the metric-exchange and rescaling symmetries act on the mass eigenstates and physical parameters, and identify a hierarchy in which the different bimetric parameters enter the interactions. In Section~\ref{sec:comp_amp}, we set up the computation of the tree-level scattering amplitudes by introducing the centre-of-mass kinematics, polarisation tensors and propagators, identifying the allowed cubic and quartic interaction vertices and the corresponding Feynman diagrams, and using time reversal, parity and particle-exchange symmetries to classify the inequivalent helicity amplitudes. We also briefly describe the symbolic computational pipeline used to evaluate them and discuss the high-energy limit. In Section~\ref{sec:results}, we present the main results, organised according to the external particle content, discuss representative amplitudes and their parameter dependence, compare with known limits and previous results, and analyse the high-energy behaviour and associated strong-coupling scale. We also demonstrate that our results yield all the tree-level amplitudes of massive gravity in the correct limit. We conclude in Section~\ref{sec:discussion} with a discussion of the main results and possible directions for further work.

\subsection{Conventions}

We work in $3+1$ spacetime dimensions, use natural units, $c_{\text{light}}=\hbar=1$, and the curvature convention of Wald but with the mostly-minus Minkowski metric,
\begin{align}
    \eta_{\mu\nu}=\operatorname{diag}(1,-1,-1,-1),
\end{align}
in agreement with the conventions used by \textsc{FeynCalc}. Greek indices $\mu,\nu,\rho,\ldots=0,1,2,3$ denote spacetime indices, and repeated indices are summed over. Since bimetric theory contains two metrics, $g_{\mu\nu}$ and $f_{\mu\nu}$, raising or lowering an index with the two metrics generally gives different tensors. We therefore keep indices in their canonical positions unless the metric used for raising or lowering them is clear from the context. In the perturbative expansion around the Minkowski background, indices on the perturbations and mass eigenstates are raised and lowered with $\eta_{\mu\nu}$. Square brackets denote normalised antisymmetrisation, e.g. $T_{[\mu\nu]}=\tfrac{1}{2}(T_{\mu\nu}-T_{\nu\mu})$.

\section{Ghost-free bimetric theory}
\label{sec:bim}

Bimetric theory, or bi-gravity, is a theory of two interacting spin-2 fields represented by two distinct spacetime metrics $g_{\mu \nu}(x)$ and $f_{\mu \nu}(x)$. The action is constructed from Einstein--Hilbert terms, one for each metric, accompanied by a non-derivative interaction $V$,
\begin{align}
   \label{bim_Action}
   S = -\int \text{d}^4x\, \Big[m_g^2\sqrt{-g}R(g)+m_f^2\sqrt{-f}R(f)+V\big(g, f;\beta_n\big)\Big] + S_\mathrm{m}(g;\psi^a_g) + S_\mathrm{m}(f;\psi^a_f).
\end{align}
The constants $m_g$ and $m_f$ are Planck-like masses for the corresponding metrics $g$ and $f$, and the interaction potential takes the form,
\begin{align}
\label{pot}
    V(g,f;\beta_n) =2m^4\sqrt{-g}\sum_{n=0}^4 \beta_n e_n(S), && S= \sqrt{g^{-1}f},
\end{align}
where $\beta_n$ are constant dimensionless theory parameters, $m$ a mass scale and $e_n(S)$ are the elementary symmetric polynomials of the eigenvalues of $S$, explicitly given by,
\begin{subequations}
\label{Symmetric_Polynomials_Part_1}
\begin{align}
    e_0(S) &= 1, \qquad e_1(S) = \text{Tr}(S), 
    \qquad e_2(S) = \tfrac{1}{2}\big[\text{Tr}(S)^2-\text{Tr}(S^2)\big], \\
    e_3(S) &= \tfrac{1}{6}\big[\text{Tr}(S)^3-3\text{Tr}(S) \text{Tr}(S^2)+2\text{Tr}(S^3) \big], \qquad e_4(S)=\text{det}(S),
\end{align}
\end{subequations}
or alternatively,
\begin{align}
    e_n(S) = \frac{1}{n!}\delta^{\mu_1\ldots \mu_n}_{\,\nu_1\ldots \nu_n}S^{\nu_1}_{~\mu_1}\ldots \,S^{\nu_n}_{~\mu_n}, && \delta^{\mu_1\ldots\mu_n}_{\,\nu_1\ldots\nu_n}=n!\,\delta^{[\mu_1}_{\,\nu_1}\ldots\,\delta^{\mu_n]}_{\,\nu_n}.
\end{align}
The square root matrix $S$ is defined by $(S^2)^\alpha_{~\beta}=g^{\alpha\rho}f_{\rho\beta}$. Although matrix square roots are generally not unique, for metrics whose null cones admit a common timelike direction and common spacelike hypersurface, the principal square root provides a unique real covariant $S^\alpha_{~\beta}$ \cite{Hassan:2017ugh}.

In addition to the spin-2 interaction \eqref{pot}, each metric may couple minimally to a distinct matter sector through the matter actions $S_\mathrm{m}(g;\psi^a_g) $ and $ S_\mathrm{m}(f;\psi^a_f)$. A given matter field can only couple to one metric, and there can be no direct interactions between the different matter sectors $\{\psi^a_g\}$ and $\{\psi^a_f\}$ if the theory is to remain ghost-free \cite{Yamashita:2014fga,deRham:2014naa}. Since we are interested in the scattering of the spin-2 fields, we will not consider matter couplings here, but stress that, as we will see below, the massive and massless spin-2 fields are linear combinations of the perturbations of $g_{\mu\nu}$ and $f_{\mu\nu}$, so any matter field will necessarily couple to both the massive and massless spin-2 fields.

Using the identity for the elementary symmetric polynomials,
\begin{align}
    \label{Symmetric_Polynomial_Identity}
    \sqrt{-g}\,e_n\left(\sqrt{g^{-1}f}\right) = \sqrt{-f}\,e_{4-n}\left(\sqrt{f^{-1}g}\right),
\end{align}
it follows that,
\begin{align}
    V\big(g,f;\beta_n\big) = V\big(f,g;\beta_{4-n}\big),
\end{align}
and hence the action \eqref{bim_Action} possesses the discrete exchange symmetry,
\begin{align}
\label{exchange}
    g \leftrightarrow f, && m_g \leftrightarrow m_f, && \beta_n \leftrightarrow \beta_{4-n}.
\end{align}
There is furthermore a continuous redundancy in the parametrisation of the theory associated with the relative normalisation of the two metrics. A useful representation of this redundancy, which leaves the normalisation of $g_{\mu\nu}$ fixed, is given by the constant rescaling,
\begin{align}
    \label{rescale}
    f_{\mu\nu} &\to \omega^2 f_{\mu\nu}, & m_f &\to \omega^{-1}m_f, & \beta_n &\to \omega^{-n}\beta_n,
\end{align}
with $g_{\mu\nu}$ and $m_g$ unchanged. Indeed, under \eqref{rescale}, $m_f^2\sqrt{-f}R(f)$ and $\sqrt{-g}\beta_n e_n(S)$ are invariant, while the $g$-sector Einstein--Hilbert term is trivially unchanged. We will later see that it is useful to combine the discrete exchange symmetry with a symmetric form of the rescaling redundancy. This gives the equivalent representation,
\begin{align}
\label{exchange_rescale}
    g_{\mu\nu} &\to \omega^{-1}f_{\mu\nu}, & f_{\mu\nu} &\to \omega g_{\mu\nu}, & m_g &\to \omega^{1/2}m_f, & m_f &\to \omega^{-1/2}m_g, & \beta_n &\to \omega^{2-n}\beta_{4-n}.
\end{align}
Parameter sets related by these transformations therefore describe the same physics, so that one continuous combination of the parameters is redundant. While the exchange symmetry is generically broken by matter couplings, the rescaling redundancy remains if matter couples to only one of the two metrics, since it can be represented in a form that leaves the matter-coupled metric unchanged, e.g. \eqref{rescale}. If both metrics couple independently to matter, this redundancy is generically lost.

The bimetric action \eqref{bim_Action} yields the field equations,
\begin{subequations}
\label{bim_EoM}
\begin{align}
    \label{Bimetric_EoM_for_g} 
    R_{\mu \nu}(g) -\tfrac{1}{2}g_{\mu \nu}R(g) - \frac{m^4}{m_g^2}V^g_{\mu \nu}(g, f; \beta_n) &= 0, \\
    \label{Bimetric_EoM_for_f}
    R_{\mu \nu}(f) -\tfrac{1}{2}f_{\mu \nu}R(f) - \frac{m^4}{m_f^2}V^f_{\mu \nu}(g, f; \beta_n) &= 0,
\end{align}    
\end{subequations}
where $V^g_{\mu \nu}$ and $V^f_{\mu \nu}$ encode the spin-2 interaction,
\begin{subequations}
\begin{align}
    \label{Bimetric_Interaction_g}
    V_{\mu \nu}^g(g, f; \beta_n) &= g_{\mu \rho} \sum_{n = 0}^3 \sum_{k = 0}^n (-1)^{n + k}\beta_{n}e_k(S)\big[S^{n-k}\big]^\rho{}_\nu, \\
    \label{Bimetric_Interaction_f}
    V_{\mu \nu}^f(g, f; \beta_n) &= f_{\mu \rho} \sum_{n = 0}^3 \sum_{k = 0}^n (-1)^{n + k}\beta_{4-n}e_k(S^{-1})\big[(S^{-1})^{n-k}\big]^\rho{}_\nu.
\end{align}
\end{subequations}
Taking the $g$-compatible covariant derivative $^{g}\nabla_\mu$ of \eqref{Bimetric_EoM_for_g} and using the contracted Bianchi identity gives,
\begin{align}
\label{bianchi}
    {}^{g}\nabla_\sigma\big(g^{\sigma \rho} V^g_{\rho \mu}\big)=0,
\end{align}
with an analogous redundant relation following from the $f$ equations. These are first order in derivatives and are therefore constraint equations. Of the $20$ components contained in the two metrics, the diagonal diffeomorphism invariance removes eight through four constraints and four gauge conditions, while the Bianchi constraints \eqref{bianchi} remove another four. A further scalar constraint, whose existence can be established in the Hamiltonian formulation, eliminates the Boulware--Deser ghost \cite{Hassan:2011zd,Hassan:2011ea,Hassan:2018mbl}. The theory therefore propagates $20-8-4-1=7$ degrees of freedom, corresponding to one massless and one massive spin-2 field, with $2+5$ degrees of freedom respectively.

\subsection{Proportional Minkowski solutions}

The metrics $g_{\mu \nu}$ and $f_{\mu \nu}$ do not themselves correspond to the massive and massless spin-2 fields. The mass eigenmodes can only be identified perturbatively around solutions where the background metrics are proportional, $f_{\mu \nu}=c^2 g_{\mu \nu}$, for some constant $c$ \cite{Hassan:2012wr}. However, we will restrict further to Minkowski backgrounds so that,
\begin{align}
    g_{\mu \nu}= \eta_{\mu\nu}, && f_{\mu \nu}= c^2 \eta_{\mu \nu}.
\end{align}
Since the Einstein tensors vanish for this Ansatz, the two interaction tensors $V^g_{\mu \nu}$ and $V^f_{\mu \nu}$ must vanish from \eqref{bim_EoM}, yielding the two conditions,
\begin{subequations}
\label{eom_flat_back}
\begin{align}
    \frac{m^4}{m_g^2}(\beta_0 + 3c\beta_1 + 3c^2\beta_2 +c^3\beta_3)\eta_{\mu \nu} &= 0,  \\
    \frac{m^4}{(m_fc)^2}(c\beta_1 + 3c^2\beta_2+3c^3\beta_3+c^4\beta_4)\eta_{\mu \nu} &= 0.
\end{align}    
\end{subequations}
These polynomial equations are typically solved for the proportionality constant $c$ and one of the $\beta_n$, but for our purpose, we will eliminate $\beta_0$ and $\beta_4$, and thus retain the theory parameters $\beta_1, \beta_2, \beta_3$ and $c$. Under the rescaling redundancy \eqref{rescale}, the proportional background relation $f_{\mu\nu}=c^2g_{\mu\nu}$ is preserved provided that $c\to\omega c$. Thus, $c$ itself is parametrisation dependent and may, for example, be fixed to $c=1$ by an appropriate choice of $\omega$. We will nevertheless retain $c$ explicitly in what follows, but we note that the combination $c^n \beta_n$ is invariant,
\begin{align}
\label{nbeta_inv}
    c^n \beta_n \to c^n \beta_n.
\end{align}

\section{Expansion of the bimetric action}
\label{sec:bim_expansion}

In order to compute bimetric scattering amplitudes, we expand the metrics about the above proportional Minkowski backgrounds, 
\begin{align}
\label{met_pert}
    g_{\mu \nu} = \eta_{\mu \nu} + \delta g_{\mu \nu}, &&    f_{\mu \nu} = c^2 \eta_{\mu \nu}+\delta f_{\mu \nu}.
\end{align}
The perturbative expansion should be understood as an effective field theory description, valid only below its strong-coupling scale. We return to this point below, where the high-energy behaviour of the scattering amplitudes is used to estimate this cutoff and compare it with the corresponding massive-gravity result \cite{Cheung:2016yqr,Scargill:2015wxs}.

The Einstein--Hilbert terms can be expanded by standard methods. For completeness and to make our conventions explicit, we perform this expansion using the \textsc{Mathematica} package \textsc{xAct} \cite{xact,xcoba,xperm,xpert,xtensor,texact}. The expressions beyond quadratic order are lengthy and are therefore provided in the accompanying GitHub repository \cite{FelixHerber_TreeLevelBimetricCalculations_2025}, in the notebook \texttt{xActExpansionOfBimetric.nb}, while the quadratic action is discussed explicitly below. The elementary symmetric polynomials entering the interaction potential are treated similarly. Their explicit expansions are provided in the accompanying notebooks, but a few features of their structure are worth noting.

To expand the square root, we start by expanding $g^{\mu \rho}f_{\rho\nu}$, which in matrix notation takes the form,
\begin{align}
    \label{Taylor_of_S_Squared}
    g^{-1} f= c^2\id+\sum_{n = 1}^\infty (-1)^n \Big[c^2 \big(\eta^{-1}\delta g \big)^n-  \big(\eta^{-1}\delta g \big)^{n-1} \eta^{-1} \delta f\Big].  
\end{align}
We can then use the standard expansion of the square root $\sqrt{1+x}=\sum_n \binom{1/2}{n}x^n$, so that,
\begin{align}
    \label{S}
    S = c\id+c\sum_{n = 1}^\infty \binom{1/2}{n}\Bigg[\sum_{m = 1}^\infty (-1)^m \Big(\big(\eta^{-1}\delta g\big)^m-\frac{1}{c^2}\big(\eta^{-1}\delta g\big)^{m-1}\eta^{-1}\delta f\Big)\Bigg]^n.
\end{align}
Note that to first order,
\begin{align}
\label{etaS=f+g}
    \eta_{\mu\sigma}\delta S^\sigma{}_\nu = \frac{1}{2c}\left(\delta f_{\mu\nu}-c^2\delta g_{\mu\nu}\right).
\end{align}
Consequently, after imposing the background equations, the quadratic potential \eqref{pot} depends only on the combination $\delta f_{\mu\nu}-c^2\delta g_{\mu\nu}$, which we will see below precisely corresponds to the massive spin-2 field.

To expand the potential terms $e_1(S)$ and $e_2(S)$, it is easiest to take the trace of \eqref{Taylor_of_S_Squared} and \eqref{S} and explicitly construct the $e_1(S)$ and $e_2(S)$ from \eqref{Symmetric_Polynomials_Part_1}. However, instead of constructing $e_3(S)$ and $e_4(S)$ this way, it is convenient to use the identity \eqref{Symmetric_Polynomial_Identity} to construct the remaining terms directly from the $e_0(S)$ and $e_1(S)$ expressions through,
\begin{align}
    \sqrt{-g}e_{4-n}(\sqrt{g^{-1}f})
    =c^{4-2n}\left.\sqrt{-g}e_n(\sqrt{g^{-1}f})\right|_{\delta g\leftrightarrow\delta f/c^2},
    \qquad n=0,1.
\end{align}
This also makes the exchange symmetry of the metric perturbations manifest.

With that, we can obtain the full expansion of the bimetric action \eqref{bim_Action} to any order in $\delta g$ and $\delta f$. The explicit results up to fourth order can be found in the accompanying \textsc{Mathematica} notebooks \cite{FelixHerber_TreeLevelBimetricCalculations_2025} and Appendix D of \cite{Herber2006232}. With both the Einstein--Hilbert terms and the potential expanded, the quadratic action takes the form,
\begin{align}
\label{Expansion_of_Bimetric_Quadratic}
    S^{(2)} = &\tfrac{1}{4}\int \text{d}^4x \Big[m_g^2\delta g_{\mu \nu}K^{\mu \nu \rho \sigma}\delta g_{\rho \sigma} +m_f^2c^{-2}\delta f_{\mu \nu}K^{\mu \nu \rho \sigma}\delta f_{\rho \sigma}\nonumber \\
    &\hspace{15mm} - (-\delta g_{\mu \nu} + c^{-2}\delta f_{\mu \nu})\mathcal{M}^{\mu \nu \rho \sigma}(-\delta g_{\rho \sigma} + c^{-2}\delta f_{\rho \sigma})\Big],
\end{align}
where we have defined the kinetic and mass operators,
\begin{subequations}
\begin{align}
    K^{\mu\nu\rho\sigma} &= 3!\,\delta^{[\mu}_{\lambda}\delta^\sigma_{\kappa}\delta^{\alpha]}_{\beta}\eta^{\lambda\nu}\eta^{\kappa\rho}\partial_\alpha\partial^\beta\\
\label{beta_FP}
    \mathcal{M}^{\mu \nu \rho \sigma} &= m^4\beta_{\text{FP}}(\eta^{\mu \rho}\eta^{\nu\sigma} - \eta^{\mu \nu}\eta^{\rho \sigma}), \qquad\quad \beta_{\text{FP}}= c\beta_1 + 2c^2\beta_2 + c^3\beta_3,
\end{align}    
\end{subequations}
and indices are raised and lowered with $\eta^{\mu \nu}$ and $\eta_{\mu \nu}$.

\subsection{Mass eigenstates}
\label{sec:mass_eigen}
The action \eqref{Expansion_of_Bimetric_Quadratic} is not diagonal in the fields $\delta g_{\mu \nu}$ and $\delta f_{\mu \nu}$, but the interaction term depends solely on the combination \eqref{etaS=f+g}, so if we define,
\begin{subequations}
\label{mass_eigenstates}
\begin{align}
\label{G_in_terms_of_f_and_g} 
    G_{\mu \nu}(x) &= \frac{m_{\text{Pl}}}{\sqrt{2}(1+q^2)}\Big[\delta g_{\mu \nu}(x) +\frac{q^2}{c^2} \delta f_{\mu \nu}(x)\Big], \\
 \label{M_in_terms_of_f_and_g}
    M_{\mu \nu}(x) &= \frac{q m_{\text{Pl}}}{\sqrt{2}(1+q^2)}\Big[-\delta g_{\mu \nu}(x) + \frac{1}{c^2}\delta f_{\mu \nu}(x)\Big],
\end{align}    
\end{subequations}
with $m_{\text{Pl}} = \sqrt{m_g^2+c^2m_f^2}$ and $q = \frac{cm_f}{m_g}$,\footnote{Note that our $m_{\text{Pl}}$ differs from the convention of \cite{Babichev:2016bxi}. We use $m_{\text{Pl}} = \frac{1}{\sqrt{16\pi G}}$ while they use $m_{\text{Pl}} = \frac{1}{\sqrt{8\pi G}}$. This difference follows from the factor $\frac{1}{\sqrt{2}}$ in \eqref{G_in_terms_of_f_and_g} and \eqref{M_in_terms_of_f_and_g}, which we introduce to match the quadratic terms of the $G_{\mu \nu}$ field to those of the graviton in eq. (3) of \cite{Berends:1974gk}, thereby obtaining the propagator \eqref{Massless_Graviton_Propagator}. Consequently, our normalisation differs from \cite{Schmidt-May:2015vnx} by a factor of $\frac{1}{2}$.} the action becomes diagonal,
\begin{align}
    \label{Expansion_of_Bimetric_In_G_and_M}
    \mathcal{L}^{(2)} = &  \tfrac{1}{2}G_{\mu \nu} K^{\mu \nu \rho \sigma}G_{\rho \sigma}+\tfrac{1}{2}M_{\mu \nu} K^{\mu \nu \rho \sigma}M_{\rho \sigma} -\tfrac{1}{2}m_{\text{FP}}^2\Big[M_{\mu \nu} M^{\mu \nu} - M^{\mu}{}_\mu M^{\nu}{}_\nu\Big],
\end{align}
where we have also defined the Fierz--Pauli mass $m_{\text{FP}}$,
\begin{align}
    \label{Fierz_Pauli_Mass_for_Bimetric}
    m_{\text{FP}}^2 = \frac{m^4}{q^2m_g^4}m_{\text{Pl}}^2\beta_{\text{FP}}.
\end{align}
Under the rescaling redundancy \eqref{rescale},  $q$ and $m_{\text{Pl}}$ are invariant. Moreover, since $\beta_{\text{FP}}$ is unchanged \eqref{nbeta_inv}, the Fierz--Pauli mass $m_{\text{FP}}$ is likewise invariant, as expected for a physical mass parameter. The rescaling \eqref{rescale} leaves the massless and massive perturbations $G_{\mu \nu}$ and $M_{\mu \nu}$ unchanged, making the perturbative action rescaling invariant.

The ratio $q=cm_f/m_g$ can be identified with the tangent of the mixing angle relating $(G_{\mu\nu},M_{\mu\nu})$ to $(\delta g_{\mu\nu},\delta f_{\mu\nu})$,
\begin{align}
    \begin{pmatrix}
        G_{\mu\nu}\\
        M_{\mu\nu}
    \end{pmatrix}
    &=
    \frac{1}{\sqrt{2}}
    \begin{pmatrix}
        \cos\gamma & \sin\gamma\\
        -\sin\gamma & \cos\gamma
    \end{pmatrix}
    \begin{pmatrix}
        m_g\,\delta g_{\mu\nu}\\
        m_f c^{-1}\delta f_{\mu\nu}
    \end{pmatrix},
    \qquad
    \tan\gamma=q.
\end{align}
The metric perturbations $\delta g_{\mu\nu}$ and $\delta f_{\mu\nu}$ thus play the role of interaction (flavour) eigenstates, while $G_{\mu\nu}$ and $M_{\mu\nu}$ are the spin-2 mass eigenstates. Since $m_{\mathrm{Pl}}^2=m_g^2+c^2m_f^2$ is the perturbative parameter \eqref{mass_eigenstates}, the two Einstein--Hilbert scales around the proportional background can equivalently be written as,
\begin{align}
    m_g &= \frac{m_{\mathrm{Pl}}}{\sqrt{1+q^2}}, & cm_f &= \frac{qm_{\mathrm{Pl}}}{\sqrt{1+q^2}}.
\end{align}
The limits $q\to0$ and $q\to\infty$ determine the alignment of the mass and interaction eigenstates, but do not by themselves specify which dimensional parameters are held fixed. For finite nonzero $c$ and matter coupled only to $g_{\mu\nu}$, the general relativity limit is,
\begin{subequations}
\begin{align}
    m_f&\to0 \qquad \text{at fixed }m_g, \qquad \text{or,}\\
    q&\to0, \qquad \mpl\to m_g, \qquad \frac{q}{\mpl}\to0. \label{GR_Limit}
\end{align}
\end{subequations}
In the exact limit, the $f_{\mu\nu}$ kinetic term disappears and its equation becomes algebraic, whereas the $g_{\mu\nu}$ equation reduces to the Einstein equation with an effective cosmological constant \cite{Akrami:2015qga,Schmidt-May:2015vnx}.

Conversely, the standard massive-gravity limit is,
\begin{subequations}
\begin{align}
    m_f&\to\infty \qquad \text{at fixed }m_g, \qquad \text{or},\\
    q&\to\infty, \qquad \mpl\to\infty, \qquad \frac{q}{\mpl}\to\frac{1}{m_g}. \label{Massive_Gravity_Limit}
\end{align}
\end{subequations}
The nonlinear fluctuations of $f_{\mu\nu}$ then decouple, leaving its background value as the fixed reference metric of the resulting massive-gravity theory. This limit applies to background branches that remain regular as $m_f/m_g\to\infty$ \cite{Baccetti:2012bk,Hassan:2014vja,Schmidt-May:2015vnx}.

Inverting the relations \eqref{mass_eigenstates} gives,
\begin{align}
\label{inv_fields}
    \delta g_{\mu\nu} &= \frac{\sqrt{2}}{m_{\mathrm{Pl}}}\Big[G_{\mu\nu}-qM_{\mu\nu}\Big], & \frac{\delta f_{\mu\nu}}{c^2} &= \frac{\sqrt{2}}{m_{\mathrm{Pl}}}\Big[G_{\mu\nu}+\frac{1}{q}M_{\mu\nu}\Big].
\end{align}
In the GR limit \eqref{GR_Limit}, the massive mode decouples from $\delta g_{\mu\nu}$, so that matter coupled to $g_{\mu\nu}$ interacts only with the massless mode.\footnote{The perturbation $\delta f_{\mu\nu}$ does not itself vanish in this limit. Holding $M_{\mu\nu}$ fixed would make the second relation in \eqref{inv_fields} singular as $q\to0$, while keeping $\delta f_{\mu\nu}$ finite requires $M_{\mu\nu}=\mathcal{O}(q)$. In the exact limit, the $f_{\mu\nu}$ kinetic term disappears and its equation becomes algebraic, so $f_{\mu\nu}$ carries no independent propagating fluctuations. On the regular proportional branch, the pure GR sector satisfies $\delta f_{\mu\nu}/c^2=\delta g_{\mu\nu}$.} Conversely, the massive-gravity limit \eqref{Massive_Gravity_Limit} gives \mbox{$\delta g_{\mu\nu}\to-\sqrt{2}M_{\mu\nu}/m_g$} and $\delta f_{\mu\nu}/c^2\to0$, explicitly showing that the massless mode and the fluctuations of $f_{\mu\nu}$ decouple.

The field redefinition \eqref{mass_eigenstates} is exact for finite nonzero $q$. As $q\to0$, however, the massive contribution to $\delta f_{\mu\nu}/c^2$ is enhanced by $1/q$, while as $q\to\infty$ its contribution to $\delta g_{\mu\nu}$ is enhanced by $q$. The expansion in the mass eigenstates remains within the weak-field regime of the original metrics only if these enhanced contributions remain small. Equation~\eqref{inv_fields} therefore requires parametrically,
\begin{align}
    q\ll1:\quad \frac{G}{m_{\mathrm{Pl}}}\ll1,\quad\frac{M}{qm_{\mathrm{Pl}}}\ll1, && q\gg1:\quad \frac{G}{m_{\mathrm{Pl}}}\ll1,\quad\frac{qM}{m_{\mathrm{Pl}}}\ll1.
\end{align}
For scattering configurations characterised by an energy scale $E$, a sufficient condition for the perturbative expansion is therefore,
\begin{align}
\label{E_scale}
    E\ll\min(m_g,cm_f)=\frac{m_{\mathrm{Pl}}}{\sqrt{1+q^2}}\min(1,q)\sim
    \begin{cases}
        qm_{\mathrm{Pl}}, & q\ll1,\\
        m_{\mathrm{Pl}}/q, & q\gg1.
    \end{cases}
\end{align}
This reproduces the small-$q$ perturbativity condition derived in Refs.~\cite{Babichev:2016hir,Babichev:2016bxi}.

The scale above controls the weak-field expansion of the original metrics. It should not be confused with the strong-coupling scale associated with the longitudinal polarisations of the massive spin-2 field. The latter also depends on $m_{\mathrm{FP}}$ and the potential parameters, and its behaviour in a limit of $q$ depends on which physical quantities are held fixed. It will be determined directly from the scattering amplitudes in Section~\ref{sec:results}.

\subsection{Linearised field equations}

The massless perturbation $G_{\mu \nu}$ has a gauge redundancy, so that,
\begin{align}
    G_{\mu \nu} \to G_{\mu \nu} + \partial_\mu \xi_\nu + \partial_\nu \xi_\mu,
\end{align}
leaves the Lagrangian \eqref{Expansion_of_Bimetric_In_G_and_M} invariant to second order. This transformation corresponds to the diagonal linear diffeomorphism acting simultaneously on both metric perturbations $\delta g_{\mu\nu}$ and $\delta f_{\mu\nu}$. Due to the relative sign in \eqref{M_in_terms_of_f_and_g}, the massive mode $M_{\mu\nu}$ is invariant under the diagonal diffeomorphism, $M_{\mu\nu}\to M_{\mu\nu}$. Without the interaction potential, each metric sector possesses its own linearised diffeomorphism invariance, but the potential breaks these down to the diagonal subgroup, leaving $G_{\mu\nu}$ massless while the orthogonal combination $M_{\mu\nu}$ acquires the Fierz--Pauli mass.

To compute the scattering amplitudes below, we will further impose the de Donder (harmonic gauge) condition for $G_{\mu\nu}$,\footnote{Since the diagonal diffeomorphism acts entirely on $G_{\mu\nu}$ while leaving $M_{\mu\nu}$ invariant, the de Donder condition may be imposed on $G_{\mu\nu}$ alone. A gauge condition imposed on $\delta g_{\mu\nu}$ would, by contrast, necessarily induce a simultaneous transformation of $\delta f_{\mu\nu}$.}
\begin{align}
\label{deDonder}
    \partial_\mu G^{\mu \nu}-\tfrac{1}{2}\partial^\nu G^\mu_{~\mu} = 0.
\end{align}
Using the residual gauge freedom to further impose tracelessness $G^\mu_{\,\mu}=0$, the on-shell conditions for the massless and massive spin-2 fields can be written as,
\begin{subequations}
\label{MG_EoM}
\begin{align}
\label{G_EoM}
    \Box G_{\mu \nu}&=0, &\partial_\mu G^\mu_{~\nu}&=0, &G^\mu_{~\mu}&=0,\\
\label{M_EoM}
    (\Box+m^2_{\mathrm{FP}}) M_{\mu \nu}&=0, &\partial_\mu M^\mu_{~\nu}&=0, &M^\mu_{~\mu}&=0.
\end{align}
\end{subequations}
For the massless field $G_{\mu\nu}$, these conditions together with the residual gauge freedom leave only the two helicity $\pm2$-polarisations. The massive field $M_{\mu\nu}$ has no corresponding gauge freedom, and the four transverse conditions together with tracelessness reduce its ten components to five physical polarisations ($\pm2,\pm1,0$). 

\subsection{Metric exchange and scale symmetry}
\label{subsec:Metric_Exchange}

The exchange symmetry \eqref{exchange} has a simple realisation in the perturbative theory. This can be seen by choosing $\omega=c^2$ in \eqref{exchange_rescale}, which preserves our background solutions $g_{\mu\nu}=\eta_{\mu\nu}$ and $f_{\mu\nu}=c^2\eta_{\mu\nu}$, while the perturbations transform as,
\begin{align}
    \label{new_exchange}
    \delta g_{\mu\nu}\to\frac{1}{c^2}\delta f_{\mu\nu}, && \delta f_{\mu\nu}\to c^2\delta g_{\mu\nu}.
\end{align}
In this representation $c$ is unchanged, while $\beta_n \to c^{4-2n}\beta_{4-n}$. It follows immediately that,
\begin{align}
    m_{\text{Pl}}\to m_{\text{Pl}}, &&q\to\frac{1}{q}, && \gamma\to\frac{\pi}{2}-\gamma.
\end{align}
Substituting \eqref{new_exchange}, together with the parameter transformations following from \eqref{exchange_rescale}, into the field redefinitions \eqref{mass_eigenstates}, we find,
\begin{align}
    G_{\mu\nu}\to G_{\mu\nu}, && M_{\mu\nu}\to-M_{\mu\nu}.
\end{align}
Thus, the massless spin-2 field is even under metric exchange, while the massive spin-2 field is odd.

The transformation also leaves $\beta_2$ invariant, so it will be useful in the higher-order expansion to eliminate $\beta_2$ in favour of the physical parameter $m_{\text{FP}}$, without obscuring the exchange symmetry. In terms of the remaining parameters, the nontrivial transformations are,
\begin{align}
\label{M_q_exchange}
    M_{\mu\nu}\to-M_{\mu\nu}, && q\to\frac{1}{q}, && c\beta_1\leftrightarrow c^3\beta_3,
\end{align}
while $G_{\mu\nu}$, $m_{\text{Pl}}$, $m_{\text{FP}}$, and $c$ are invariant. This symmetry will provide a useful check on the interaction vertices and help relate the scattering amplitudes. We stress, however, that the exchange symmetry is generically broken once matter couplings distinguish the two metrics.

\subsection{Parameter hierarchy of the interactions}
\label{subsec:parameter_structure}

The flat-background conditions \eqref{eom_flat_back} leave three independent combinations of the potential parameters $\beta_n$. In view of the rescaling invariance \eqref{nbeta_inv} and the metric-exchange transformation discussed above, it is useful to choose these as,
\begin{align}
\label{Bimetric_Parameter_Combinations}
    \beta_{\mathrm{FP}} &= c\beta_1+2c^2\beta_2+c^3\beta_3, & \beta_\pm &= c\beta_1\pm c^3\beta_3,
\end{align}
and to introduce,
\begin{align}
\label{Q_pm_Definition}
    Q_\pm= q\pm q^{-1}.
\end{align}
Using the definitions of $m_{\mathrm{Pl}}$ and $q$, the Fierz--Pauli mass \eqref{Fierz_Pauli_Mass_for_Bimetric} can equivalently be written as,
\begin{align}
\label{FP_Mass_Interaction_Combination}
    m_{\mathrm{FP}}^2=\frac{m^4Q_+^2}{m_{\mathrm{Pl}}^2}\beta_{\mathrm{FP}}.
\end{align}
Thus, $\beta_{\mathrm{FP}}$ determines the universal quadratic Fierz--Pauli mass term, while we will see below that $\beta_\pm$ parametrise the remaining freedom in its nonlinear ghost-free completion.

The exchange properties of the combinations (\ref{Bimetric_Parameter_Combinations}--\ref{Q_pm_Definition}) follow from the transformations \eqref{M_q_exchange},
\begin{subequations}
\label{QQ_exchange}
\begin{align}
    Q_+&\to Q_+, & \beta_{\mathrm{FP}}&\to\beta_{\mathrm{FP}}, & \beta_+&\to\beta_+,\\
    Q_-&\to-Q_-, & \beta_-&\to-\beta_-, & M_{\mu\nu}&\to-M_{\mu\nu}.
\end{align}
\end{subequations}
Thus, $Q_-$ and $\beta_-$ are the exchange-odd parameter combinations. Interactions containing an odd number of massive fields must consequently vanish at the exchange-symmetric point $q=1$ and $\beta_-=0$. We will elaborate further on this in Section \ref{sec:results}.

We now show that $\beta_{\mathrm{FP}}$, $\beta_\pm$, and $q$ enter the interactions according to a clear hierarchy in powers of $M_{\mu\nu}$. To identify the lowest order at which each parameter appears, we introduce the rescaling-invariant relative matrices,
\begin{align}
\label{Relative_Matrix_Definition}
    \mathbb{X}=\frac{1}{c}S=\frac{1}{c}\sqrt{g^{-1}f}, && \mathbb{Y}= \mathbb{X}-\id.
\end{align}
The potential \eqref{pot} can then be expressed as,
\begin{align}
    V=2m^4\sqrt{-g}\,\mathcal{U}(\mathbb{X}), &&\mathcal{U}(\mathbb{X})=\sum_{n=0}^4c^n\beta_ne_n(\mathbb{X}).
\end{align}
We can now use $\mathbb{X}=\id+\mathbb{Y}$ and the identity,
\begin{align}
    e_n(\id + \mathbb{Y}) = \sum_{k=0}^n\binom{4-k}{n-k}e_k(\mathbb{Y}),
\end{align}
together with the flat-background conditions \eqref{eom_flat_back} to eliminate $\beta_0$ and $\beta_4$, and rewrite the potential entirely in terms of $\mathbb{Y}$ and \eqref{Bimetric_Parameter_Combinations},
\begin{align}
\label{Potential_Parameter_Hierarchy}
    \mathcal{U}(\mathbb{X})=-\beta_{\mathrm{FP}}e_2(\mathbb{Y})+\tfrac{1}{2}(\beta_--3\beta_{\mathrm{FP}})e_3(\mathbb{Y})+\tfrac{1}{2}(2\beta_--\beta_+-3\beta_{\mathrm{FP}})e_4(\mathbb{Y}).
\end{align}
We stress that this is an exact rewriting of the potential and not a truncation of its perturbative expansion. 

Using the square-root perturbation \eqref{etaS=f+g} together with the field redefinitions \eqref{inv_fields}, it is clear that $\mathbb{Y}$ is proportional to the massive mode at linear order,
\begin{align}
    \mathbb{Y}^\mu{}_\nu=\frac{Q_+}{\sqrt{2}m_{\mathrm{Pl}}}M^\mu{}_\nu+\mathcal{O}(GM,M^2).
    \label{Y_Massive_Expansion}
\end{align}
More generally, setting $M_{\mu\nu}=0$ in the inverse mass eigenstate transformation gives $f_{\mu\nu}=c^2g_{\mu\nu}$ and hence $\mathbb{X}=\id$ and $\mathbb{Y}=0$. Every term in the perturbative expansion of $\mathbb{Y}$ must therefore contain at least one massive field.

The elementary symmetric polynomials $e_n(\mathbb{Y})$ are homogeneous of degree $n$. Consequently, every monomial in $e_n(\mathbb{Y})$ contains at least $n$ massive fields, so that,
\begin{align}
\label{e(Y)_M}
    e_n(\mathbb{Y})=\mathcal{O}(M^n),
\end{align}
where the order counts the number of massive fields rather than the total number of perturbations. Equation \eqref{Potential_Parameter_Hierarchy} therefore makes the parameter hierarchy manifest,
\begin{align}
    \beta_{\mathrm{FP}} &: \ \mathcal{O}(M^2), & \beta_- &: \ \mathcal{O}(M^3), & \beta_+ &: \ \mathcal{O}(M^4).
    \label{Bimetric_Parameter_Hierarchy}
\end{align}
The combination $\beta_{\mathrm{FP}}$ fixes the quadratic Fierz--Pauli term and parts of its nonlinear completion. The first new potential parameter appears through $\beta_-$ at cubic order in the massive field, while $\beta_+$ first introduces new freedom at quartic order in the massive perturbation $M$.

The mixing parameter $q$ also does not appear at every orders. After substituting the field redefinition \eqref{inv_fields} into the two Einstein--Hilbert terms \eqref{bim_Action}, their tensor structures are identical, while their relative coefficients are determined by the number of massive fields. Apart from the common normalisation, an Einstein--Hilbert vertex containing $k$ massive fields is proportional to,\footnote{These can equivalently be expressed entirely in terms of $Q_-$ through the recursion relation,
\begin{align*}
    C_0=1, \qquad C_1=0, \qquad C_{k+2} &= -Q_-C_{k+1}+C_k.
\end{align*}}
\begin{align}
\label{EH_q_Coefficients}
    C_k(q)=\frac{(-q)^k+q^{2-k}}{1+q^2}.
\end{align}
In particular,
\begin{align}
\label{EH_q_Coefficients_Explicit}
    C_0=1, \qquad C_1=0, \qquad C_2=1, \qquad C_3=-Q_-, \qquad C_4=1+Q_-^2.
\end{align}
The relation $C_1=0$ explains the absence of interaction vertices containing a single massive field, while $C_2=1$ shows that the Einstein--Hilbert vertices containing two massive fields are independent of $q$. For the potential contribution at quadratic order, the remaining $Q_+$ dependence from \eqref{Y_Massive_Expansion} is absorbed into the physical mass through \eqref{FP_Mass_Interaction_Combination}. Consequently, the complete action containing at most two massive fields is independent of both $q$ and $\beta_\pm$.

From the above it also follows that there are no vertices containing a single massive field to all orders in $G_{\mu\nu}$. The coefficient $C_1=0$ is independent of the number of massless fields and therefore eliminates every term of the form $MG^n$ from the sum of the Einstein--Hilbert actions. The potential cannot generate such terms either, since \eqref{Potential_Parameter_Hierarchy} does not contain $e_1(\mathbb{Y})\propto M$ and the remaining terms are all at least quadratic in $M$ \eqref{e(Y)_M}. Consequently, the complete perturbative expansion contains no $G^nM$ vertices, as previously noted in \cite{Babichev:2016bxi}.

The preceding results separate the universal Fierz--Pauli sector from the genuinely nonlinear bimetric interactions. Consider any two-metric theory with Einstein--Hilbert kinetic terms and a diagonal-diffeomorphism-invariant interaction, expanded around the same proportional Minkowski background. Once its quadratic interaction is fixed to be the Fierz--Pauli mass term, all vertices containing exactly two massive fields are fixed by its covariant completion. Assuming that no additional derivative interactions are present, the $GMM$ and $GGMM$ vertices are therefore universal at fixed $m_{\mathrm{Pl}}$ and $m_{\mathrm{FP}}$, independently of whether the higher-order potential satisfies the ghost-free tuning.

This universality directly determines the structure of $GG\rightarrow MM$ and $MG\to MG$ scattering. Their tree-level diagrams consist of a massless exchange constructed from $GGG$ and $GMM$, massive exchanges constructed from two $GMM$ vertices, and the $GGMM$ contact interaction. This list is exhaustive because all $G^nM$ vertices are absent. Each diagram therefore belongs to the universal sector and is individually insensitive to the nonlinear bimetric parameters such as $\beta_{\pm}$ and $q$. These processes consequently probe the gravitational interactions of a Fierz--Pauli massive spin-2 field, but not its nonlinear completion in ghost-free bimetric theory.

The nonlinear bimetric structure is first probed by vertices containing three massive external states. At $\mathcal{O}(M^3)$, the Einstein--Hilbert terms depend on the mixing through $Q_-$, while $\beta_-$ is the only new potential parameter beyond the contributions fixed by $\beta_{\mathrm{FP}}$. The parameter $\beta_+$ first appears at $\mathcal{O}(M^4)$, where the vertices can depend on the full set of parameters $q$, $\beta_{\mathrm{FP}}$, and $\beta_\pm$. Amplitudes involving four massive fields therefore probe the complete parameter dependence of the ghost-free bimetric interaction.

\section{Computation of scattering amplitudes}
\label{sec:comp_amp}

Having established the perturbative mass eigenstates and their parameter symmetries, we now turn to the tree-level $2\to2$ scattering amplitudes of bimetric theory. We will denote a generic process by $AB\to CD$, where each of $A,B,C,D$ is either the massless spin-2 particle $G$ or the massive spin-2 particle $M$. For $X=A,B,C,D$, we denote the corresponding four-momentum and mass by $p_X^\mu=(p_X^0, \bar p^{\,}_X)$ and $m_X$, with $m_X=0$ for $X=G$ and $m_X=m_{\text{FP}}$ for $X=M$. The helicity amplitudes are denoted by,
\begin{align}
    \mathcal{A}^{AB\to CD}_{h_A,h_B;h_C,h_D}(s,t,u),
\end{align}
where the helicities of each state are $h_X= \pm 2$ for $X=G$ and $h_X = \pm 2, \pm 1,0$ for $X=M$, and,
\begin{align}
    s=(p_A+p_B)^2, && t=(p_A-p_C)^2, && u=(p_A-p_D)^2,
\end{align}
are the Mandelstam variables, satisfying,
\begin{align}
\label{Mandelstam_identity}
    s+t+u=m_A^2+m_B^2+m_C^2+m_D^2.
\end{align}
For simplicity, the amplitudes are computed in the centre-of-mass frame, where the spatial momenta satisfy $\bar p_A+\bar p_B=\bar p_C+\bar p_D=0$, and the helicity of each external state is defined as its spin projection along its spatial momentum in this frame. The four-momenta can then be written as \cite{Sannan:1986tz},
\begin{subequations}
\label{CoM_momenta}
\begin{align}
    \label{Momenta_Incoming}
    p_A^\mu &= \big(E_A,0,0,p_{i}\big), \qquad &p_B^\mu &= \big(E_B,0,0,-p_{i}\big), \\
    \label{Outgoing_Momenta}
    p_C^\mu &= \big(E_C,p_{f}\sin\theta,0,p_{f}\cos\theta\big), \qquad &p_D^\mu &= \big(E_D,-p_{f}\sin\theta,0,-p_{f}\cos\theta\big).
\end{align}
\end{subequations}
Here and throughout, $\theta\in[0,\pi]$ denotes the angle between the incoming momentum $\bar p_A$ and the outgoing momentum $\bar p_C$,
\begin{align}
    \cos\theta=\frac{\bar p_A\cdot\bar p_C}{p_i p_f}.
\end{align}
Thus, $\theta=0$ corresponds to forward scattering from $A$ to $C$, while exchanging the two outgoing particles sends $\theta\to\pi-\theta$.

Introducing the Källén function,
\begin{align}
    \lambda(x,y,z)=x^2+y^2+z^2-2xy-2xz-2yz,
\end{align}
the magnitudes of the initial and final three-momenta are,
\begin{align}
\label{p_in_p_out}
    p_{i} &= \frac{\sqrt{\lambda(s,m_A^2,m_B^2)}}{2\sqrt{s}}, &
    p_{f} &= \frac{\sqrt{\lambda(s,m_C^2,m_D^2)}}{2\sqrt{s}},
\end{align}
while the scattering angle is given by,
\begin{align}
\label{CoM_angle}
    \cos\theta=\frac{s(t-u)+(m_A^2-m_B^2)(m_C^2-m_D^2)}{\sqrt{\lambda(s,m_A^2,m_B^2)\lambda(s,m_C^2,m_D^2)}}.
\end{align}
The corresponding energies are,
\begin{subequations} 
\label{E_A}
\begin{align}
    E_A &= \frac{s+m_A^2-m_B^2}{2\sqrt{s}}, & E_B &= \frac{s+m_B^2-m_A^2}{2\sqrt{s}}, \\
    E_C &= \frac{s+m_C^2-m_D^2}{2\sqrt{s}}, & E_D &= \frac{s+m_D^2-m_C^2}{2\sqrt{s}}.
\end{align}
\end{subequations}
For later convenience, whenever either the incoming or outgoing state consists of two massive particles, we denote their common centre-of-mass speed by,
\begin{align}
\label{CoM_speed}
    v=\frac{p_M}{E_M}
    =\frac{\sqrt{\lambda(s,m_{\mathrm{FP}}^2,m_{\mathrm{FP}}^2)}}{s}
    =\sqrt{1-\frac{4m_{\mathrm{FP}}^2}{s}}.
\end{align}
Thus, $0\leq v<1$ in the physical region, with $v=0$ at the two-particle threshold. The same definition applies whether the massive pair is incoming or outgoing.

Together, these relations show that all kinematic quantities entering the centre-of-mass calculation can be expressed in terms of the Lorentz-invariant Mandelstam variables and the external masses. The choice of centre-of-mass frame is therefore only a computational convenience, while the helicity labels refer to the corresponding centre-of-mass helicity basis. As we shall see, however, some amplitudes take a relatively compact form when expressed directly in terms of the appropriate centre-of-mass speed and the scattering angle.

\subsection{Polarisation tensors}

To evaluate the helicity amplitudes in this frame, we must specify the corresponding polarisation tensors for the external spin-2 states. For an on-shell momentum $p^\mu$, the massless and massive momentum-space fields can be decomposed in a basis of polarisation tensors $\epsilon^h_{\mu \nu}(p)$ with definite helicity $h$ as,
\begin{align}
    G_{\mu\nu}(p) &= \sum_{h=\pm2}G_h(p)\epsilon^{h}_{\mu\nu}(p), &
    M_{\mu\nu}(p) &= \sum_{h=-2}^{2}M_h(p)\epsilon^{h}_{\mu\nu}(p),
\end{align}
where the massless field has the two helicity states $h=\pm2$, while the massive field has the five helicity states $h=\pm2,\pm1,0$.

The spin-2 polarisation tensors $\epsilon^h_{\mu \nu}(p)$ can conveniently be constructed from polarisation vectors $\epsilon^\mu_\lambda(p)$, with $\lambda=\pm1,0$, according to \cite{Gleisberg:2003ue},
\begin{align}
\label{pol_decomp}
    \epsilon_0^{\mu\nu} &= \tfrac{1}{\sqrt{6}}\left[\epsilon_{+1}^{\mu}\epsilon_{-1}^{\nu}+\epsilon_{+1}^{\nu}\epsilon_{-1}^{\mu}-2\epsilon_0^{\mu}\epsilon_0^{\nu}\right], &
    \epsilon_{\pm1}^{\mu\nu} &= \tfrac{1}{\sqrt{2}}\left[\epsilon_{\pm1}^{\mu}\epsilon_0^{\nu}+\epsilon_0^{\mu}\epsilon_{\pm1}^{\nu}\right], &
    \epsilon_{\pm2}^{\mu\nu} &= \epsilon_{\pm1}^{\mu}\epsilon_{\pm1}^{\nu}.
\end{align}
With the momentum conventions in \eqref{CoM_momenta}, we can choose the polarisation vectors as \cite{Gleisberg:2003ue},
\begin{subequations}
\label{particle_pol}
\begin{align}
    \label{Particle_1_Polarisations}
    \epsilon_{\pm1}^{\mu}(\bar p_A) &= \frac{1}{\sqrt{2}}\big(0,1,\pm i,0\big), \qquad &\epsilon_0^\mu(\bar p_A) &= \frac{1}{m_A}\big(p_{i},0,0,E_A\big), \\
    \label{Particle_2_Polarisations}
    \epsilon_{\pm1}^{\mu}(\bar p_B) &= \frac{1}{\sqrt{2}}\big(0,-1,\pm i,0\big), \qquad &\epsilon_0^\mu(\bar p_B) &= \frac{1}{m_B}\big(p_{i},0,0,-E_B\big), \\
    \label{Particle_3_Polarisations}
    \epsilon_{\pm1}^{\mu}(\bar p_C) &= \frac{1}{\sqrt{2}}\big(0,\cos\theta,\pm i,-\sin\theta\big), \qquad &\epsilon_0^\mu(\bar p_C) &= \frac{1}{m_C}\big(p_{f},E_C\sin\theta,0,E_C\cos\theta\big), \\
    \label{Particle_4_Polarisations}
    \epsilon_{\pm1}^{\mu}(\bar p_D) &= \frac{1}{\sqrt{2}}\big(0,-\cos\theta,\pm i,\sin\theta\big), \qquad &\epsilon_0^\mu(\bar p_D) &= \frac{1}{m_D}\big(p_{f},-E_D\sin\theta,0,-E_D\cos\theta\big),
\end{align}
\end{subequations}
where the mass-dependent vectors $\epsilon^\mu_0$ do not appear in the massless states $G$.

The resulting spin-2 polarisation tensors are transverse and traceless,
\begin{align}
    \label{Transversality_and_Tracelessness}
    p_\mu\epsilon_h^{\mu\nu}(p)=0, && \eta_{\mu\nu}\epsilon_h^{\mu\nu}(p)=0.
\end{align}
For the massive field, these conditions follow directly from the Fierz--Pauli equations of motion \eqref{M_EoM}, while for the massless field, the de Donder gauge condition \eqref{deDonder} gives,
\begin{align}
    p_\mu\epsilon^{\mu\nu}(p)-\tfrac{1}{2}p^\nu\epsilon^\mu_{~\mu}(p)=0.
\end{align}
The residual gauge freedom preserving the de Donder condition can then be used to impose $\epsilon^\mu_{~\mu}=0$, after which transversality follows. This leaves precisely the two physical helicities $h=\pm2$.

The same residual gauge freedom implies that amplitudes containing an external massless spin-2 particle must be invariant under \cite{weinberg_photons_1964, kugo_massless_1981},
\begin{align}
    \label{Mass_Shell_Gauge_Invariance}
    \epsilon_{\mu\nu}(p)\to\epsilon_{\mu\nu}(p)+p_\mu\xi_\nu(p)+p_\nu\xi_\mu(p),
\end{align}
for an on-shell gauge parameter $\xi_\mu(p)$. This on-shell gauge invariance provides a useful consistency check on the scattering amplitudes containing $G$.

\subsection{Propagators}

The quadratic action \eqref{Expansion_of_Bimetric_In_G_and_M} also determines the propagators of the two spin-2 fields. Inverting the quadratic Fierz--Pauli operator gives the massive propagator \cite{vanDam:1970vg,Hinterbichler:2011tt},
\begin{align}
    \label{Massive_Graviton_Propagator}
    D^{(M)}_{\mu\nu\rho\sigma}(p) &= \frac{1}{p^2-m_{\mathrm{FP}}^2}\Big[\tfrac{1}{2}\left(\Pi_{\mu\rho}\Pi_{\nu\sigma}+\Pi_{\mu\sigma}\Pi_{\nu\rho}\right)-\tfrac{1}{3}\Pi_{\mu\nu}\Pi_{\rho\sigma}\Big], &
    \Pi_{\mu\nu}(p) &= \eta_{\mu\nu}-\frac{p_\mu p_\nu}{m_{\mathrm{FP}}^2}.
\end{align}
For the massless field, the quadratic kinetic operator is not invertible before fixing the gauge. Imposing the de Donder condition \eqref{deDonder}, its inverse is \cite{vanDam:1970vg,Berends:1974gk,Hinterbichler:2011tt},
\begin{align}
    \label{Massless_Graviton_Propagator}
    D^{(G)}_{\gamma\delta\mu\nu}(p)=\frac{1}{2p^2}\Big[\eta_{\gamma\mu}\eta_{\delta\nu}+\eta_{\gamma\nu}\eta_{\delta\mu}-\eta_{\gamma\delta}\eta_{\mu\nu}\Big].
\end{align}
A useful difference between the two propagators is seen in the limit $m_{\mathrm{FP}}\to0$. Consider the tree-level exchange between two conserved matter sources $T_{\mu\nu}$ and $T'_{\mu\nu}$, mediated either by the massive field $M$ or the massless field $G$. Contracting the sources with the corresponding propagators, the terms proportional to $p_\mu p_\nu/m_{\text{FP}}^2$ in the massive propagator vanish by $p_\mu T^{\mu\nu}=p_\mu T'^{\mu\nu}=0$. The numerator therefore has a finite $m_{\text{FP}}\to0$ limit, but differs from that of the massless propagator, giving the two exchange amplitudes,
\begin{align}
    \mathcal{A}_{M} &\propto T_{\mu\nu}T'^{\mu\nu}-\tfrac{1}{3}TT', &
    \mathcal{A}_{G} &\propto T_{\mu\nu}T'^{\mu\nu}-\tfrac{1}{2}TT'.
\end{align}
Hence, the massive propagator does not continuously approach the massless one as $m_{\mathrm{FP}}\to0$. This is the van Dam--Veltman--Zakharov (vDVZ) discontinuity \cite{vanDam:1970vg,Zakharov:1970cc}, which originates from the helicity-$0$ mode of the massive spin-2 field remaining coupled in the massless limit.

It is important, however, that the general relativity limit of bimetric theory, $q\to0$ for finite $c$, is not the same as simply taking $m_{\mathrm{FP}}\to0$ at fixed coupling. In the limit \eqref{GR_Limit}, the massless mode aligns with $\delta g_{\mu\nu}$, while the massive mode decouples from the metric $g_{\mu\nu}$ coupled to matter. Thus, although $g_{\mu\nu}$ contains an admixture of the massive mode at finite $q$, the helicity-$0$ contribution disappears together with the remaining massive helicities as $q\to0$. 

\subsection{Interaction vertices and tree-level diagrams}

Having specified the external polarisation states and the propagators, it remains to identify the interaction vertices in the amplitudes. Since we are restricting ourselves to $1 \to 2$ and $2 \to 2$ processes at tree level, only the cubic and quartic terms in the perturbative expansion are required. 

The non-vanishing cubic interactions are of the form,
\begin{align}
\label{cubic_vertex}
    GGG, \qquad GMM, \qquad MMM,
\end{align}
while the quartic interactions are,
\begin{align}
\label{quartic_vertex}
    GGGG, \qquad GMMM, \qquad GGMM, \qquad MMMM.      
\end{align}
Their explicit tensor expressions are lengthy and will not be displayed here, but are presented in the accompanying material \cite{FelixHerber_TreeLevelBimetricCalculations_2025} (they can also be found in Appendix D of \cite{Herber2006232}).

As we saw in Section \ref{subsec:parameter_structure}, an important feature of the expansion is that there are no terms with precisely one massive field $M_{\mu \nu}$. Thus, for example, all terms of the form $GGM$ and $GGGM$ are absent, implying that a single massive spin-2 particle cannot decay purely into massless spin-2 particles through the bimetric self-interactions at tree level. Note, however, that when matter fields are coupled to, for example, the metric $g_{\mu\nu}$, both $G$ and $M$ couple to the matter sector, allowing the massive field to decay into kinematically accessible matter fields \cite{Babichev:2016hir}.

\begin{figure}
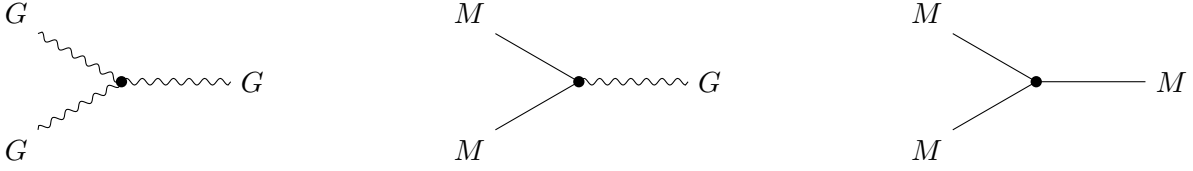

    \centering
    \ThreeVertex{G}{G}{G}
    \hspace{2cm}
    \ThreeVertex{M}{M}{G}
    \hspace{2cm}
    \ThreeVertex{M}{M}{M}
    \caption{Three-point vertices of bimetric theory. Wavy lines denote the massless spin-2 field $G$, while straight lines denote the massive spin-2 field $M$.}
    \label{fig:bimetric_3pt_vertices}
\end{figure}

At tree level, the $1\to2$ amplitudes follow directly from the cubic vertices in \eqref{cubic_vertex} represented by Figure \ref{fig:bimetric_3pt_vertices}. For real on-shell momenta in flat spacetime, however, these do not describe nontrivial decays among the spin-2 states. In particular, the decay $M\to GG$ is absent because there is no $GGM$ vertex, while decays involving a massive particle in the final state are kinematically forbidden for particles of equal mass. The three-point amplitudes nevertheless enter as building blocks of the four-point amplitudes.

\begin{figure}
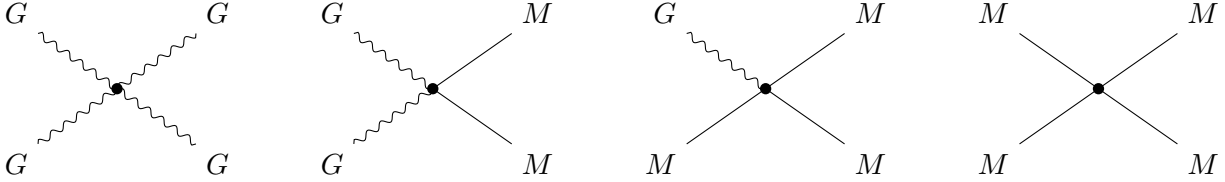

    \centering
    \FourVertex{G}{G}{G}{G}
    \hspace{0.7cm}
    \FourVertex{G}{G}{M}{M}
    \hspace{0.7cm}
    \FourVertex{G}{M}{M}{M}
    \hspace{0.7cm}
    \FourVertex{M}{M}{M}{M}
    \caption{Four-point vertices of bimetric theory.}
    \label{fig:bimetric_4pt_vertices}
\end{figure}

The tree-level $2\to2$ amplitudes receive contributions from the quartic interaction vertices in \eqref{quartic_vertex}, see Figure \ref{fig:bimetric_4pt_vertices}, as well as from exchange diagrams involving the cubic interactions and the propagators given above. Which internal fields can appear is therefore determined by the available cubic vertices. For $GG\to GG$, only $G$ exchange occurs, while for $MM\to MM$ both $G$ and $M$ exchange diagrams contribute. Mixed processes contain the corresponding allowed $G$ and $M$ exchanges, with the distinction between the $t$- and $u$-channels depending on the ordering of the distinct external particles. The complete set of $2 \to 2$ tree-level diagrams is shown in Figure \ref{fig:Bimetric_Scattering_Sectors}.

\begin{figure}
    \centering

    \SChannel{G}{G}{G}{G}{G}
    \hspace{0.7cm}
    \TChannel{G}{G}{G}{G}{G}
    \hspace{0.7cm}
    \UChannel{G}{G}{G}{G}{G}
    \hspace{0.7cm}
    \ContactDiagram{G}{G}{G}{G}

    \par\vspace{0.6cm}
    a) $GG\to GG$

    \vspace{1cm}

    \SChannel{G}{G}{M}{M}{G}
    \hspace{0.7cm}
    \TChannel{G}{G}{M}{M}{M}
    \hspace{0.7cm}
    \UChannel{G}{G}{M}{M}{M}
    \hspace{0.7cm}
    \ContactDiagram{G}{G}{M}{M}

    \par\vspace{0.6cm}
    b) $GG\to MM$

    \vspace{1cm}

    \SChannel{G}{M}{G}{M}{M}
    \hspace{0.7cm}
    \UChannel{G}{M}{G}{M}{M}
    \hspace{0.7cm}
    \TChannel{G}{M}{G}{M}{G}
    \hspace{0.7cm}
    \ContactDiagram{G}{M}{G}{M}

    \par\vspace{0.6cm}
    c) $GM\to GM$
    
    \vspace{1cm}

    \SChannel{G}{M}{M}{M}{M}
    \hspace{0.7cm}
    \TChannel{G}{M}{M}{M}{M}
    \hspace{0.7cm}
    \UChannel{G}{M}{M}{M}{M}
    \hspace{0.7cm}
    \ContactDiagram{G}{M}{M}{M}

    \par\vspace{0.6cm}
    d) $GM\to MM$

    \vspace{1cm}

    \SChannel{M}{M}{M}{M}{G}
    \hspace{0.7cm}
    \SChannel{M}{M}{M}{M}{M}
    \hspace{0.7cm}
    \TChannel{M}{M}{M}{M}{G}
    \hspace{0.7cm}
    \TChannel{M}{M}{M}{M}{M}

    \par\vspace{0.5cm}

    \UChannel{M}{M}{M}{M}{G}
    \hspace{0.7cm}
    \UChannel{M}{M}{M}{M}{M}
    \hspace{0.7cm}
    \ContactDiagram{M}{M}{M}{M}

    \par\vspace{0.6cm}
    e) $MM\to MM$

    \caption{Tree-level diagrams contributing to the five $2\to2$ scattering sectors of bimetric theory. Wavy lines denote the massless spin-2 field $G$, while straight lines denote the massive spin-2 field $M$.}
    \label{fig:Bimetric_Scattering_Sectors}
\end{figure}

\subsection{Structure and symmetry simplifications}
\label{sec:Symmetry_Simplifications}

Before presenting the complete set of $2\to2$ tree-level helicity amplitudes, we first identify some structures and symmetries that reduce the number of independent amplitudes. Let us denote the set of potentially non-vanishing $2\to2$ helicity amplitudes in bimetric theory by $\Acal^{2\to2}$ and determine how many are inequivalent under the symmetries of the theory.

For a generic amplitude $\Acal_{h_A,h_B;h_C,h_D}^{AB\to CD}$, each external state can be one of 7 particle-helicity states, 2 for $G$ and 5 for $M$, giving naively $7^4=2401$ ordered $2\to2$ helicity configurations. However, as we saw in Section \ref{subsec:parameter_structure}, amplitudes containing precisely one massive field vanish. There are $\binom{4}{1}$ choices for the massive external leg, five choices for its helicity, and two choices for each of the remaining three massless helicities, giving,
\begin{align}
\label{nr_MGGG}
    \binom{4}{1}\times5\times2^3=160.
\end{align}
Removing these leaves,
\begin{align}
    |\Acal^{2\to2}|=2401-160=2241,
\end{align}
potentially non-vanishing amplitudes. We will now see that the symmetries of the action and the Minkowski background further reduce the number of independent amplitudes considerably.

The bimetric action \eqref{bim_Action} and the Minkowski background are invariant under time reversal,
\begin{align}
    x^\mu=(t,x^i)\to(\tau x)^\mu=(-t,x^i).
\end{align}
Time reversal therefore relates amplitudes with the initial and final states interchanged. For the polarisation conventions \eqref{particle_pol}, this relation takes the form,\footnote{Time reversal implies microreversibility of the $S$-matrix element, $\braket{f|S|i}=\braket{\tau i|S|\tau f}$. In our conventions, $\tau a_h(\bar p)\tau^{-1}=a_h(-\bar p)$ and $\tau a_h^\dagger(\bar p)\tau^{-1}=a_h^\dagger(-\bar p)$, so time reversal reverses all three-momenta while preserving helicity. To return the time-reversed process to the standard centre-of-mass frame configuration used to define our amplitudes, we rotate the momenta back to their standard orientation and apply the same rotations to the polarisation tensors. The phases acquired by the polarisation tensors under these rotations give the phase relating the two helicity amplitudes. In particular, for the final rotation $R$ by $\pi$ about the $z$-axis, our conventions give $R^\mu{}_\alpha R^\nu{}_\beta\epsilon_h^{\alpha\beta}(R\bar p)=(-1)^h\epsilon_h^{\mu\nu}(\bar p)$.}
\begin{align}
\label{Time_Reversal}
    \mathcal{A}_{h_A,h_B;h_C,h_D}^{AB\to CD}(s,t,u) = \zeta\mathcal{A}_{h_C,h_D;h_A,h_B}^{CD\to AB}(s,t,u),
\end{align}
where $\zeta = (-1)^{h_A+h_B+h_C+h_D}$. 

Furthermore, the bimetric action \eqref{bim_Action} and the Minkowski background are invariant under parity,
\begin{align}
    x^\mu=(t,x^i)\to(\pi x)^\mu=(t,-x^i).
\end{align}
Parity transformations therefore relate amplitudes with all helicities reversed \cite{Jacob:1959at}. For the polarisation conventions \eqref{particle_pol}, we find,\footnote{Under parity, our polarisation vectors satisfy $(\pi\epsilon(\bar p))^\mu_h=\epsilon^\mu_{-h}(-\bar p)$. The parity-transformed momenta are returned to the standard centre-of-mass frame configuration by a rotation by $\pi$ in the scattering plane, represented by $R=\operatorname{diag}(1,-1,1,-1)$, for which $R^\mu{}_\nu p_X^\nu=(\pi p_X)^\mu$ and $R^\mu{}_\nu\epsilon_h^\nu(\bar p_X)=\epsilon_h^\mu(-\bar p_X)$ for $X=A,B,C,D$. With these conventions no additional phase is generated.}
\begin{align}
    \label{Parity_Bimetric}
    \mathcal{A}^{AB\to CD}_{-h_A,-h_B;-h_C,-h_D}(s,t,u)
    =\mathcal{A}^{AB\to CD}_{h_A,h_B;h_C,h_D}(s,t,u).
\end{align}
In general, parity may introduce a spin- and intrinsic-parity-dependent phase \cite{Jacob:1959at}, but for the polarisation conventions \eqref{particle_pol} this phase is unity.

Lastly, since all external spin-2 states are bosonic, exchanging the two incoming or the two outgoing external states leaves the amplitude unchanged, up to the corresponding interchange $t\leftrightarrow u$ and phase factor $\zeta= (-1)^{h_A+ h_B+ h_C+ h_D}$.
\begin{align}
\label{Swap_Symmetry}
    \mathcal{A}^{AB \to CD}_{h_A, h_B; h_C, h_D}(s,t,u) = \zeta\mathcal{A}^{BA \to CD}_{h_B, h_A; h_C, h_D}(s,u,t) = \zeta\mathcal{A}^{AB \to DC}_{h_A, h_B; h_D, h_C}(s,u,t) = \mathcal{A}^{BA \to DC}_{h_B, h_A; h_D, h_C}(s,t,u).
\end{align}
The time-reversal, parity, and particle-exchange relations \eqref{Time_Reversal}, \eqref{Parity_Bimetric}, and \eqref{Swap_Symmetry} generate a finite group $\Gcal$ acting on $\Acal^{2\to2}$. Amplitudes related by this action belong to the same orbit in $\Acal^{2\to2}/\Gcal$.

The group $\Gcal$ is generated by $s_i$, $s_f$, $\pi$, and $\tau$, corresponding respectively to the exchange of the initial states, the exchange of the final states, parity, and time reversal. Since $\pi$ reverses all helicities \eqref{Parity_Bimetric}, it commutes with $s_i$, $s_f$, and $\tau$, and together with $\pi^2=e$ generates a central subgroup $\mathbb{Z}_2$. The subgroup generated by $s_i$, $s_f$, and $\tau$ is isomorphic to the dihedral group $D_8$ of order eight, describing the symmetries of a square, and does not contain $\pi$. Hence,
\begin{align}
    \Gcal\simeq D_8\times\mathbb{Z}_2,
\end{align}
which has order $16$.

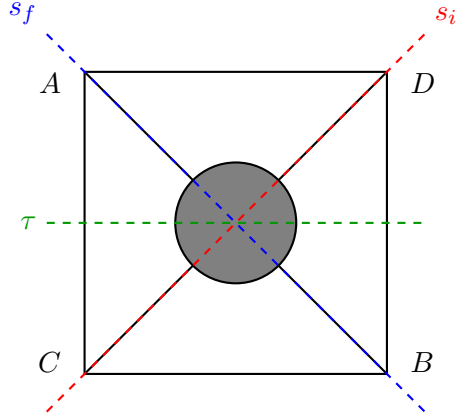
\begin{figure}
    \centering
    \begin{tikzpicture}[scale=1]

        \coordinate (A) at (0,4);
        \coordinate (D) at (4,4);
        \coordinate (B) at (4,0);
        \coordinate (C) at (0,0);
        \coordinate (V) at (2,2);

        \draw[thick] (A) -- (D) -- (B) -- (C) -- cycle;

        \node[above left, yshift=-12pt, xshift=-5pt]  at (A) {$A$};
        \node[above right, yshift=-12pt, xshift=5pt] at (D) {$D$};
        \node[below right, yshift=12pt, xshift=5pt]  at (B) {$B$};
        \node[below left, yshift=12pt, xshift=-5pt]   at (C) {$C$};

        \draw[thick] (A) -- (V);
        \draw[thick] (B) -- (V);
        \draw[thick] (C) -- (V);
        \draw[thick] (D) -- (V);

        \filldraw[fill=gray, thick] (V) circle (0.8);

        \draw[dashed, blue, thick] (-0.5,4.5) -- (4.5,-0.5);
        \node[blue, above left] at (-0.5,4.5) {$s_f$};

        \draw[dashed, red, thick] (-0.5,-0.5) -- (4.5,4.5);
        \node[red, above right] at (4.5,4.5) {$s_i$};

        \draw[dashed, green!60!black, thick] (-0.5,2) -- (4.5,2);
        \node[green!60!black, left] at (-0.5,2) {$\tau$};

    \end{tikzpicture}
    \caption{A generic $2\to2$ scattering process arranged on a square, illustrating the action of the dihedral group $D_8=\langle s_i, s_f, \tau\rangle$ on the amplitudes. The scattering process should still be read as $A,B$ incoming and $C,D$ outgoing. The ordering of the labels is chosen so that $s_i$, $s_f$, and $\tau$ act as simple reflections of the square.}
    \label{fig:Dihedral_Group_Square}
\end{figure}

The dihedral structure is illustrated in Figure \ref{fig:Dihedral_Group_Square}, where each particle-helicity pair is regarded as a single external-state label and the states are arranged so that $s_i$, $s_f$, and $\tau$ act as reflections of the square. Acting with $\Gcal$ partitions the full set of amplitudes into $199$ orbits, or symmetry-inequivalent amplitudes,
\begin{align}
    \left|\Acal^{2\to2}/\Gcal\right|=199.
\end{align}
This result is derived in Appendix \ref{app:amplitude_counting} using Burnside's lemma, by counting the amplitudes left fixed by each element of $\Gcal$. It is therefore sufficient to compute one representative from each orbit. These representatives are grouped according to their external particle content in Tables \ref{tab:Amps_GGGG}--\ref{tab:HighEnergyMMMM}.

\subsection{Computation of tree-level scattering amplitudes}

The 199 symmetry-inequivalent scattering amplitudes of bimetric theory were computed using \textsc{Mathematica} \cite{mathematica}, together with the packages \textsc{xAct} \cite{xact,xcoba,xtensor,xperm,xpert,texact}, \textsc{FeynRules} \cite{Alloul:2013bka}, \textsc{FeynArts} \cite{Hahn:2000kx}, and \textsc{FeynCalc} \cite{Shtabovenko:2020gxv}. While these packages provide the basic tools for generating vertices, diagrams, and tensor contractions, they are not directly adapted to the spin-2 polarisation structure and kinematic conventions required here. The calculation therefore required a tailored symbolic workflow in which the standard package output was reorganised before contraction with the external polarisation tensors and subsequently simplified using the on-shell, transversality and traceless conditions \eqref{Transversality_and_Tracelessness}, together with the centre-of-mass forms \eqref{particle_pol}. This is particularly important because the intermediate expressions rapidly become too large and algebraically complicated for direct application of general-purpose simplification routines, requiring the relevant kinematic structure to be imposed at intermediate stages of the calculation.

We now briefly outline the pipeline used to compute the scattering amplitudes, from the expansion of the bimetric action to fourth order in \textsc{xAct} to the final evaluation and simplification of the amplitudes using \textsc{FeynCalc}.
\begin{enumerate}
    \item \textbf{Expansion in} \textsc{xAct}:\\
    The bimetric action was expanded to fourth order about the proportional Minkowski backgrounds in terms of $\delta g$ and $\delta f$, and subsequently rewritten in terms of the mass eigenstates $G$ and $M$. The resulting Lagrangian was then converted to a form suitable for input into \textsc{FeynRules}.\footnote{At this stage, several consistency checks were performed on the perturbative expansion, including verification of the symmetric-polynomial identity \eqref{Symmetric_Polynomial_Identity} and of the defining relation $S^\mu{}_\nu S^\nu{}_\rho=g^{\mu\alpha}f_{\alpha\rho}$ at each order retained in the expansion.} 

    \item \textbf{Model generation with} \textsc{FeynRules}:\\
    The \textsc{FeynRules} Lagrangian, together with the definitions of the fields $G$ and $M$, was used to generate a \textsc{FeynArts} model file through \textsc{FeynRules}.

    \item \textbf{Massless spin-2 propagator}:\\
    The \textsc{FeynArts} model contains only the massive spin-2 propagator by default. However, the massless propagator \eqref{Massless_Graviton_Propagator} cannot be obtained from the massive propagator \eqref{Massive_Graviton_Propagator} by simply taking $m_{\mathrm{FP}}\to0$, as reflected by the vDVZ discontinuity \cite{vanDam:1970vg,Zakharov:1970cc}. The propagator numerator was therefore replaced by \texttt{FAPropagatorNumerator}, following the procedure described in the accompanying \texttt{README} file \cite{FelixHerber_TreeLevelBimetricCalculations_2025}. The resulting substitution yields \eqref{Massless_Graviton_Propagator} for internal $G$ lines and \eqref{Massive_Graviton_Propagator} for internal $M$ lines.

    \item \textbf{Computations in} \textsc{FeynCalc}:\\
    The tree-level amplitudes were evaluated in \textsc{FeynCalc}. Because the bimetric vertices contain hundreds to thousands of terms, direct contractions with propagators and spin-2 polarisation tensors can become computationally expensive, especially for amplitudes involving helicity-0 massive modes.

    To reduce the intermediate expressions, the spin-2 polarisation tensors are first decomposed into outer products of auxiliary polarisation vectors according to \eqref{pol_decomp}. The contractions are then carried out while imposing transversality and tracelessness, before substituting the explicit polarisation vectors for the physical helicity states.

    A separate simplification is required for expressions containing square roots, e.g. $\sqrt{s-4m_{\text{FP}}^2}$, for which the standard \textsc{Mathematica} routines retain general branch information. These terms are therefore simplified under the appropriate physical assumptions on the Mandelstam variables and masses using the procedure described in the accompanying \texttt{README} file.
    
    \item \textbf{Computations with HPC}:\\
    Even after the simplifications described above, the amplitude calculations remain computationally demanding. The workflow was therefore parallelised and run on an HPC cluster, allowing all $199$ tree-level scattering amplitudes listed in Appendix~\ref{sec:List_of_Orbits} to be computed.
\end{enumerate}

\subsection{High-energy limit}
\label{sec:high_energy}
The high-energy behaviour of an effective field theory is crucial for determining the regime in which its predictions remain reliable. One restriction was already encountered at the end of Section~\ref{sec:mass_eigen}, where the perturbative expansion of the Einstein--Hilbert terms in the mass eigenstates was found to be controlled by,
\begin{align}
    \Lambda_{\mathrm{EH}}\sim\min(m_g,cm_f)=\frac{\mpl}{\sqrt{1+q^2}}\min(1,q).
\end{align}
The expansion therefore requires $E\ll\Lambda_{\mathrm{EH}}$. In particular, $\Lambda_{\mathrm{EH}}\sim q\mpl$ for $q\ll1$ and $\Lambda_{\mathrm{EH}}\sim\mpl/q$ for $q\gg1$, such that the perturbative scale is lowered in terms of the mass eigenstates in the extreme mixing limits. Massive gravity and bimetric theory possess an additional restriction, often associated with, but not restricted to, the helicity-0 modes of the massive spin-2 field. As we will see, their high-energy enhancement can lead to a strong-coupling scale parametrically well below $\Lambda_{\mathrm{EH}}$. The regime in which the perturbative amplitude calculation is reliable is therefore ultimately determined by the smaller of these two scales.

The origin of the high-energy scaling can be understood heuristically from the external states and propagators entering the amplitudes. The states acquire their energy dependence through the polarisation tensors, and in particular the helicity-0 mode grows most rapidly with energy. From \eqref{particle_pol},
\begin{align}
    \epsilon_{0}^{\mu \nu} \sim \frac{p^\mu p^\nu}{\mFP^2}
    \qquad \Longrightarrow \qquad
    \epsilon_{0}^{\mu \nu} \sim \frac{E^2}{\mFP^2},
    \qquad E\gg \mFP.
\end{align}
By comparison, the helicity-1 and helicity-2 polarisations scale as,
\begin{align}
    \epsilon_{\pm1}^{\mu \nu} \sim \frac{E}{\mFP},
    &&
    \epsilon_{\pm2}^{\mu \nu} \sim 1.
\end{align}
A massless spin-2 propagator introduces no analogous enhancement, since its physical propagating modes are purely helicity-2 and the propagator scales as $E^{-2}$. We therefore focus on the massive propagator \eqref{Massive_Graviton_Propagator}, which contains terms with additional inverse powers of $\mFP$. The most singular term behaves schematically as,
\begin{align}
    \frac{1}{p^2-\mFP^2}\frac{p_\mu p_\nu p_\rho p_\sigma}{\mFP^4}
    \qquad \Longrightarrow \qquad
    \frac{E^2}{\mFP^4},
    \qquad E\gg \mFP .
\end{align}
Including the two cubic Einstein--Hilbert vertices, each scaling as $E^2/\mpl$, a naive power counting of an $MM\to MM$ exchange diagram with an internal massive spin-2 field then gives \cite{Aubert:2003je},
\begin{align}
    \mathcal{A}^{MM\to MM}
    \sim
    \left(\frac{E^2}{\mFP^2}\right)^4
    \left(\frac{E^2}{\mpl}\right)^2
    \left(\frac{E^2}{\mFP^4}\right)
    =
    \frac{E^{14}}{\mpl^2\mFP^{12}}.
\end{align}
Introducing the conventional hierarchy of scales $\Lambda_k=\left(\mpl^{\,}\mFP^{k-1}\right)^{1/k}$, this naive estimate would correspond to,
\begin{align}
    \Acal^{MM\to MM}\sim\left(\frac{E}{\Lambda_7}\right)^{14}.
\end{align}
Such a low scale would be problematic, since the perturbative expansion would cease to be reliable already at energies of order $\Lambda_7$, far below the Planck scale when $\mFP\ll\mpl$ and, away from the extreme mixing limits, also below the Einstein--Hilbert scale $\Lambda_{\mathrm{EH}}$. This estimate, however, substantially overstates the high-energy growth even of a generic theory with a massive graviton, which has been argued to have a strong-coupling scale no worse than $\Lambda_5 = (\mpl \mFP^4)^{1/5}$ \cite{Arkani-Hamed:2002bjr,Aubert:2003je,Bonifacio:2018aon}. The actual scaling must, however, be determined from the complete amplitudes, where cancellations between the different contributions can improve the high-energy behaviour.

To extract this behaviour systematically, one can use the following procedure:
\begin{enumerate}[label=\roman*)]
    \item Use $u =\sum m_X^2 -s-t $ \eqref{Mandelstam_identity} to eliminate $u$ in the amplitudes in favour of $s$, $t$, and the particle masses.
    
    \item Introduce the high-energy scaling,
    \begin{align}
        s \to \frac{s}{\epsilon^2},
        \qquad
        t \to \frac{t}{\epsilon^2},
    \end{align}
    such that $\epsilon^{-1}$ scales with the characteristic energy $E$.
    
    \item Expand the resulting amplitude for $\epsilon\to0^+$. If the leading term scales as $\epsilon^{-n}$, then,
    \begin{align}
        \mathcal{A}(\epsilon)
        =
        \epsilon^{-n}\mathcal{A}_{\mathrm{HE}}
        +\mathcal{O}\!\left(\epsilon^{-n+1}\right),
        \qquad
        \mathcal{A}_{\mathrm{HE}}
        =
        \lim_{\epsilon\to0^+}\epsilon^n\mathcal{A}(\epsilon),
    \end{align}
    corresponding to the high-energy behaviour $\mathcal{A}\sim E^n$.
    
    \item Finally, use $s+t+u\approx0$ to simplify $\mathcal{A}_{\mathrm{HE}}$ in the high-energy limit and extract the corresponding coefficients.
    
\end{enumerate}
As we will see in Section~\ref{sec:strong_coupling}, the amplitudes are softer than the generic $\Lambda_5$ estimate, resulting in a still higher strong-coupling scale.

\section{Results}
\label{sec:results}

The main results of this paper are all tree-level $2\to2$ scattering amplitudes of bimetric theory. As described above, these correspond to $199$ symmetry-inequivalent amplitudes. Most of the amplitudes are rather lengthy, and their direct inspection does not by itself provide much insight into their structure. We therefore refrain from presenting them all explicitly here. Their complete symbolic expressions are available in \textsc{Mathematica} format, as well as in plain-text \texttt{.wl} files, in the accompanying repository \cite{FelixHerber_TreeLevelBimetricCalculations_2025}. In the remainder of this section, we instead discuss the overall structure of the amplitudes, present some explicitly where they have a compact analytic form, compare our results with previous literature, consider several limits, and analyse the high-energy behaviour of the theory and the associated strong-coupling scale.

Before turning to the amplitudes, it is useful to show how the $199$ amplitude representatives are distributed among the different scattering sectors. Here, a scattering sector denotes a fixed assignment of the massless and massive spin-2 fields to the initial and final states. The number of representatives in each sector is summarised in Table~\ref{tab:Amplitude_Representatives}. The table also emphasises the results discussed above: the purely massless sector reproduces general relativity, while amplitudes involving two massive fields probe the universal interactions of a non-self-interacting Fierz--Pauli massive spin-2 field coupled to gravity. Amplitudes involving three or four massive fields, by contrast, begin to probe the additional interaction structure specific to ghost-free bimetric theory.

Several of the representative amplitudes vanish identically at tree level, for generic kinematics and generic values of the theory parameters. Interestingly, all such amplitudes occur within the first two rows of Table~\ref{tab:Amplitude_Representatives}. We will see, however, that there also exists a particularly interesting parameter point at which the entire $GMMM$ sector vanishes.

\begin{table}
    \centering
    \begin{tabularx}{\linewidth}{lr>{\raggedright\arraybackslash}X}
        \toprule
        Sector & Representatives & Structure probed \\
        \midrule
        $GGGG$ & 4 & General relativity \\
        $GGMM$ & 24 & Universal gravitational coupling of a Fierz--Pauli field \\
        $GMGM$ & 30 & Universal gravitational coupling of a Fierz--Pauli field \\
        $GMMM$ & 75 & First non-universal bimetric structure: $q$ and $\beta_-$, but no $\beta_+$ \\
        $MMMM$ & 66 & Full nonlinear bimetric structure: $q$, $\beta_-$ and $\beta_+$ \\
        \midrule
        Total & 199 & \\
        \bottomrule
    \end{tabularx}
    \caption{Number of representative helicity amplitudes and the interaction structure probed by each scattering sector. The parameter dependence shown in the final two rows is additional to the universal dependence fixed by $m_{\mathrm{Pl}}$ and $m_{\mathrm{FP}}$.}
    \label{tab:Amplitude_Representatives}
\end{table}

\subsection{\texorpdfstring{$GGGG$}{GGGG} amplitudes}

The purely massless sector coincides with general relativity at tree level. Indeed, a massive exchange contribution would require a vertex containing a single massive field, such as $GGM$. As shown in section~\ref{subsec:parameter_structure}, however, all such vertices vanish. More generally, since every vertex involving the massive field contains at least two massive legs, any internal massive lines in a process with only external massless gravitons necessarily form a closed loop. The massive mode therefore cannot contribute to a tree-level $GG\to GG$ diagram. The remaining vertices are precisely those obtained from the Einstein--Hilbert action for $G_{\mu\nu}$, and the amplitude agrees with the standard result from general relativity \cite{Berends:1974gk,Grisaru:1975bx,Sannan:1986tz}.

The four symmetry-inequivalent representatives are,
\begin{subequations}
\label{Massless_Spin_2_Amplitudes}
\begin{align}
    \mathcal{A}^{GG\to GG}_{2,2;2,2}(s,t,u)
    &=\frac{1}{2m_{\mathrm{Pl}}^2}\frac{s^3}{tu},\\
    \mathcal{A}^{GG\to GG}_{2,2;2,-2}(s,t,u)
    &=0,\\[2mm]
    \mathcal{A}^{GG\to GG}_{2,2;-2,-2}(s,t,u)
    &=0,\\
    \mathcal{A}^{GG\to GG}_{2,-2;2,-2}(s,t,u)
    &=\frac{1}{2m_{\mathrm{Pl}}^2}\frac{u^3}{st}.
\end{align}
\end{subequations}
Using parity, time reversal, and particle exchange, these results can be written as,
\begin{align}
    \mathcal{A}^{GG\to GG}_{h_A,h_B;h_C,h_D}(s,t,u)
    =\frac{1}{2m_{\mathrm{Pl}}^2}
    \begin{cases}
        \dfrac{s^3}{tu},
        & h_A=h_B=h_C=h_D,\\[2mm]
        \dfrac{u^3}{st},
        & h_A=-h_B,\quad h_C=-h_D,\quad h_A=h_C,\\[2mm]
        \dfrac{t^3}{su},
        & h_A=-h_B,\quad h_C=-h_D,\quad h_A=h_D,\\[2mm]
        0,
        & \text{otherwise},
    \end{cases}
\end{align}
where all helicities take the values $h_X=\pm2$.

This equivalence with general relativity holds only at tree level. At one loop, the massive spin-2 field can circulate through vertices of the form $G^nM^2$, producing corrections that depend on $m_{\mathrm{FP}}$. For diagrams constructed solely from the physical spin-2 fields, these corrections remain independent of $q$ and $\beta_\pm$ at fixed $m_{\mathrm{Pl}}$ and $m_{\mathrm{FP}}$, while vertices containing three or more massive fields can first contribute to a purely massless amplitude at two loops.\footnote{A complete loop-level quantisation must also incorporate the second-class constraints that eliminate the Boulware--Deser mode. In the phase-space path integral, these constraints generate the Senjanovi\'c determinant $\sqrt{\det\{\chi_a,\chi_b\}}$, which can be represented by auxiliary ghost fields \cite{Senjanovic:1976br}. These fields contribute only through closed loops and can therefore be omitted at tree level. The stated loop-level hierarchy refers only to diagrams constructed from $G_{\mu\nu}$ and $M_{\mu\nu}$.} 

\subsection{\texorpdfstring{$GGMM$}{GGMM} and \texorpdfstring{$GMGM$}{GMGM} amplitudes}

At tree level, all amplitudes in this sector are constructed from vertices containing at most two massive fields. These vertices coincide with those obtained by coupling the quadratic Fierz--Pauli action covariantly to Einstein gravity. Consequently, these amplitudes do not probe the independent nonlinear structure of the ghost-free bimetric potential, and depend only on $m_{\mathrm{Pl}}$, $m_{\mathrm{FP}}$, and the Mandelstam variables. In contrast to processes involving three or four massive fields, their expressions are of manageable length and can be written in relatively compact form.

Among the $24$ $GGMM$ representatives, $10$ amplitudes vanish identically. They are,
\begin{align}
    \Acal^{GG \to MM}_{2,2;h_C,h_D}(s,t,u) &= 0, \qquad h_C>h_D.
\end{align}
Using parity, particle exchange, and time reversal, this extends to,
\begin{subequations}
\begin{align}
    \Acal^{GG \to MM}_{h,h;h_C,h_D}(s,t,u) &= 0, \qquad h_C\neq h_D, \\
    \Acal^{MM \to GG}_{h_A,h_B;h,h}(s,t,u) &= 0, \qquad h_A\neq h_B,
\end{align}
\end{subequations}
where $h=\pm2$, while the massive helicities take values $h_X=0,\pm1,\pm2$. Thus, a pair of equal-helicity massless gravitons can produce, or be produced from, only an equal-helicity pair of massive spin-2 particles.

The remaining $GG\to MM$ amplitudes share a common structure when expressed in terms of the centre-of-mass speed $v$ in \eqref{CoM_speed} and the scattering angle $\theta$, for which \eqref{CoM_angle} reduces to,
\begin{align}
    \cos\theta&=\frac{t-u}{sv}.
\end{align}
In terms of these variables, every $GG\to MM$ amplitude can be written as,
\begin{align}
\label{GG_to_MM}
    \Acal^{GG\to MM}_{h_A,h_B;h_C,h_D}(s,t,u)
    =\frac{s}{m_{\mathrm{Pl}}^2}
    \frac{\Fcal_{h_A,h_B;h_C,h_D}(v,\theta)}
    {1-v^2\cos^2\theta},
\end{align}
where the dimensionless helicity-dependent functions $\Fcal_{h_A,h_B;h_C,h_D}(v,\theta)$ are collected in Table~\ref{tab:GGMM_Functions}. By \eqref{Time_Reversal}, the same expressions also determine the inverse process $MM\to GG$, with $v$ interpreted as the centre-of-mass speed of either incoming massive particle.

While these amplitudes are expressed in terms of $v$ and $\theta$, the relations \eqref{CoM_speed} and \eqref{CoM_angle} are invertible and can thus be used to write \eqref{GG_to_MM} entirely in terms of the covariant Mandelstam variables. For example,
\begin{align}
\label{GGMMhhhtht}
    \Acal^{GG\to MM}_{h,h;\tilde h,\tilde h}(s,t,u)
    &=\frac{s}{2m_{\mathrm{Pl}}^2}\frac{ m_{\mathrm{FP}}^4}
    {
    \left(m_{\mathrm{FP}}^2-t\right)
    \left(m_{\mathrm{FP}}^2-u\right)}
    \left[
        \frac{1+\sqrt{1-4m_{\mathrm{FP}}^2/s}}
        {1-\sqrt{1-4m_{\mathrm{FP}}^2/s}}
    \right]^{h\tilde h/2},
\end{align}
where $h= \pm 2$ and $\tilde h= \pm 2, \pm 1, 0$.

\begin{table}[t]
    \centering
    \begin{tabular}{ccc}
        \toprule
        $(h_A,h_B)$ & $(h_C,h_D)$ &
        $\Fcal_{h_A,h_B;h_C,h_D}(v,2\vartheta)$ \\
        \midrule
        $(2,2)$ & $(\tilde h,\tilde h)$
        & $\displaystyle
        \tfrac{1}{8}(1+v)^{2+\tilde h}(1-v)^{2-\tilde h}$ \\[2mm]

        $(2,2)$ & $h_C\neq h_D$
        & $0$ \\

        \midrule
        $(2,-2)$ & $(2,2)$
        & $\displaystyle
        2(1-v^2)^2s_\vartheta^4c_\vartheta^4$ \\[2mm]

        $(2,-2)$ & $(2,1)$
        & $\displaystyle
        -4(1-v^2)^{3/2}s_\vartheta^3c_\vartheta^5$ \\[2mm]

        $(2,-2)$ & $(2,0)$
        & $\displaystyle
        -2\sqrt{6}\,(1-v^2)s_\vartheta^2c_\vartheta^6$ \\[2mm]

        $(2,-2)$ & $(2,-1)$
        & $\displaystyle
        4\sqrt{1-v^2}\,s_\vartheta c_\vartheta^7$ \\[2mm]

        $(2,-2)$ & $(2,-2)$
        & $\displaystyle
        2c_\vartheta^8$ \\[2mm]

        $(2,-2)$ & $(1,1)$
        & $\displaystyle
        -2(1-v^2)(4-v^2)s_\vartheta^4c_\vartheta^4$ \\[2mm]

        $(2,-2)$ & $(1,0)$
        & $\displaystyle
        -2\sqrt{6}\,\sqrt{1-v^2}\,(2-v^2)
        s_\vartheta^3c_\vartheta^5$ \\[2mm]

        $(2,-2)$ & $(1,-1)$
        & $\displaystyle
        2(4-3v^2)s_\vartheta^2c_\vartheta^6$ \\[2mm]

        $(2,-2)$ & $(0,0)$
        & $\displaystyle
        2(v^4-6v^2+6)s_\vartheta^4c_\vartheta^4$ \\
        \bottomrule
    \end{tabular}
    \caption{The functions $\Fcal$ appearing in \eqref{GG_to_MM}. For notational convenience, we write the scattering angle as $\theta=2\vartheta$ and use $s_\vartheta=\sin\vartheta$ and $c_\vartheta=\cos\vartheta$. In the first row, $\tilde h=-2,-1,0,1,2$. The remaining helicity amplitudes follow from parity, time reversal, and particle exchange.}
    \label{tab:GGMM_Functions}
\end{table}

The individual $MG\to MG$ helicity amplitudes cannot be obtained from \eqref{GG_to_MM} by a simple permutation of the Mandelstam variables, but require an analytic crossing continuation together with the corresponding Wigner rotations of the helicity states. Nevertheless, the amplitudes themselves admit a common form. Whereas $v$ in \eqref{CoM_speed} denotes the speed of either particle in an equal-mass $MM$ pair, we use $w$ for the speed of the massive particle in $MG\to MG$ kinematics. We additionally introduce the dimensionless mass ratio $x$,
\begin{align}
    w&=\frac{p_i}{E_A}=\frac{p_f}{E_C}
    =\frac{s-m_{\mathrm{FP}}^2}{s+m_{\mathrm{FP}}^2}, &
    x&=\frac{1-w}{1+w}=\frac{m_{\mathrm{FP}}^2}{s}.
\end{align}
For this process, the general expression \eqref{CoM_angle} for the scattering angle simplifies to,
\begin{align}
    \cos\theta=1+\frac{2t}{s(1-x)^2}.
\end{align}
In the centre-of-mass frame all $MG\to MG$ amplitudes can then be written in the form,
\begin{align}
\label{MG_to_MG_Common_Form}
    \Acal^{MG\to MG}_{h_A,h_B;h_C,h_D}(s,t,u)
    =
    \frac{s}{m_{\mathrm{Pl}}^2}
    \frac{(1+w)}
    {(1-\cos\theta)(1+w\cos\theta)}\Hcal_{h_A,h_B;h_C,h_D}(x,\theta).
\end{align}
Although the functions $\Hcal$ are more involved than $\Fcal$, all 30 representatives are collected in Table~\ref{tab:MGGM_Helicity_Functions}.
\begin{table}
    \centering
    \setlength{\tabcolsep}{3pt}
    \renewcommand{\arraystretch}{1.45}
    \begin{tabular}{@{}ccc@{}}
        \toprule
        $(h_A,h_C)$
        & $\Hcal_{h_A,2;h_C,2}(x,2\vartheta)$
        & $\Hcal_{h_A,2;-h_C,-2}(x,2\vartheta)$ \\
        \midrule
        $(2,2)$
        & $\displaystyle \left(1-xs_\vartheta^2\right)^4$
        & $\displaystyle x^4s_\vartheta^8$ \\

        $(2,1)$
        & $\displaystyle 2\sqrt{x}\,c_\vartheta s_\vartheta
        \left(1-xs_\vartheta^2\right)^3$
        & $\displaystyle -2x^{7/2}c_\vartheta s_\vartheta^7$ \\

        $(2,0)$
        & $\displaystyle -\sqrt{6}\,x c_\vartheta^2s_\vartheta^2
        \left(1-xs_\vartheta^2\right)^2$
        & $\displaystyle -\sqrt{6}\,x^3c_\vartheta^2s_\vartheta^6$ \\

        $(2,-1)$
        & $\displaystyle -2x^{3/2}c_\vartheta^3s_\vartheta^3
        \left(1-xs_\vartheta^2\right)$
        & $\displaystyle 2x^{5/2}c_\vartheta^3s_\vartheta^5$ \\

        $(2,-2)$
        & $\displaystyle x^2c_\vartheta^4s_\vartheta^4$
        & $\displaystyle x^2c_\vartheta^4s_\vartheta^4$ \\

        $(1,1)$
        & $\displaystyle c_\vartheta^2
        \left(1-xs_\vartheta^2\right)^2
        \left(1-4xs_\vartheta^2\right)$
        & $\displaystyle x^3s_\vartheta^6
        \left(4s_\vartheta^2-3\right)$ \\

        $(1,0)$
        & $\displaystyle -\sqrt{6x}\,c_\vartheta^3s_\vartheta
        \left(1-xs_\vartheta^2\right)
        \left(1-2xs_\vartheta^2\right)$
        & $\displaystyle \sqrt{6}\,x^{5/2}c_\vartheta s_\vartheta^5
        \left(2s_\vartheta^2-1\right)$ \\

        $(1,-1)$
        & $\displaystyle -x c_\vartheta^4s_\vartheta^2
        \left(3-4xs_\vartheta^2\right)$
        & $\displaystyle x^2c_\vartheta^2s_\vartheta^4
        \left(1-4s_\vartheta^2\right)$ \\

        $(1,-2)$
        & $\displaystyle 2x^{3/2}c_\vartheta^5s_\vartheta^3$
        & $\displaystyle -2x^{3/2}c_\vartheta^3s_\vartheta^5$ \\

        $(0,0)$
        & $\displaystyle c_\vartheta^4
        \left(1-6xs_\vartheta^2+6x^2s_\vartheta^4\right)$
        & $\displaystyle x^2s_\vartheta^4
        \left(1-6s_\vartheta^2+6s_\vartheta^4\right)$ \\

        $(0,-1)$
        & $\displaystyle \sqrt{6x}\,c_\vartheta^5s_\vartheta
        \left(1-2xs_\vartheta^2\right)$
        & $\displaystyle \sqrt{6}\,x^{3/2}c_\vartheta s_\vartheta^5
        \left(1-2s_\vartheta^2\right)$ \\

        $(0,-2)$
        & $\displaystyle -\sqrt{6}\,x c_\vartheta^6s_\vartheta^2$
        & $\displaystyle -\sqrt{6}\,x c_\vartheta^2s_\vartheta^6$ \\

        $(-1,-1)$
        & $\displaystyle c_\vartheta^6
        \left(1-4xs_\vartheta^2\right)$
        & $\displaystyle x s_\vartheta^6
        \left(4s_\vartheta^2-3\right)$ \\

        $(-1,-2)$
        & $\displaystyle -2\sqrt{x}\,c_\vartheta^7s_\vartheta$
        & $\displaystyle 2\sqrt{x}\,c_\vartheta s_\vartheta^7$ \\

        $(-2,-2)$
        & $\displaystyle c_\vartheta^8$
        & $\displaystyle s_\vartheta^8$ \\
        \bottomrule
    \end{tabular}
    \caption{The functions governing the graviton-helicity-preserving and graviton-helicity-flipping $MG\to MG$ representatives. The representative helicity pairs in the two sectors are related by $h_C\mapsto-h_C$, which accounts for the argument $-h_C$ in the third column. For notational convenience, we write the scattering angle as $\theta=2\vartheta$ and use $s_\vartheta=\sin\vartheta$ and $c_\vartheta=\cos\vartheta$.}
    \label{tab:MGGM_Helicity_Functions}
\end{table}

\newpage

\subsection{\texorpdfstring{$GMMM$}{GMMM} amplitudes}

The representative amplitudes in the $GMMM$ sector are generically lengthy and, in addition to $\mFP$ and $\mpl$, depend on $\beta_-$ and $q$. Their dependence on the ghost-free potential parameter $\beta_-$ and the mixing parameter $q$ makes this the first sector that probes the nonlinear bimetric structure beyond the covariant Fierz--Pauli completion. As expected from the parameter hierarchy, $\beta_+$ does not enter this sector. All amplitudes take the form,
\begin{align}
\label{A=QA+BA}
    \Acal^{GM\to MM}
    =Q_-\Acal_Q^{GM\to MM}
    +\beta_-\Acal_\beta^{GM\to MM}.
\end{align}
The absence of $MGG$ vertices means that this sector contains no massless exchange, as illustrated in Figure~\ref{fig:Bimetric_Scattering_Sectors}d.

Due to the form \eqref{A=QA+BA}, all these processes vanish identically at the exchange-symmetric parameter choice $q=1$ and $c\beta_1=c^3\beta_3$,
\begin{align}
\label{self_dual}
    Q_-=0, && \beta_-=0.
\end{align}
This also follows directly from \eqref{QQ_exchange} as at this parameter choice the transformation \mbox{$M_{\mu\nu}\to-M_{\mu\nu}$} is a symmetry of the action, and hence every vertex containing an odd number of massive fields must vanish.

Out of the $75$ representative amplitudes in this sector, precisely five are independent of $\beta_-$. Choosing the incoming graviton to have helicity $+2$, these amplitudes admit a relatively compact form. Using the centre-of-mass speed \eqref{CoM_speed} and the scattering angle \eqref{CoM_angle}, which here simplifies to,
\begin{align}
    \cos\theta&=\frac{t-u}{(s-m_{\mathrm{FP}}^2)v},
\end{align}
the five $\beta_-$-independent representatives can be written in a closed form,
\begin{align}
    \Acal^{GM\to MM}_{2,h;2,-2}
    &=
    (-1)^{h(h+1)/2}\sqrt{\binom{4}{2+h}}\,
    \frac{Q_-s(3+v^2)}
    {2m_{\mathrm{Pl}}^2\left(1-v^2\cos^2\theta\right)}
    \left(
        \frac{\sqrt{1-v^2}}{2}\sin\frac{\theta}{2}
    \right)^{2+h}
    \cos^{6-h}\frac{\theta}{2},
\end{align}
where $h=\pm2,\pm1,0$. Combining parity \eqref{Parity_Bimetric} with exchange of the two outgoing massive particles \eqref{Swap_Symmetry} gives the corresponding amplitudes with incoming graviton helicity $-2$.\footnote{For these amplitudes, the corresponding expression is obtained from the displayed result by $h\to-h$ and $\sin(\theta/2)\leftrightarrow\cos(\theta/2)$ and an additional $(-1)^h$.}

The amplitudes with $h_G=2$ admit the common factorisation,
\begin{align}
\label{GMMM_Common_Form}
    \Acal^{GM\to MM}_{2,h_B;h_C,h_D} &= \frac{s}{\mpl^2}\frac{\sin^{H_+}(\frac{\theta}{2})\cos^{H_-}(\frac{\theta}{2})}{1-v^2\cos^2\theta}\left[Q_-\mathcal{Q}^{(h_B)}_{h_C,h_D}(v,\theta)+Q_+\frac{\beta_-}{\beta_{\mathrm{FP}}}\mathcal{B}^{(h_B)}_{h_C,h_D}(v,\theta)\right],
\end{align}
where the exponents are defined by $H_\pm = |2-h_B\mp h_C\pm h_D|$. For generic $h_B$, the functions $\mathcal{Q}^{(h_B)}$ and $\mathcal{B}^{(h_B)}$ depend nontrivially on both $v$ and $\theta$ and are too lengthy to display here. For opposite incoming helicities, $(h_G,h_M)=(2,-2)$, however, the explicit prefactor in \eqref{GMMM_Common_Form} contains all the angular dependence. The remaining functions depend only on $v$; denoting them by $\mathcal{Q}_{h_C,h_D}(v)$ and $\mathcal{B}_{h_C,h_D}(v)$, we collect them in Table~\ref{tab:GMMM_hMinus2_Functions}. The amplitudes with $h_A=-2$ for the incoming $G$ can then be obtained from \eqref{Parity_Bimetric} with the appropriate change of the remaining helicities, while the amplitudes for the time-reversed process $MM\to MG$ follow directly from \eqref{Time_Reversal} and particle exchange \eqref{Swap_Symmetry}. 

\begin{table}
    \centering
    \renewcommand{\arraystretch}{1.5}
    \begin{tabular}{ccc}
        \toprule
        $(h_C,h_D)$
        & $\mathcal{Q}_{h_C,h_D}(v)$
        & $\mathcal{B}_{h_C,h_D}(v)$ \\
        \midrule
        $(2,2)$
        & $\displaystyle -\frac{9(1-v^2)^2}{2(3+v^2)}$
        & $\displaystyle \frac{v(1-v)(1+v)^2(v+3)}{2(3+v^2)}$ \\

        $(2,1)$
        & $\displaystyle \frac{9(1-v^2)^{3/2}}{3+v^2}$
        & $\displaystyle -\frac{v(1+v)\sqrt{1-v^2}(v^2+2v+9)}
        {4(3+v^2)}$ \\

        $(2,0)$
        & $\displaystyle \frac{\sqrt{6}(1-v^2)(v^2+9)}
        {2(3+v^2)}$
        & $\displaystyle \frac{\sqrt{6}v(1+v)(v^3+v^2-v-9)}
        {12(3+v^2)}$ \\

        $(2,-1)$
        & $\displaystyle -3\sqrt{1-v^2}$
        & $\displaystyle \frac{v(1+v)\sqrt{1-v^2}}{4}$ \\

        $(2,-2)$
        & $\displaystyle -\frac{3+v^2}{2}$
        & $0$ \\

        $(1,1)$
        & $\displaystyle \frac{6(1-v^2)(3-v^2)}{3+v^2}$
        & $\displaystyle -\frac{v(1+v)(v+3)}{3+v^2}$ \\

        $(1,0)$
        & $\displaystyle \frac{\sqrt{6}\sqrt{1-v^2}(9-5v^2)}
        {3+v^2}$
        & $\displaystyle \frac{\sqrt{6}v\sqrt{1-v^2}
        (v^4+8v^2-8v-9)}
        {12(1-v)(3+v^2)}$ \\

        $(1,-1)$
        & $\displaystyle \frac{2(v^4+6v^2-9)}{3+v^2}$
        & $\displaystyle \frac{2v^2}{3+v^2}$ \\

        $(1,-2)$
        & $\displaystyle -3\sqrt{1-v^2}$
        & $\displaystyle -\frac{v(1-v)\sqrt{1-v^2}}{4}$ \\

        $(0,0)$
        & $\displaystyle -\frac{(3-v^2)(9-7v^2)}{3+v^2}$
        & $\displaystyle \frac{4v^2}{3+v^2}$ \\

        $(0,-1)$
        & $\displaystyle \frac{\sqrt{6}\sqrt{1-v^2}(9-5v^2)}
        {3+v^2}$
        & $\displaystyle -\frac{\sqrt{6}v\sqrt{1-v^2}
        (v^4+8v^2+8v-9)}
        {12(1+v)(3+v^2)}$ \\

        $(0,-2)$
        & $\displaystyle \frac{\sqrt{6}(1-v^2)(v^2+9)}
        {2(3+v^2)}$
        & $\displaystyle \frac{\sqrt{6}v(1-v)(v^3-v^2-v+9)}
        {12(3+v^2)}$ \\

        $(-1,-1)$
        & $\displaystyle \frac{6(1-v^2)(3-v^2)}{3+v^2}$
        & $\displaystyle \frac{v(1-v)(3-v)}{3+v^2}$ \\

        $(-1,-2)$
        & $\displaystyle \frac{9(1-v^2)^{3/2}}{3+v^2}$
        & $\displaystyle \frac{v(1-v)\sqrt{1-v^2}(v^2-2v+9)}
        {4(3+v^2)}$ \\

        $(-2,-2)$
        & $\displaystyle -\frac{9(1-v^2)^2}{2(3+v^2)}$
        & $\displaystyle -\frac{v(1-v)^2(1+v)(3-v)}
        {2(3+v^2)}$ \\
        \bottomrule
    \end{tabular}
    \caption{The functions $\mathcal{Q}_{h_C,h_D}(v)$ and
    $\mathcal{B}_{h_C,h_D}(v)$ appearing in
    \eqref{GMMM_Common_Form} for $(h_A,h_B)=(2,-2)$. The table contains the representatives
    with $h_C\geq h_D$; the remaining helicity configurations follow from
    particle exchange and parity.}
    \label{tab:GMMM_hMinus2_Functions}
\end{table}

\newpage

\subsection{\texorpdfstring{$MMMM$}{MMMM} amplitudes}

The $MMMM$ sector contains $66$ representative amplitudes and is the first sector that can probe the complete parameter dependence of the ghost-free bimetric interaction. Their complete finite-energy expressions are generically lengthy, and we have not found a common representation that makes their form more transparent. We therefore restrict the discussion to their generic structure, while their complete expressions are included in the accompanying symbolic files.

At tree level, the amplitudes receive contributions from both massless and massive exchange, together with the $MMMM$ contact interaction, as shown in Figure~\ref{fig:Bimetric_Scattering_Sectors}e. The massless-exchange diagrams are constructed from two $GMM$ vertices and belong to the universal Fierz--Pauli sector. By contrast, the massive-exchange diagrams are constructed from two $MMM$ vertices and can consequently depend quadratically on $\beta_-$. The $MMMM$ contact vertex is the only contribution containing $\beta_+$, which therefore appears at most linearly in the tree-level amplitudes.

The exchange symmetry further constrains this dependence. Since the amplitudes contain an even number of massive fields, they are invariant under the simultaneous transformations \mbox{$Q_-\to-Q_-$} and $\beta_-\to-\beta_-$. Terms linear in $\beta_-$ must therefore be accompanied by an exchange-odd dependence on $Q_-$, while terms quadratic in $\beta_-$ are allowed independently. However, individual helicity projections need not probe every parameter, e.g. among the $66$ representatives, only $35$ depend explicitly on $\beta_+$.

A particularly informative result is provided by the fully longitudinal amplitude. Although its exact finite-energy expression is lengthy, its leading high-energy behaviour is compact and displays the above structure. At fixed scattering angle it reads,
\begin{align}
\label{High_Energy_Helicity_Zero_Amplitude}
    \Acal^{MM\to MM}_{0,0;0,0}(s,t,u)
    &=
    -\frac{Q_+^2}{144}
    \left[
        3\left(1+\frac{\beta_-^2}{\beta_{\mathrm{FP}}^2}\right)
        +2\frac{Q_-}{Q_+}\frac{\beta_-}{\beta_{\mathrm{FP}}}
        -8\frac{\beta_+}{\beta_{\mathrm{FP}}}
    \right]
    \frac{stu}{\mpl^2\mFP^4}
    +\mathcal{O}\left(\frac{E^4}{m_{\mathrm{Pl}}^2m_{\mathrm{FP}}^2}\right).
\end{align}
For $E\gg m_{\mathrm{FP}}$, the longitudinal equivalence theorem identifies the helicity-0 mode with the Stückelberg scalar, and \eqref{High_Energy_Helicity_Zero_Amplitude} reproduces the characteristic $stu$ dependence of four-Galileon scattering \cite{Bonifacio:2018aon,Cheung:2016yqr}. The coefficient also exhibits the expected exchange symmetry: $\beta_-$ appears either quadratically or together with the exchange-odd combination $Q_-$, whereas $\beta_+$ enters linearly. Since $stu\sim E^6$ at fixed angle, this amplitude provides a first example of the $\Lambda_3$-type high-energy growth. 

\subsection{High-energy limit and strong coupling}
\label{sec:strong_coupling}
We now turn to the high-energy behaviour of the complete set of amplitudes and its implications for the strong-coupling scale.

Applying the procedure outlined in Section~\ref{sec:high_energy} to all of the 199 representatives, one can see that the fastest-growing amplitudes include, as expected, those involving the massive helicity-0 mode, such as \eqref{High_Energy_Helicity_Zero_Amplitude},
\begin{align}
    \Acal^{MM \to MM}_{0,0;0,0}  \sim \left( \frac{E}{\Lambda_3} \right)^6, && \Lambda_3 = (\mpl^{\,} \mFP^2)^{1/3}.
\end{align}
This is two steps higher in the hierarchy of strong-coupling scales than the $\Lambda_5$ scaling of a generic massive gravity theory. The ghost-free bimetric interactions therefore not only guarantee the absence of the Boulware--Deser ghost, but also raise the strong-coupling scale from $\Lambda_5$ to $\Lambda_3$.

The $\Lambda_3$ scaling has previously been argued to be the case in massive gravity \cite{Bonifacio:2018vzv,Bonifacio:2018aon,Bonifacio:2019mgk,Cheung:2016yqr,Bonifacio:2014rba} and bimetric theory \cite{Fasiello:2013woa,Bonifacio:2018aon,Scargill:2015wxs}, but has not previously been shown for all processes in the latter. Interestingly, there is one process that scales as $E^6$ despite having no helicity-0 external states, namely,
\begin{align}
    \Acal^{MM\to MM}_{1,1;1,1} \sim \left(\frac{E}{\Lambda_3}\right)^6.
\end{align}
The high-energy behaviours of all 199 representatives are presented in Tables \ref{tab:Amps_GGGG}--\ref{tab:HighEnergyMMMM}, where one can see that the actual strong-coupling scale also depends on the parameters of the theory. This is also seen explicitly in \eqref{High_Energy_Helicity_Zero_Amplitude}, for which the leading behaviour would instead grow at most as $\mathcal{O}(E^4)$ if the multiplicative coefficient vanished. Among the complete set of amplitudes, there are 8 channels that exhibit $\Lambda_3$ scaling, but there is not enough parameter freedom to cancel the leading contributions in all of them and raise the strong-coupling scale to $\Lambda_2$ without decoupling the two metrics. This is in line with the result of \cite{Schwartz:2003vj}, which established that $\Lambda_3$ is the highest parametrically attainable strong-coupling scale for a theory of an interacting massive graviton.

\subsection{Massive-gravity limit and  scattering amplitudes}
\label{sec:massive_gravity_limit}

As discussed in Section~\ref{sec:mass_eigen}, the massive-gravity limit is obtained by the limit \eqref{Massive_Gravity_Limit}, and the decoupling of the massless modes follows directly from,
\begin{align}
    \delta g_{\mu\nu}&=\frac{\sqrt{2}}{\mpl}\Big[G_{\mu\nu}-qM_{\mu\nu}\Big]\longrightarrow-\frac{\sqrt{2}}{m_g}M_{\mu\nu}, & \frac{\delta f_{\mu\nu}}{c^2}&=\frac{\sqrt{2}}{\mpl}\Big[G_{\mu\nu}+\frac{1}{q}M_{\mu\nu}\Big]\longrightarrow0.
\end{align}
Thus, $f_{\mu\nu} = c^2\eta_{\mu \nu}$ becomes a fixed reference metric, while the fluctuation of the dynamical metric $g_{\mu\nu}$ is entirely described by $M_{\mu\nu}$. The field $G_{\mu\nu}$ retains its quadratic kinetic term but decouples from all nonlinear interactions, so the limiting theory consists of  massive gravity together with a free massless spin-2 field.

The limit $\mpl\to\infty$ does not remove the self-interactions of $M_{\mu\nu}$, since their coefficients contain compensating powers of $q$. In particular,
\begin{align}
    \frac{Q_+}{\mpl}&\longrightarrow\frac{1}{m_g}, & \frac{C_3(q)}{\mpl}=-\frac{Q_-}{\mpl}&\longrightarrow-\frac{1}{m_g}, & \frac{C_4(q)}{\mpl^2}=\frac{1+Q_-^2}{\mpl^2}&\longrightarrow\frac{1}{m_g^2}.
\end{align}
Consequently, both the cubic and quartic massive self-interactions of the Einstein--Hilbert term remain finite and are controlled by the Planck scale $m_g$ of the dynamical metric. Similarly, the bimetric interaction in the form \eqref{Potential_Parameter_Hierarchy} depends only on $M$ and reproduces the expansion of the massive gravity potential.

The decoupling is also manifest directly in the amplitudes. Since the $GGGG$ \eqref{Massless_Spin_2_Amplitudes}, $GGMM$ \eqref{GG_to_MM}, and $GMGM$ \eqref{MG_to_MG_Common_Form} amplitudes scale as $\mpl^{-2}$, while the $GMMM$ amplitudes scale at most as $Q_\pm/\mpl^2$, all amplitudes containing an external $G$ vanish,
\begin{align}
    \Acal^{GG\to GG}&\longrightarrow0, & \Acal^{GG\to MM}&\longrightarrow0, & \Acal^{GM\to GM}&\longrightarrow0, & \Acal^{GM\to MM}&\longrightarrow0,
\end{align}
together with their crossings. Amplitudes containing three massless fields and one massive field already vanish identically because the perturbative action contains no vertices of the form $G^nM$.

Within the $MMMM$ amplitudes, every massless-exchange contribution contains two $GMM$ vertices. Since each such vertex scales as $\mpl^{-1}$ without any compensating $Q_\pm$ dependence, these contributions vanish as $\mpl^{-2}$. Therefore, the surviving massive-exchange and contact contributions are precisely those obtained from massive gravity. The complete massive-gravity amplitudes can therefore be obtained from,
\begin{align}
\label{MG_amplitudes}
    \Acal^{\mathrm{MG}}_{h_A,h_B;h_C,h_D}(s,t,u)=\lim_{q\to\infty}\left[\left.\mathcal{A}^{MM\to MM}_{h_A,h_B;h_C,h_D}(s,t,u)\right|_{\mpl=m_g\sqrt{1+q^2}}\right],
\end{align}
with $m_g$, $\mFP$, and $\beta_\pm/\beta_{\mathrm{FP}}$ held fixed. The $66$ symmetry-inequivalent $MMMM$ representatives consequently generate all $5^4=625$ ordered tree-level helicity amplitudes of massive gravity. We compared all $66$ representatives with the general tree-level massive-gravity amplitudes of \cite{Cheung:2016yqr} and found exact agreement after converting between polarisation, momentum, normalisation, and parameter conventions.\footnote{Reference~\cite{Cheung:2016yqr} uses the mostly-plus signature and a real scalar--vector--tensor basis. With Lorentz indices suppressed, the polarisation map is $\epsilon_{\pm2}=(\epsilon_{T_1}\pm i\epsilon_{T_2})/\sqrt{2}$, $\epsilon_{+1}=(\epsilon_{V_2}-i\epsilon_{V_1})/\sqrt{2}$, $\epsilon_{-1}=-(\epsilon_{V_2}+i\epsilon_{V_1})/\sqrt{2}$, and $\epsilon_0=-\epsilon_S$, with outgoing polarisations complex conjugated. Their all-incoming momenta are identified as $(k_1,k_2,k_3,k_4)=(p_A,p_B,-p_C,-p_D)$. The parameter map is $m_{\mathrm{CR}}=\mFP$, $\kappa_{\mathrm{CR}}=1/(\sqrt{2}m_g)$, and $(\beta_1,\beta_2,\beta_3)=(3-18c_3-48d_5,-1+12c_3+48d_5,-6c_3-48d_5)$ after setting $c=1$ and taking $q\to\infty$ with $\mpl^2=m_g^2(1+q^2)$ and $m^4=m_g^2\mFP^2$.}

The comparison also extends to the high-energy limit: after applying the same conversion of conventions, the leading high-energy behaviour of our amplitudes agrees with all explicit examples presented in \cite{Cheung:2016yqr}. In particular, the maximal growth remains $E^6$, confirming that the bimetric $\Lambda_3$ scaling survives the massive-gravity limit with $\mpl$ replaced by $m_g$,
\begin{align}
    \Lambda_{3}^{\mathrm{MG}}=(m_g\mFP^2)^{1/3}.
\end{align}

\subsection{General relativity and massless limits}
\label{sec:GR_and_massless_limits}

For matter coupled to $g_{\mu\nu}$, the conventional general relativity limit \eqref{GR_Limit} is obtained by taking $q\to0$ through $m_f\to0$ while keeping $m_g$ and the $\beta_n$ fixed. In this limit,
\begin{align}
    \mpl&\longrightarrow m_g, & \mFP^2=\frac{m^4Q_+^2}{\mpl^2}\beta_{\mathrm{FP}}&\sim\frac{m^4\beta_{\mathrm{FP}}}{q^2m_g^2}\longrightarrow\infty, & \delta g_{\mu\nu}&\longrightarrow\frac{\sqrt{2}}{m_g}G_{\mu\nu}.
\end{align}
The massive mode therefore becomes infinitely heavy and decouples at fixed energy, while the surviving $G$ sector reduces to perturbative general relativity with Planck scale $m_g$ \cite{Akrami:2015qga,Schmidt-May:2015vnx},
\begin{align}
    \lim_{q\to0}\Acal^{GG\to GG}_{h_A,h_B;h_C,h_D}(s,t,u)&=\Acal^{\mathrm{GR}}_{h_A,h_B;h_C,h_D}(s,t,u;m_g).
\end{align}
The same limit should not be imposed on amplitudes with external $M$ states while holding all Mandelstam variables fixed, since \eqref{Mandelstam_identity} cannot be fulfilled at constant energy while $\mFP \to \infty$. Any apparent singularities from such a substitution are therefore unphysical: the massive states become inaccessible at fixed energy. Accordingly, the GR and high-energy limits do not commute, since they correspond respectively to $E/\mFP\to0$ and $E/\mFP\to\infty$.

The limit $\mFP\to0$ does not define a unique theory. Scaling away the entire potential yields two decoupled Einstein--Hilbert sectors. The former massive field must then be reinterpreted as a massless spin-2 field, for which the helicity $\pm1$ and $0$ states are unphysical. If instead $\beta_{\mathrm{FP}}\to0$ at fixed nonzero $\beta_\pm$, the quadratic mass term vanishes while nonlinear interactions remain,
\begin{align}
    \mathcal{U}(\mathbb{X})&\longrightarrow\tfrac{1}{2}\beta_-e_3(\mathbb{Y})+\tfrac{1}{2}(2\beta_--\beta_+)e_4(\mathbb{Y}).
\end{align}
The quadratic theory then contains two apparently massless spin-2 fields, but the surviving interactions break the additional linearised diffeomorphism invariance. The would-be helicity $\pm1$ and $0$ modes therefore lack a quadratic description and are infinitely strongly coupled.
\vspace{-1mm}

\section{Discussion}
\label{sec:discussion}

We have computed the complete set of $2\to 2$ tree-level scattering amplitudes of bimetric theory in the centre-of-mass helicity basis, with the exact expressions provided in the accompanying material \cite{FelixHerber_TreeLevelBimetricCalculations_2025}. The amplitudes reveal a clear parameter hierarchy, where processes with one massive state vanish, those with two are universal, and the nonlinear bimetric parameters first enter with three massive external states. The results reproduce the general relativity and massive gravity limits, and establish the characteristic $\Lambda_3$ high-energy scale. A more detailed summary is given in Section~\ref{sec:summary}.

The agreement between our amplitudes and those of Ref.~\cite{Cheung:2016yqr} provides a nontrivial verification of our calculation. Our amplitudes also provide a benchmark for independent amplitude constructions using massive spinor-helicity methods \cite{Arkani-Hamed:2017jhn,Chung:2018kqs} and massive double-copy constructions \cite{Momeni:2020vvr}, while the underlying cubic vertices can be compared directly with string-theoretic calculations \cite{Lust:2021jps,Lust:2023sfk}. Such comparisons may also inform how ghost-free spin-2 interactions are realised in these alternative approaches.

The complete amplitudes also provide a basis for a more detailed analysis of the regime of validity of bimetric theory. Although their high-energy behaviour identifies a $\Lambda_3$ strong-coupling scale, we have only considered the parametric dependence; a deeper analysis is needed to determine its precise numerical value and the helicity and angular-momentum channels that first violate perturbative unitarity. The exact momentum dependence can similarly be used to derive dispersion relations and positivity bounds, although the massless $t$-channel pole requires particular care \cite{deRham:2018qqo,Alberte:2019xfh,Alberte:2020jsk}. Finally, the eikonal limit provides a complementary probe of causality through time delays and possible asymptotic superluminality \cite{Hinterbichler:2017qyt,Bonifacio:2017nnt}.

Extending the calculation to include matter couplings would give access to massive-spin-2 production and decay, as well as matter annihilation through massless and massive spin-2 exchange \cite{Wood:2026zau}. These processes, together with the self-scattering amplitudes obtained here, are particularly relevant when the massive mode is interpreted as dark matter \cite{Aoki:2016zgp,Babichev:2016hir,Babichev:2016bxi,Marzola:2017lbt}, and we will return to this in future work. Although our amplitudes are derived around Minkowski space, they provide a valid local approximation on astrophysical and cosmological backgrounds whenever the backgrounds are close to proportional and the characteristic length and time scales of the scattering process are short compared with the curvature and evolution scales, making them directly applicable to processes relevant for dark-matter searches.

Another natural extension is to ghost-free multi-gravity theories \cite{Hassan:2018mcw,Flinckman:2025bje}. Around a proportional background, multi-gravity with $\mathcal{N}$ interacting metrics contains one massless and $\mathcal{N}-1$ massive spin-2 modes \cite{Flinckman:2024zpb}, so the external states carry both species and helicity labels. Distinct massive modes introduce conversion processes, mixed exchanges, and kinematic thresholds that are absent in bimetric theory. Their complete amplitudes would reveal how the parameter hierarchy and selection rules found here generalise, and whether the $\Lambda_3$ scaling is retained or 
whether additional channels introduce a lower parameter-dependent strong-coupling scale.

\appendix

\section{Counting inequivalent amplitude representatives}
\label{app:amplitude_counting}

In this Appendix, we show how the group structure relating the $2\to2$ amplitudes reduces the number of amplitudes that must be computed explicitly. Rather than evaluating every ordered helicity configuration separately, we identify $199$ inequivalent amplitudes and show how all remaining amplitudes can be generated from these by acting with the relevant group transformations. Particle exchange and time reversal generate the dihedral group $D_8$ of order eight, while parity generates an additional commuting cyclic group of order two, $\mathbb{Z}_2$. Together, these transformations generate the group,
\begin{align}
    \Gcal \cong D_8\times\mathbb{Z}_2,
\end{align}
of order $16$.

A single external particle can occupy one of seven particle-helicity states,
\begin{align}
    G_{+2},\qquad G_{-2},\qquad
    M_{+2},\qquad M_{+1},\qquad M_0,\qquad M_{-1},\qquad M_{-2},
\end{align}
so there are initially $7^4=2401$ ordered $2\to2$ helicity configurations. Let $\Acal^{2\to2}$ denote this full set. The $\Gcal$-inequivalent configurations are the orbits of $\Gcal$ acting on $\Acal^{2\to2}$, and their number is therefore the size of the quotient set,
\begin{align}
    |\Acal^{2\to2}/\Gcal|.
\end{align}
Rather than constructing these orbits explicitly, we count them using Burnside's lemma. For a group element $g\in\Gcal$, define the set of amplitudes fixed by the transformation $g$ as,
\begin{align}
    \operatorname{Fix}(g)
    =\left\{\Acal\in\Acal^{2\to2}\,\middle|\,g\cdot\Acal=\Acal\right\},
\end{align}
so that $|\operatorname{Fix}(g)|$ is the number of configurations left unchanged by the transformation $g$. Burnside's lemma then gives the size of the quotient set,
\begin{align}
    |\Acal^{2\to2}/\Gcal|
    =\frac{1}{|\Gcal|}\sum_{g\in\Gcal}|\operatorname{Fix}(g)|.
\end{align}
To make the action of $\Gcal$ explicit and to compute $\operatorname{Fix}(g)$, we denote the exchange of the two incoming particles by $s_i$, the exchange of the two outgoing particles by $s_f$, time reversal by $\tau$, and parity by $\pi$. The transformations $s_i$, $s_f$, and $\tau$ generate the subgroup $D_8$ and satisfy,
\begin{align}
    \tau s_i\tau &= s_f, &
    \tau s_f\tau &= s_i, &
    s_is_f &= s_fs_i, &
    \tau^2=s_i^2=s_f^2 &= e,
\end{align}
while parity satisfies,
\begin{align}
    \pi^2=e, \qquad \pi g=g\pi,
\end{align}
for every $g\in D_8$.

In principle, we could compute $\operatorname{Fix}(g)$ for every element in $\Gcal$. But to reduce the number of fixed-point counts that must be evaluated, we group together elements of $D_8$ that are conjugate, i.e. related by $g' =h g h^{-1}$ for some $h \in D_8$. Elements that are conjugate form a conjugacy class, and all elements in such a class have the same number of fixed points.\footnote{Indeed, if $g'=hgh^{-1}$, then $\Acal\mapsto h\cdot\Acal$ gives a bijection between $\operatorname{Fix}(g)$ and $\operatorname{Fix}(g')$, so that, $|\operatorname{Fix}(g)|=|\operatorname{Fix}(g')|$.}
It is therefore sufficient to evaluate one representative from each conjugacy class. For the $D_8$ subgroup, there are five such classes,
\begin{align}
    \{e\}, \qquad
    \{s_is_f\}, \qquad
    \{s_i,s_f\}, \qquad
    \{\tau,s_is_f\tau\}, \qquad
    \{\tau s_i,\tau s_f\}.
\end{align}
Since $\pi$ commutes with $D_8$, multiplication by parity gives five further classes,
\begin{align}
    \{\pi\}, \qquad
    \{\pi s_is_f\}, \qquad
    \{\pi s_i,\pi s_f\}, \qquad
    \{\pi\tau,\pi s_is_f\tau\}, \qquad
    \{\pi\tau s_i,\pi\tau s_f\}.
\end{align}
Burnside's sum can consequently be written using one representative from each class, for example,
\begin{align}
\label{burnside_sum}
    |\Acal^{2\to2}/\Gcal|
    =\frac{1}{16}\Big[
    &|\operatorname{Fix}(e)|
    +|\operatorname{Fix}(s_is_f)|
    +2|\operatorname{Fix}(s_i)|
    +2|\operatorname{Fix}(\tau)|
    +2|\operatorname{Fix}(\tau s_i)| \nonumber\\
    +&|\operatorname{Fix}(\pi)|
    +|\operatorname{Fix}(\pi s_is_f)|
    +2|\operatorname{Fix}(\pi s_i)|
    +2|\operatorname{Fix}(\pi\tau)|
    +2|\operatorname{Fix}(\pi\tau s_i)|
    \Big].
\end{align}
We first consider the elements of $D_8$. Their action is simply a permutation of the four external particle-helicity states. Using the ordering $(A,B,C,D)$, the relevant representatives act as the permutations written in cycle notation,\footnote{In this notation, a cycle such as $(AB)$ denotes the exchange $A\leftrightarrow B$, while a $1$-cycle such as $(A)$ denotes a fixed external leg. Disjoint cycles act simultaneously. We adopt the convention that products of permutations are applied from left to right, so that, for example, $\tau s_i=(ACBD)$.}
\begin{align}
\label{perm_cycles}
    e &= (A)(B)(C)(D), &
    s_i &= (AB)(C)(D), &
    \tau &= (AC)(BD), \nonumber\\
    s_is_f &= (AB)(CD), &
    \tau s_i &= (ACBD).
\end{align}
A configuration is fixed precisely when all positions belonging to the same cycle carry the same particle-helicity state. Since there are seven possible states, a permutation with $k$ cycles has $7^k$ fixed configurations. Hence,
\begin{subequations}
\begin{align}
    |\operatorname{Fix}(e)| &= 7^4=2401, \\
    |\operatorname{Fix}(s_i)| &= 7^3=343, \\
    |\operatorname{Fix}(\tau)| &= 7^2=49, \\
    |\operatorname{Fix}(s_is_f)| &= 7^2=49, \\
    |\operatorname{Fix}(\tau s_i)| &= 7,
\end{align}
\end{subequations}
giving the first line in \eqref{burnside_sum}. For the second line, the group elements contain parity in addition to a permutation of the external legs. Parity does not change the particle species, but reverses the helicity, $h\to-h$. A configuration fixed by such a transformation must therefore reproduce itself after both the permutation and the helicity reversal have been applied.

Consider first a $2$-cycle, for example $(AB)$. If the state at $A$ is $(X,h)$, where $X$ denotes either $G$ or $M$, then invariance under the parity transformation requires the state at $B$ to be $(X,-h)$. Acting once more returns to $A$ and reverses the helicity again, giving $(X,h)$, so there is no further restriction. Thus, an even cycle can start with any of the seven particle-helicity states, with the helicities alternating as,
\begin{align}
    (X,h),\,(X,-h),\,(X,h),\ldots
\end{align}
around the cycle.

For an odd cycle, however, after going once around the cycle the state returns to its starting position with its helicity reversed. Invariance therefore requires,
\begin{align}
    h=-h,
\end{align}
and hence $h=0$. Of the seven possible particle-helicity states, only the massive state $M_0$ has helicity 0. Consequently, every even cycle contributes seven possible fixed assignments, whereas every odd cycle is forced to contain the single state $M_0$. Using the cycle decompositions \eqref{perm_cycles} therefore gives,
\begin{subequations}
\begin{align}
    |\operatorname{Fix}(\pi)| &= 1, \\
    |\operatorname{Fix}(\pi s_i)| &= 7, \\
    |\operatorname{Fix}(\pi\tau)| &= 7^2=49, \\
    |\operatorname{Fix}(\pi s_is_f)| &= 7^2=49, \\
    |\operatorname{Fix}(\pi\tau s_i)| &= 7.
\end{align}
\end{subequations}
Substituting these results into Burnside's lemma gives,
\begin{align}
    |\Acal^{2\to2}/\Gcal|
    &=\frac{1}{16}\Big[
    2401+49+2\cdot343+2\cdot49+2\cdot7
    +1+49+2\cdot7+2\cdot49+2\cdot7
    \Big]=214.
\end{align}
Thus, before using any dynamical information about bimetric theory, the $2401$ ordered helicity configurations fall into $214$ symmetry-inequivalent classes.

We now remove the amplitudes containing precisely one massive external field, which we saw in Section \ref{subsec:parameter_structure} vanish identically. Let $\Acal_M$ denote the subset of amplitudes with one $M$. Since $\Gcal$ preserves the number of massive particles, $\Acal_M$ is invariant under the group action and can be counted separately. There are,
\begin{align}
    |\Acal_M|=4\times5\times2^3=160,
\end{align}
such ordered configurations \eqref{nr_MGGG}.

Applying Burnside's lemma to $\Acal_M$, the identity fixes all $160$ configurations, while $s_i$ and $s_f$ each fix,
\begin{align}
    |\operatorname{Fix}_{1M}(s_i)|&=2\times5\times2^2=40,\\
    |\operatorname{Fix}_{1M}(s_f)|&=2\times5\times2^2=40.
\end{align}
No other group element has fixed points in $\Acal_M$. Hence,
\begin{align}
    |\Acal_M/\Gcal|
    =\frac{1}{16}\left(160+2\times40\right)
    =15.
\end{align}
These $15$ representatives vanish identically, so the number of potentially non-vanishing $\Gcal$-inequivalent amplitudes is,
\begin{align}
    |\Acal^{2\to2}/\Gcal|
    =214-15
    =199.
\end{align}
Thus, it is sufficient to compute one representative from each of these $199$ orbits.

\subsection{Construction of orbit representatives}
Representatives for each orbit are listed in Tables~\ref{tab:Amps_GGGG}--\ref{tab:HighEnergyMMMM}. They were obtained with \textsc{Mathematica} using the notebook \texttt{GetOrbitsBinAmps.nb} provided in \cite{FelixHerber_TreeLevelBimetricCalculations_2025}. To construct the orbits, each symmetry in $\Gcal$ is represented by a $4\times4$ matrix $Z$, whose action encodes both permutations of the external legs and possible sign flips of the helicities $(h_A,h_B,h_C,h_D)$.

We identify each element of $\mathcal{A}^{2\to2}$ with a tuple,
\begin{align}
    \big((A,h_A),(B,h_B),(C,h_C),(D,h_D)\big).
\end{align}
The symmetry transformation associated with $Z$ maps this tuple to,
\begin{align}
    \big((A',h_A'),(B',h_B'),(C',h_C'),(D',h_D')\big),
\end{align}
with,
\begin{align}
    (h_A',h_B',h_C',h_D') &= Z(h_A,h_B,h_C,h_D), \\
    (A',B',C',D') &= \widetilde Z(A,B,C,D),
\end{align}
where $\widetilde Z$ is obtained from $Z$ by taking the absolute value of each matrix element, $\widetilde Z_{ij}=|Z_{ij}|$. Thus, $Z$ acts on the helicities by permutation and possible sign reversal, while $\widetilde Z$ implements the corresponding permutation of the particle labels.

Representing the symmetries in \eqref{Time_Reversal}, \eqref{Parity_Bimetric} and \eqref{Swap_Symmetry} in this way, repeated matrix multiplication generates the full matrix representation of $\Gcal$. For each $x\in\mathcal{A}^{2\to2}$, its orbit is then constructed as,
\begin{align}
    \Gcal\cdot x=\{g\cdot x\}_{g\in\Gcal}.
\end{align}
These orbits partition $\mathcal{A}^{2\to2}$ into $199$ disjoint equivalence classes, from each of which one representative is chosen.

\section{Amplitudes at high energy}
\label{sec:List_of_Orbits}
In the tables below, we list the high energy behavior of all 199 inequivalent amplitude classes for $2 \to 2$ processes at tree-level in bimetric theory. To make equations less cluttered, we use the notation $\beta_{\pm}$ and $Q_{\pm}$ defined by \eqref{Bimetric_Parameter_Combinations} and \eqref{Q_pm_Definition}. An entry equal to $0$ denotes an amplitude that vanishes at tree-level. For $GG \to GG$, the listed high-energy expressions coincide with the full amplitudes, see Table \ref{tab:Amps_GGGG}. A vanishing leading high-energy term for a particular parameter choice does not in general imply that the full amplitude vanishes, since subleading terms may remain. The high energy behaviours for the sectors $GG \to MM$, $GM \to GM$, $GM \to MM$ and $MM \to MM$ are in Tables \ref{tab:HighEnergyGGMM}, \ref{tab:HighEnergyGMGM}, \ref{tab:HighEnergyGMMM} and \ref{tab:HighEnergyMMMM} respectively.

\begingroup
\footnotesize
\renewcommand{\arraystretch}{2}
\begin{longtable}{m{0.3\textwidth}c}\\
\toprule
\textbf{Amplitude} & \textbf{Scaling} \\
\midrule
\endfirsthead
\textbf{Amplitude} & \textbf{Scaling} \\
\endhead
$\displaystyle \mathcal{A}^{GG\to GG}_{2,2;2,2}(s,t,u) = \frac{s^3}{2 t u m_{\text{Pl}}^2}$ & $E^{2}$ \\
$\displaystyle \mathcal{A}^{GG\to GG}_{2,2;2,-2}(s,t,u) = 0$ & $\text{--}$ \\
$\displaystyle \mathcal{A}^{GG\to GG}_{2,2;-2,-2}(s,t,u) = 0$ & $\text{--}$ \\
$\displaystyle \mathcal{A}^{GG\to GG}_{2,-2;2,-2}(s,t,u) = \frac{u^3}{2 s t m_{\text{Pl}}^2}$ & $E^{2}$ \\
\bottomrule
\caption{The $GG\to GG$ helicity amplitudes, which are identical in their high energy limit.}
\label{tab:Amps_GGGG}
\end{longtable}
\endgroup

\begingroup
\footnotesize
\renewcommand{\arraystretch}{2.3}
\begin{longtable}{m{0.31\textwidth}c}\\
\toprule
\textbf{Amplitude} & \textbf{Scaling} \\
\midrule

\endfirsthead

\textbf{Amplitude} & \textbf{Scaling} \\

\endhead
$\displaystyle \mathcal{A}^{GG\to MM}_{2,2;2,2}(s,t,u) = \frac{s^3}{2 t u m_{\text{Pl}}^2}$ & $E^{2}$ \\

$\displaystyle \mathcal{A}^{GG\to MM}_{2,2;2,1}(s,t,u) = 0$ & $\text{--}$ \\

$\displaystyle \mathcal{A}^{GG\to MM}_{2,2;2,0}(s,t,u) = 0$ & $\text{--}$ \\

$\displaystyle \mathcal{A}^{GG\to MM}_{2,2;2,-1}(s,t,u) = 0$ & $\text{--}$ \\

$\displaystyle \mathcal{A}^{GG\to MM}_{2,2;2,-2}(s,t,u) = 0$ & $\text{--}$ \\

$\displaystyle \mathcal{A}^{GG\to MM}_{2,2;1,1}(s,t,u) = \frac{s^2 m_{\text{FP}}^2}{2 t u m_{\text{Pl}}^2}$ & $E^{0}$ \\

$\displaystyle \mathcal{A}^{GG\to MM}_{2,2;1,0}(s,t,u) = 0$ & $\text{--}$ \\

$\displaystyle \mathcal{A}^{GG\to MM}_{2,2;1,-1}(s,t,u) = 0$ & $\text{--}$ \\

$\displaystyle \mathcal{A}^{GG\to MM}_{2,2;1,-2}(s,t,u) = 0$ & $\text{--}$ \\

$\displaystyle \mathcal{A}^{GG\to MM}_{2,2;0,0}(s,t,u) = \frac{s m_{\text{FP}}^4}{2 t u m_{\text{Pl}}^2}$ & $E^{-2}$ \\

$\displaystyle \mathcal{A}^{GG\to MM}_{2,2;0,-1}(s,t,u) = 0$ & $\text{--}$ \\

$\displaystyle \mathcal{A}^{GG\to MM}_{2,2;0,-2}(s,t,u) = 0$ & $\text{--}$ \\

$\displaystyle \mathcal{A}^{GG\to MM}_{2,2;-1,-1}(s,t,u) = \frac{m_{\text{FP}}^6}{2 t u m_{\text{Pl}}^2}$ & $E^{-4}$ \\

$\displaystyle \mathcal{A}^{GG\to MM}_{2,2;-1,-2}(s,t,u) = 0$ & $\text{--}$ \\

$\displaystyle \mathcal{A}^{GG\to MM}_{2,2;-2,-2}(s,t,u) = \frac{m_{\text{FP}}^8}{2 s t u m_{\text{Pl}}^2}$ & $E^{-6}$ \\

$\displaystyle \mathcal{A}^{GG\to MM}_{2,-2;2,2}(s,t,u) = \frac{8 t u m_{\text{FP}}^4}{s^3 m_{\text{Pl}}^2}$ & $E^{-2}$ \\

$\displaystyle \mathcal{A}^{GG\to MM}_{2,-2;2,1}(s,t,u) = -\frac{8 m_{\text{FP}}^3 \sqrt{t u^3}}{s^{5/2} m_{\text{Pl}}^2}$ & $E^{-1}$ \\

$\displaystyle \mathcal{A}^{GG\to MM}_{2,-2;2,0}(s,t,u) = -\frac{2 \sqrt{6} u^2 m_{\text{FP}}^2}{s^2 m_{\text{Pl}}^2}$ & $E^{0}$ \\

$\displaystyle \mathcal{A}^{GG\to MM}_{2,-2;2,-1}(s,t,u) = \frac{2 (-u)^{5/2} m_{\text{FP}}}{m_{\text{Pl}}^2 \sqrt{-s^3 t}}$ & $E^{1}$ \\

$\displaystyle \mathcal{A}^{GG\to MM}_{2,-2;2,-2}(s,t,u) = \frac{u^3}{2 s t m_{\text{Pl}}^2}$ & $E^{2}$ \\

$\displaystyle \mathcal{A}^{GG\to MM}_{2,-2;1,1}(s,t,u) = -\frac{6 t u m_{\text{FP}}^2}{s^2 m_{\text{Pl}}^2}$ & $E^{0}$ \\

$\displaystyle \mathcal{A}^{GG\to MM}_{2,-2;1,0}(s,t,u) = -\frac{\sqrt{6} m_{\text{FP}} \sqrt{t u^3}}{s^{3/2} m_{\text{Pl}}^2}$ & $E^{1}$ \\

$\displaystyle \mathcal{A}^{GG\to MM}_{2,-2;1,-1}(s,t,u) = \frac{u^2}{2 s m_{\text{Pl}}^2}$ & $E^{2}$ \\

$\displaystyle \mathcal{A}^{GG\to MM}_{2,-2;0,0}(s,t,u) = \frac{t u}{2 s m_{\text{Pl}}^2}$ & $E^{2}$ \\
\bottomrule 
\caption{Leading high-energy behavior of the $GG\to MM$ helicity amplitudes.}
\label{tab:HighEnergyGGMM}
\end{longtable}
\endgroup

\begingroup
\footnotesize
\renewcommand{\arraystretch}{2.3}
\begin{longtable}{m{0.45\textwidth}c}\\
\toprule
\textbf{Amplitude} & \textbf{Scaling} \\
\midrule
\endfirsthead

\textbf{Amplitude} & \textbf{Scaling} \\

\endhead
$\displaystyle \mathcal{A}^{GM\to GM}_{2,2;2,2}(s,t,u) = \frac{s^3}{2 t u m_{\text{Pl}}^2}$ & $E^{2}$ \\

$\displaystyle \mathcal{A}^{GM\to GM}_{2,2;2,1}(s,t,u) = \frac{s^2 m_{\text{FP}}}{m_{\text{Pl}}^2 \sqrt{s t u}}$ & $E^{1}$ \\

$\displaystyle \mathcal{A}^{GM\to GM}_{2,2;2,0}(s,t,u) = -\frac{\sqrt{\frac{3}{2}} m_{\text{FP}}^2}{m_{\text{Pl}}^2}$ & $E^{0}$ \\

$\displaystyle \mathcal{A}^{GM\to GM}_{2,2;2,-1}(s,t,u) = -\frac{m_{\text{FP}}^3 \sqrt{\frac{t u}{s^3}}}{m_{\text{Pl}}^2}$ & $E^{-1}$ \\

$\displaystyle \mathcal{A}^{GM\to GM}_{2,2;2,-2}(s,t,u) = \frac{t u m_{\text{FP}}^4}{2 s^3 m_{\text{Pl}}^2}$ & $E^{-2}$ \\

$\displaystyle \mathcal{A}^{GM\to GM}_{2,2;-2,2}(s,t,u) = \frac{t u m_{\text{FP}}^4}{2 s^3 m_{\text{Pl}}^2}$ & $E^{-2}$ \\

$\displaystyle \mathcal{A}^{GM\to GM}_{2,2;-2,1}(s,t,u) = -\frac{t m_{\text{FP}}^5 \sqrt{\frac{t u}{s^7}}}{m_{\text{Pl}}^2}$ & $E^{-3}$ \\

$\displaystyle \mathcal{A}^{GM\to GM}_{2,2;-2,0}(s,t,u) = -\frac{\sqrt{\frac{3}{2}} t^2 m_{\text{FP}}^6}{s^4 m_{\text{Pl}}^2}$ & $E^{-4}$ \\

$\displaystyle \mathcal{A}^{GM\to GM}_{2,2;-2,-1}(s,t,u) = \frac{t^3 m_{\text{FP}}^7}{m_{\text{Pl}}^2 \sqrt{s^9 t u}}$ & $E^{-5}$ \\

$\displaystyle \mathcal{A}^{GM\to GM}_{2,2;-2,-2}(s,t,u) = \frac{t^3 m_{\text{FP}}^8}{2 s^5 u m_{\text{Pl}}^2}$ & $E^{-6}$ \\

$\displaystyle \mathcal{A}^{GM\to GM}_{2,1;2,1}(s,t,u) = -\frac{s^2}{2 t m_{\text{Pl}}^2}$ & $E^{2}$ \\

$\displaystyle \mathcal{A}^{GM\to GM}_{2,1;2,0}(s,t,u) = \frac{\sqrt{\frac{3}{2}} s u m_{\text{FP}}}{m_{\text{Pl}}^2 \sqrt{s t u}}$ & $E^{1}$ \\

$\displaystyle \mathcal{A}^{GM\to GM}_{2,1;2,-1}(s,t,u) = \frac{3 u m_{\text{FP}}^2}{2 s m_{\text{Pl}}^2}$ & $E^{0}$ \\

$\displaystyle \mathcal{A}^{GM\to GM}_{2,1;2,-2}(s,t,u) = \frac{m_{\text{FP}}^3 \sqrt{s t u^3}}{s^3 m_{\text{Pl}}^2}$ & $E^{-1}$ \\

$\displaystyle \mathcal{A}^{GM\to GM}_{2,1;-2,2}(s,t,u) = \frac{t m_{\text{FP}}^3 \sqrt{\frac{t u}{s^5}}}{m_{\text{Pl}}^2}$ & $E^{-1}$ \\

$\displaystyle \mathcal{A}^{GM\to GM}_{2,1;-2,1}(s,t,u) = -\frac{t m_{\text{FP}}^4 (s+4 t)}{2 s^3 m_{\text{Pl}}^2}$ & $E^{-2}$ \\

$\displaystyle \mathcal{A}^{GM\to GM}_{2,1;-2,0}(s,t,u) = -\frac{\sqrt{\frac{3}{2}} t^2 m_{\text{FP}}^5 (s+2 t)}{m_{\text{Pl}}^2 \sqrt{s^7 t u}}$ & $E^{-3}$ \\

$\displaystyle \mathcal{A}^{GM\to GM}_{2,1;-2,-1}(s,t,u) = \frac{t^2 m_{\text{FP}}^6 (3 s+4 t)}{2 s^4 u m_{\text{Pl}}^2}$ & $E^{-4}$ \\

$\displaystyle \mathcal{A}^{GM\to GM}_{2,0;2,0}(s,t,u) = \frac{s u}{2 t m_{\text{Pl}}^2}$ & $E^{2}$ \\

$\displaystyle \mathcal{A}^{GM\to GM}_{2,0;2,-1}(s,t,u) = \frac{\sqrt{\frac{3}{2}} m_{\text{FP}} \left(\sqrt{\frac{s^3 u}{t}}-\sqrt{s t u}\right)}{s m_{\text{Pl}}^2}$ & $E^{1}$ \\

$\displaystyle \mathcal{A}^{GM\to GM}_{2,0;2,-2}(s,t,u) = -\frac{\sqrt{\frac{3}{2}} u^2 m_{\text{FP}}^2}{s^2 m_{\text{Pl}}^2}$ & $E^{0}$ \\

$\displaystyle \mathcal{A}^{GM\to GM}_{2,0;-2,2}(s,t,u) = -\frac{\sqrt{\frac{3}{2}} t^2 m_{\text{FP}}^2}{s^2 m_{\text{Pl}}^2}$ & $E^{0}$ \\

$\displaystyle \mathcal{A}^{GM\to GM}_{2,0;-2,1}(s,t,u) = \frac{\sqrt{\frac{3}{2}} t^2 m_{\text{FP}}^3 (s+2 t)}{m_{\text{Pl}}^2 \sqrt{s^5 t u}}$ & $E^{-1}$ \\

$\displaystyle \mathcal{A}^{GM\to GM}_{2,0;-2,0}(s,t,u) = \frac{t m_{\text{FP}}^4 \left(s^2+6 s t+6 t^2\right)}{2 s^3 u m_{\text{Pl}}^2}$ & $E^{-2}$ \\

$\displaystyle \mathcal{A}^{GM\to GM}_{2,-1;2,-1}(s,t,u) = -\frac{u^2}{2 t m_{\text{Pl}}^2}$ & $E^{2}$ \\

$\displaystyle \mathcal{A}^{GM\to GM}_{2,-1;2,-2}(s,t,u) = \frac{t u^4 m_{\text{FP}}}{m_{\text{Pl}}^2 (s t u)^{3/2}}$ & $E^{1}$ \\

$\displaystyle \mathcal{A}^{GM\to GM}_{2,-1;-2,2}(s,t,u) = -\frac{t^4 u m_{\text{FP}}}{m_{\text{Pl}}^2 (s t u)^{3/2}}$ & $E^{1}$ \\

$\displaystyle \mathcal{A}^{GM\to GM}_{2,-1;-2,1}(s,t,u) = \frac{t^2 m_{\text{FP}}^2 (3 s+4 t)}{2 s^2 u m_{\text{Pl}}^2}$ & $E^{0}$ \\

$\displaystyle \mathcal{A}^{GM\to GM}_{2,-2;2,-2}(s,t,u) = \frac{u^3}{2 s t m_{\text{Pl}}^2}$ & $E^{2}$ \\

$\displaystyle \mathcal{A}^{GM\to GM}_{2,-2;-2,2}(s,t,u) = \frac{t^3}{2 s u m_{\text{Pl}}^2}$ & $E^{2}$ \\
\bottomrule
\caption{Leading high-energy behaviour of the $GM\to GM$ helicity amplitudes.}
\label{tab:HighEnergyGMGM}
\end{longtable}
\endgroup

\begingroup
\footnotesize
\renewcommand{\arraystretch}{3}
\begin{longtable}{m{0.6\textwidth}c}\\
\toprule
\textbf{Amplitude} & \textbf{Scaling} \\
\midrule
\endfirsthead

\textbf{Amplitude} & \textbf{Scaling} \\

\endhead
$\displaystyle \mathcal{A}^{GM\to MM}_{2,2;2,2}(s,t,u) = -\frac{Q_- s^3}{2 t u m_{\text{Pl}}^2}$ & $E^{2}$ \\

$\displaystyle \mathcal{A}^{GM\to MM}_{2,2;2,1}(s,t,u) = \frac{t u \left(\beta _- m^4 Q_+^3 u \left(s^2+2 s t+2 t^2\right)-2 Q_- s^3 m_{\text{FP}}^2 m_{\text{Pl}}^2\right)}{4 m_{\text{FP}} m_{\text{Pl}}^4 (s t u)^{3/2}}$ & $E^{1}$ \\

$\displaystyle \mathcal{A}^{GM\to MM}_{2,2;2,0}(s,t,u) = \frac{\beta _- m^4 Q_+^3 u^2}{2 \sqrt{6} s m_{\text{FP}}^2 m_{\text{Pl}}^4}$ & $E^{2}$ \\

$\displaystyle \mathcal{A}^{GM\to MM}_{2,2;2,-1}(s,t,u) = \frac{\beta _- m^4 Q_+^3 \left(\sqrt{s^3 t u}+t \sqrt{s t u}\right)}{4 s^2 m_{\text{FP}} m_{\text{Pl}}^4}$ & $E^{1}$ \\

$\displaystyle \mathcal{A}^{GM\to MM}_{2,2;2,-2}(s,t,u) = -\frac{Q_- t u m_{\text{FP}}^4}{2 s^3 m_{\text{Pl}}^2}$ & $E^{-2}$ \\

$\displaystyle \mathcal{A}^{GM\to MM}_{2,2;1,1}(s,t,u) = \frac{\beta _- m^4 Q_+^3 t u}{2 s m_{\text{FP}}^2 m_{\text{Pl}}^4}$ & $E^{2}$ \\

$\displaystyle \mathcal{A}^{GM\to MM}_{2,2;1,0}(s,t,u) = -\frac{\beta _- m^4 Q_+^3 t u^2}{2 \sqrt{6} m_{\text{FP}}^3 m_{\text{Pl}}^4 \sqrt{s t u}}$ & $E^{3}$ \\

$\displaystyle \mathcal{A}^{GM\to MM}_{2,2;1,-1}(s,t,u) = \frac{\beta _- m^4 Q_+^3 t u}{4 s m_{\text{FP}}^2 m_{\text{Pl}}^4}$ & $E^{2}$ \\

$\displaystyle \mathcal{A}^{GM\to MM}_{2,2;1,-2}(s,t,u) = -\frac{\beta _- m^4 Q_+^3 m_{\text{FP}} \left(\sqrt{s^3 t u}+t \sqrt{s t u}\right)}{4 s^3 m_{\text{Pl}}^4}$ & $E^{-1}$ \\

$\displaystyle \mathcal{A}^{GM\to MM}_{2,2;0,0}(s,t,u) = -\frac{\beta _- m^4 Q_+^3 t u}{12 m_{\text{FP}}^4 m_{\text{Pl}}^4}$ & $E^{4}$ \\

$\displaystyle \mathcal{A}^{GM\to MM}_{2,2;0,-1}(s,t,u) = \frac{\beta _- m^4 Q_+^3 t \sqrt{s t u}}{4 \sqrt{6} s m_{\text{FP}}^3 m_{\text{Pl}}^4}$ & $E^{3}$ \\

$\displaystyle \mathcal{A}^{GM\to MM}_{2,2;0,-2}(s,t,u) = \frac{\beta _- m^4 Q_+^3 t u}{\sqrt{6} s^2 m_{\text{Pl}}^4}$ & $E^{0}$ \\

$\displaystyle \mathcal{A}^{GM\to MM}_{2,2;-1,-1}(s,t,u) = \frac{\beta _- m^4 Q_+^3 \left(s^2+2 s t+2 t^2\right)}{4 s^2 m_{\text{Pl}}^4}$ & $E^{0}$ \\

$\displaystyle \mathcal{A}^{GM\to MM}_{2,2;-1,-2}(s,t,u) = -\frac{\beta _- m^4 Q_+^3 t \sqrt{s t u}}{4 s^2 m_{\text{FP}} m_{\text{Pl}}^4}$ & $E^{1}$ \\

$\displaystyle \mathcal{A}^{GM\to MM}_{2,2;-2,-2}(s,t,u) = \frac{\beta _- m^4 Q_+^3 t u}{2 s^2 m_{\text{Pl}}^4}$ & $E^{0}$ \\

$\displaystyle \mathcal{A}^{GM\to MM}_{2,1;2,2}(s,t,u) = \frac{\beta _- m^4 Q_+^3 t u (s+2 t) \left(2 s^2+s t+t^2\right)}{4 m_{\text{FP}} m_{\text{Pl}}^4 (s t u)^{3/2}}$ & $E^{1}$ \\

$\displaystyle \mathcal{A}^{GM\to MM}_{2,1;2,1}(s,t,u) = \frac{2 Q_- s^3 m_{\text{FP}}^2 m_{\text{Pl}}^2+\beta _- m^4 Q_+^3 (s+2 t) \left(s^2+2 s t+2 t^2\right)}{8 s t m_{\text{FP}}^2 m_{\text{Pl}}^4}$ & $E^{2}$ \\

$\displaystyle \mathcal{A}^{GM\to MM}_{2,1;2,0}(s,t,u) = \frac{\beta _- m^4 Q_+^3 u^2 (s+2 t)}{4 \sqrt{6} m_{\text{FP}}^3 m_{\text{Pl}}^4 \sqrt{s t u}}$ & $E^{3}$ \\

$\displaystyle \mathcal{A}^{GM\to MM}_{2,1;2,-1}(s,t,u) = \frac{\beta _- m^4 Q_+^3 u^2}{4 s m_{\text{FP}}^2 m_{\text{Pl}}^4}$ & $E^{2}$ \\

$\displaystyle \mathcal{A}^{GM\to MM}_{2,1;2,-2}(s,t,u) = \frac{Q_- u m_{\text{FP}}^3 \sqrt{t u}}{s^{5/2} m_{\text{Pl}}^2}$ & $E^{-1}$ \\

$\displaystyle \mathcal{A}^{GM\to MM}_{2,1;1,1}(s,t,u) = \frac{\beta _- m^4 Q_+^3 t u (s+2 t)}{4 m_{\text{FP}}^3 m_{\text{Pl}}^4 \sqrt{s t u}}$ & $E^{3}$ \\

$\displaystyle \mathcal{A}^{GM\to MM}_{2,1;1,0}(s,t,u) = -\frac{\beta _- m^4 Q_+^3 u (s+2 t)}{4 \sqrt{6} m_{\text{FP}}^4 m_{\text{Pl}}^4}$ & $E^{4}$ \\

$\displaystyle \mathcal{A}^{GM\to MM}_{2,1;1,-1}(s,t,u) = -\frac{\beta _- m^4 Q_+^3 u \sqrt{s t u}}{4 s m_{\text{FP}}^3 m_{\text{Pl}}^4}$ & $E^{3}$ \\

$\displaystyle \mathcal{A}^{GM\to MM}_{2,1;1,-2}(s,t,u) = -\frac{\beta _- m^4 Q_+^3 u^2}{4 s^2 m_{\text{Pl}}^4}$ & $E^{0}$ \\

$\displaystyle \mathcal{A}^{GM\to MM}_{2,1;0,0}(s,t,u) = -\frac{\beta _- m^4 Q_+^3 (s+2 t) \sqrt{s t u}}{24 m_{\text{FP}}^5 m_{\text{Pl}}^4}$ & $E^{5}$ \\

$\displaystyle \mathcal{A}^{GM\to MM}_{2,1;0,-1}(s,t,u) = -\frac{\beta _- m^4 Q_+^3 t u}{4 \sqrt{6} m_{\text{FP}}^4 m_{\text{Pl}}^4}$ & $E^{4}$ \\

$\displaystyle \mathcal{A}^{GM\to MM}_{2,1;0,-2}(s,t,u) = -\frac{\beta _- m^4 Q_+^3 u \sqrt{s t u}}{\sqrt{6} s^2 m_{\text{FP}} m_{\text{Pl}}^4}$ & $E^{1}$ \\

$\displaystyle \mathcal{A}^{GM\to MM}_{2,1;-1,-1}(s,t,u) = -\frac{\beta _- m^4 Q_+^3 (s+2 t) \sqrt{s t u}}{4 s^2 m_{\text{FP}} m_{\text{Pl}}^4}$ & $E^{1}$ \\

$\displaystyle \mathcal{A}^{GM\to MM}_{2,1;-1,-2}(s,t,u) = \frac{\beta _- m^4 Q_+^3 t u}{4 s m_{\text{FP}}^2 m_{\text{Pl}}^4}$ & $E^{2}$ \\

$\displaystyle \mathcal{A}^{GM\to MM}_{2,1;-2,-2}(s,t,u) = \frac{\beta _- m^4 Q_+^3 (s+2 t) \sqrt{s t u}}{4 s^2 m_{\text{FP}} m_{\text{Pl}}^4}$ & $E^{1}$ \\

$\displaystyle \mathcal{A}^{GM\to MM}_{2,0;2,2}(s,t,u) = \frac{\beta _- m^4 Q_+^3 \left(2 s^2+s t+t^2\right)}{2 \sqrt{6} s m_{\text{FP}}^2 m_{\text{Pl}}^4}$ & $E^{2}$ \\

$\displaystyle \mathcal{A}^{GM\to MM}_{2,0;2,1}(s,t,u) = \frac{\beta _- m^4 Q_+^3 u \left(s^2+2 s t+2 t^2\right)}{4 \sqrt{6} m_{\text{FP}}^3 m_{\text{Pl}}^4 \sqrt{s t u}}$ & $E^{3}$ \\

$\displaystyle \mathcal{A}^{GM\to MM}_{2,0;2,0}(s,t,u) = \frac{\beta _- m^4 Q_+^3 u^2}{12 m_{\text{FP}}^4 m_{\text{Pl}}^4}$ & $E^{4}$ \\

$\displaystyle \mathcal{A}^{GM\to MM}_{2,0;2,-1}(s,t,u) = \frac{\beta _- m^4 Q_+^3 u^3}{4 \sqrt{6} m_{\text{FP}}^3 m_{\text{Pl}}^4 \sqrt{s t u}}$ & $E^{3}$ \\

$\displaystyle \mathcal{A}^{GM\to MM}_{2,0;2,-2}(s,t,u) = \frac{\sqrt{\frac{3}{2}} Q_- u^2 m_{\text{FP}}^2}{s^2 m_{\text{Pl}}^2}$ & $E^{0}$ \\

$\displaystyle \mathcal{A}^{GM\to MM}_{2,0;1,1}(s,t,u) = \frac{\beta _- m^4 Q_+^3 t u}{2 \sqrt{6} m_{\text{FP}}^4 m_{\text{Pl}}^4}$ & $E^{4}$ \\

$\displaystyle \mathcal{A}^{GM\to MM}_{2,0;1,0}(s,t,u) = -\frac{\beta _- m^4 Q_+^3 u \sqrt{s t u}}{12 m_{\text{FP}}^5 m_{\text{Pl}}^4}$ & $E^{5}$ \\

$\displaystyle \mathcal{A}^{GM\to MM}_{2,0;1,-1}(s,t,u) = -\frac{\beta _- m^4 Q_+^3 u^2}{4 \sqrt{6} m_{\text{FP}}^4 m_{\text{Pl}}^4}$ & $E^{4}$ \\

$\displaystyle \mathcal{A}^{GM\to MM}_{2,0;1,-2}(s,t,u) = -\frac{\beta _- m^4 Q_+^3 t u^4}{4 \sqrt{6} m_{\text{FP}} m_{\text{Pl}}^4 (s t u)^{3/2}}$ & $E^{1}$ \\

$\displaystyle \mathcal{A}^{GM\to MM}_{2,0;0,0}(s,t,u) = -\frac{\beta _- m^4 Q_+^3 s t u}{12 \sqrt{6} m_{\text{FP}}^6 m_{\text{Pl}}^4}$ & $E^{6}$ \\

$\displaystyle \mathcal{A}^{GM\to MM}_{2,0;0,-1}(s,t,u) = -\frac{\beta _- m^4 Q_+^3 u \sqrt{s t u}}{24 m_{\text{FP}}^5 m_{\text{Pl}}^4}$ & $E^{5}$ \\

$\displaystyle \mathcal{A}^{GM\to MM}_{2,0;0,-2}(s,t,u) = -\frac{\beta _- m^4 Q_+^3 u^2}{6 s m_{\text{FP}}^2 m_{\text{Pl}}^4}$ & $E^{2}$ \\

$\displaystyle \mathcal{A}^{GM\to MM}_{2,0;-1,-1}(s,t,u) = -\frac{\beta _- m^4 Q_+^3 t u}{2 \sqrt{6} s m_{\text{FP}}^2 m_{\text{Pl}}^4}$ & $E^{2}$ \\

$\displaystyle \mathcal{A}^{GM\to MM}_{2,0;-1,-2}(s,t,u) = \frac{\beta _- m^4 Q_+^3 t u^2}{4 \sqrt{6} m_{\text{FP}}^3 m_{\text{Pl}}^4 \sqrt{s t u}}$ & $E^{3}$ \\

$\displaystyle \mathcal{A}^{GM\to MM}_{2,0;-2,-2}(s,t,u) = \frac{\beta _- m^4 Q_+^3 t u}{2 \sqrt{6} s m_{\text{FP}}^2 m_{\text{Pl}}^4}$ & $E^{2}$ \\

$\displaystyle \mathcal{A}^{GM\to MM}_{2,-1;2,2}(s,t,u) = \frac{\beta _- m^4 Q_+^3 m_{\text{FP}} (s+2 t) \sqrt{s t u}}{s^3 m_{\text{Pl}}^4}$ & $E^{-1}$ \\

$\displaystyle \mathcal{A}^{GM\to MM}_{2,-1;2,1}(s,t,u) = -\frac{\beta _- m^4 Q_+^3 u}{4 m_{\text{FP}}^2 m_{\text{Pl}}^4}$ & $E^{2}$ \\

$\displaystyle \mathcal{A}^{GM\to MM}_{2,-1;2,0}(s,t,u) = \frac{\beta _- m^4 Q_+^3 s u^2}{4 \sqrt{6} m_{\text{FP}}^3 m_{\text{Pl}}^4 \sqrt{s t u}}$ & $E^{3}$ \\

$\displaystyle \mathcal{A}^{GM\to MM}_{2,-1;2,-1}(s,t,u) = \frac{u^2 \left(2 Q_- m_{\text{FP}}^2 m_{\text{Pl}}^2+\beta _- m^4 Q_+^3\right)}{8 t m_{\text{FP}}^2 m_{\text{Pl}}^4}$ & $E^{2}$ \\

$\displaystyle \mathcal{A}^{GM\to MM}_{2,-1;2,-2}(s,t,u) = -\frac{Q_- t u^4 m_{\text{FP}}}{m_{\text{Pl}}^2 (s t u)^{3/2}}$ & $E^{1}$ \\

$\displaystyle \mathcal{A}^{GM\to MM}_{2,-1;1,1}(s,t,u) = -\frac{\beta _- m^4 Q_+^3 (s+2 t) \sqrt{s t u}}{2 s^2 m_{\text{FP}} m_{\text{Pl}}^4}$ & $E^{1}$ \\

$\displaystyle \mathcal{A}^{GM\to MM}_{2,-1;1,0}(s,t,u) = \frac{\beta _- m^4 Q_+^3 t u}{2 \sqrt{6} s m_{\text{FP}}^2 m_{\text{Pl}}^4}$ & $E^{2}$ \\

$\displaystyle \mathcal{A}^{GM\to MM}_{2,-1;1,-1}(s,t,u) = \frac{t u^3 \left(2 Q_- u m_{\text{FP}}^2 m_{\text{Pl}}^2+\beta _- m^4 Q_+^3 t\right)}{4 m_{\text{FP}} m_{\text{Pl}}^4 (s t u)^{3/2}}$ & $E^{1}$ \\

$\displaystyle \mathcal{A}^{GM\to MM}_{2,-1;1,-2}(s,t,u) = \frac{u^2 \left(2 Q_- m_{\text{FP}}^2 m_{\text{Pl}}^2 (s+12 t)-\beta _- m^4 Q_+^3 s\right)}{8 s^2 t m_{\text{Pl}}^4}$ & $E^{0}$ \\

$\displaystyle \mathcal{A}^{GM\to MM}_{2,-1;0,0}(s,t,u) = -\frac{(s+2 t) \sqrt{s t u} \left(Q_- m_{\text{FP}}^2 m_{\text{Pl}}^2-\beta _- m^4 Q_+^3\right)}{4 s^2 m_{\text{FP}} m_{\text{Pl}}^4}$ & $E^{1}$ \\

$\displaystyle \mathcal{A}^{GM\to MM}_{2,-1;0,-1}(s,t,u) = -\frac{u (s+2 t) \left(12 Q_- m_{\text{FP}}^2 m_{\text{Pl}}^2-\beta _- m^4 Q_+^3\right)}{4 \sqrt{6} s^2 m_{\text{Pl}}^4}$ & $E^{0}$ \\

$\displaystyle \mathcal{A}^{GM\to MM}_{2,-1;0,-2}(s,t,u) = \frac{u^2 m_{\text{FP}} \left(12 Q_- m_{\text{FP}}^2 m_{\text{Pl}}^2 (s+5 t)-\beta _- m^4 Q_+^3 (s-4 t)\right)}{4 \sqrt{6} s^2 m_{\text{Pl}}^4 \sqrt{s t u}}$ & $E^{-1}$ \\

$\displaystyle \mathcal{A}^{GM\to MM}_{2,-1;-1,-1}(s,t,u) = \frac{m_{\text{FP}} (s+2 t) \sqrt{s t u} \left(12 Q_- m_{\text{FP}}^2 m_{\text{Pl}}^2+\beta _- m^4 Q_+^3\right)}{4 s^3 m_{\text{Pl}}^4}$ & $E^{-1}$ \\

$\displaystyle \mathcal{A}^{GM\to MM}_{2,-1;-1,-2}(s,t,u) = -\frac{u m_{\text{FP}}^2 \left(12 Q_- m_{\text{FP}}^2 m_{\text{Pl}}^2 (s+3 t)+\beta _- m^4 Q_+^3 (s+4 t)\right)}{4 s^3 m_{\text{Pl}}^4}$ & $E^{-2}$ \\

$\displaystyle \mathcal{A}^{GM\to MM}_{2,-1;-2,-2}(s,t,u) = -\frac{m_{\text{FP}}^3 (s+2 t) \sqrt{s t u} \left(9 Q_- m_{\text{FP}}^2 m_{\text{Pl}}^2+\beta _- m^4 Q_+^3\right)}{2 s^4 m_{\text{Pl}}^4}$ & $E^{-3}$ \\

$\displaystyle \mathcal{A}^{GM\to MM}_{2,-2;2,2}(s,t,u) = \frac{\beta _- m^4 Q_+^3 t u}{s^2 m_{\text{Pl}}^4}$ & $E^{0}$ \\

$\displaystyle \mathcal{A}^{GM\to MM}_{2,-2;2,1}(s,t,u) = -\frac{3 \beta _- m^4 Q_+^3 \left(\sqrt{s^3 t u}+t \sqrt{s t u}\right)}{4 s^2 m_{\text{FP}} m_{\text{Pl}}^4}$ & $E^{1}$ \\

$\displaystyle \mathcal{A}^{GM\to MM}_{2,-2;2,0}(s,t,u) = -\frac{\beta _- m^4 Q_+^3 u^2}{2 \sqrt{6} s m_{\text{FP}}^2 m_{\text{Pl}}^4}$ & $E^{2}$ \\

$\displaystyle \mathcal{A}^{GM\to MM}_{2,-2;2,-1}(s,t,u) = \frac{t u^4 \left(6 Q_- m_{\text{FP}}^2 m_{\text{Pl}}^2-\beta _- m^4 Q_+^3\right)}{4 m_{\text{FP}} m_{\text{Pl}}^4 (s t u)^{3/2}}$ & $E^{1}$ \\

$\displaystyle \mathcal{A}^{GM\to MM}_{2,-2;2,-2}(s,t,u) = -\frac{Q_- u^3}{2 s t m_{\text{Pl}}^2}$ & $E^{2}$ \\

$\displaystyle \mathcal{A}^{GM\to MM}_{2,-2;1,1}(s,t,u) = -\frac{\beta _- m^4 Q_+^3 t u}{2 s m_{\text{FP}}^2 m_{\text{Pl}}^4}$ & $E^{2}$ \\

$\displaystyle \mathcal{A}^{GM\to MM}_{2,-2;1,0}(s,t,u) = \frac{\beta _- m^4 Q_+^3 u \sqrt{s t u}}{4 \sqrt{6} s m_{\text{FP}}^3 m_{\text{Pl}}^4}$ & $E^{3}$ \\

$\displaystyle \mathcal{A}^{GM\to MM}_{2,-2;1,-1}(s,t,u) = \frac{u^2 \left(\beta _- m^4 Q_+^3-2 Q_- m_{\text{FP}}^2 m_{\text{Pl}}^2\right)}{8 s m_{\text{FP}}^2 m_{\text{Pl}}^4}$ & $E^{2}$ \\

$\displaystyle \mathcal{A}^{GM\to MM}_{2,-2;1,-2}(s,t,u) = \frac{3 Q_- t u^4 m_{\text{FP}}}{2 m_{\text{Pl}}^2 (s t u)^{3/2}}$ & $E^{1}$ \\

$\displaystyle \mathcal{A}^{GM\to MM}_{2,-2;0,0}(s,t,u) = \frac{t u \left(\beta _- m^4 Q_+^3-Q_- m_{\text{FP}}^2 m_{\text{Pl}}^2\right)}{4 s m_{\text{FP}}^2 m_{\text{Pl}}^4}$ & $E^{2}$ \\

$\displaystyle \mathcal{A}^{GM\to MM}_{2,-2;0,-1}(s,t,u) = \frac{u \sqrt{s t u} \left(\beta _- m^4 Q_+^3-12 Q_- m_{\text{FP}}^2 m_{\text{Pl}}^2\right)}{4 \sqrt{6} s^2 m_{\text{FP}} m_{\text{Pl}}^4}$ & $E^{1}$ \\

$\displaystyle \mathcal{A}^{GM\to MM}_{2,-2;0,-2}(s,t,u) = \frac{u^2 \left(15 Q_- m_{\text{FP}}^2 m_{\text{Pl}}^2+\beta _- m^4 Q_+^3\right)}{2 \sqrt{6} s^2 m_{\text{Pl}}^4}$ & $E^{0}$ \\

$\displaystyle \mathcal{A}^{GM\to MM}_{2,-2;-1,-1}(s,t,u) = \frac{t u \left(12 Q_- m_{\text{FP}}^2 m_{\text{Pl}}^2+\beta _- m^4 Q_+^3\right)}{4 s^2 m_{\text{Pl}}^4}$ & $E^{0}$ \\

$\displaystyle \mathcal{A}^{GM\to MM}_{2,-2;-1,-2}(s,t,u) = \frac{m_{\text{FP}} \left(\sqrt{s^3 t u}+t \sqrt{s t u}\right) \left(9 Q_- m_{\text{FP}}^2 m_{\text{Pl}}^2+\beta _- m^4 Q_+^3\right)}{2 s^3 m_{\text{Pl}}^4}$ & $E^{-1}$ \\

$\displaystyle \mathcal{A}^{GM\to MM}_{2,-2;-2,-2}(s,t,u) = -\frac{t u m_{\text{FP}}^2 \left(9 Q_- m_{\text{FP}}^2 m_{\text{Pl}}^2+\beta _- m^4 Q_+^3\right)}{2 s^3 m_{\text{Pl}}^4}$ & $E^{-2}$ \\
\bottomrule
\caption{Leading high-energy behaviour of the $GM\to MM$ helicity amplitudes.}
\label{tab:HighEnergyGMMM}
\end{longtable}
\endgroup

\newpage
\begingroup
\footnotesize
\renewcommand{\arraystretch}{3}
\begin{longtable}{m{0.77\textwidth} c}\\
\toprule
\textbf{Amplitude} & \textbf{Scaling} \\
\midrule
\endfirsthead

\textbf{Amplitude} & \textbf{Scaling} \\

\endhead
$\displaystyle \mathcal{A}^{MM\to MM}_{2,2;2,2}(s,t,u) = \frac{s \left(6 \left(Q_+^2-3\right) s^2 \beta _{\text{FP}}^2-\beta _-^2 Q_+^2 t u\right)}{12 t u \beta _{\text{FP}}^2 m_{\text{Pl}}^2}$ & $E^{2}$ \\

$\displaystyle \mathcal{A}^{MM\to MM}_{2,2;2,1}(s,t,u) = \frac{Q_+ m_{\text{FP}} (s+2 t) \left(2 Q_+ \beta _{\text{FP}} \left(s^2-3 s t-3 t^2\right)+3 \beta _- Q_- \left(s^2+s t+t^2\right)\right)}{8 \beta _{\text{FP}} m_{\text{Pl}}^2 \sqrt{s^3 t u}}$ & $E^{1}$ \\

$\displaystyle \mathcal{A}^{MM\to MM}_{2,2;2,0}(s,t,u) = -\frac{Q_+ \left(Q_+ t u \left(\beta _++4 \beta _{\text{FP}}\right)+3 \beta _- Q_- \left(s^2-t u\right)\right)}{4 \sqrt{6} s \beta _{\text{FP}} m_{\text{Pl}}^2}$ & $E^{2}$ \\

$\displaystyle \mathcal{A}^{MM\to MM}_{2,2;2,-1}(s,t,u) = \frac{\beta _- Q_+ m_{\text{FP}} (s+2 t) \left(\beta _- Q_+-3 Q_- \beta _{\text{FP}}\right) \sqrt{s t u}}{24 s^2 \beta _{\text{FP}}^2 m_{\text{Pl}}^2}$ & $E^{1}$ \\

$\displaystyle \mathcal{A}^{MM\to MM}_{2,2;2,-2}(s,t,u) = -\frac{\beta _- Q_+ t u m_{\text{FP}}^2 \left(6 Q_- \beta _{\text{FP}}+\beta _- Q_+\right)}{12 s^2 \beta _{\text{FP}}^2 m_{\text{Pl}}^2}$ & $E^{0}$ \\

$\displaystyle \mathcal{A}^{MM\to MM}_{2,2;1,1}(s,t,u) = -\frac{\beta _-^2 Q_+^2 s^2}{48 \beta _{\text{FP}}^2 m_{\text{FP}}^2 m_{\text{Pl}}^2}$ & $E^{4}$ \\

$\displaystyle \mathcal{A}^{MM\to MM}_{2,2;1,0}(s,t,u) = \frac{Q_+ (s+2 t) \left(Q_+ \left(4 \beta _{\text{FP}}^2-\beta _-^2\right)+4 \beta _{\text{FP}} \left(\beta _+ Q_+-\beta _- Q_-\right)\right) \sqrt{\frac{t u}{s}}}{16 \sqrt{6} \beta _{\text{FP}}^2 m_{\text{FP}} m_{\text{Pl}}^2}$ & $E^{3}$ \\

$\displaystyle \mathcal{A}^{MM\to MM}_{2,2;1,-1}(s,t,u) = \frac{Q_+ t u \left(Q_+ \left(\beta _-^2+12 \beta _{\text{FP}}^2+12 \beta _+ \beta _{\text{FP}}\right)-24 \beta _- Q_- \beta _{\text{FP}}\right)}{96 s \beta _{\text{FP}}^2 m_{\text{Pl}}^2}$ & $E^{2}$ \\

$\displaystyle \mathcal{A}^{MM\to MM}_{2,2;1,-2}(s,t,u) = \frac{\beta _- Q_+ m_{\text{FP}}^3 (s+2 t) \left(18 Q_- \beta _{\text{FP}}+\beta _- Q_+\right) \sqrt{s t u}}{48 s^3 \beta _{\text{FP}}^2 m_{\text{Pl}}^2}$ & $E^{-1}$ \\

$\displaystyle \mathcal{A}^{MM\to MM}_{2,2;0,0}(s,t,u) = -\frac{Q_+ \left(-\beta _+ Q_+ \beta _{\text{FP}} \left(s^2-4 t u\right)-2 \beta _- Q_- t u \beta _{\text{FP}}+\beta _-^2 Q_+ \left(s^2+s t+t^2\right)\right)}{24 \beta _{\text{FP}}^2 m_{\text{FP}}^2 m_{\text{Pl}}^2}$ & $E^{4}$ \\

$\displaystyle \mathcal{A}^{MM\to MM}_{2,2;0,-1}(s,t,u) = -\frac{Q_+ (s+2 t) \left(2 \beta _- Q_- \beta _{\text{FP}}-4 \beta _+ Q_+ \beta _{\text{FP}}+\beta _-^2 Q_+\right) \sqrt{\frac{t u}{s}}}{16 \sqrt{6} \beta _{\text{FP}}^2 m_{\text{FP}} m_{\text{Pl}}^2}$ & $E^{3}$ \\

$\displaystyle \mathcal{A}^{MM\to MM}_{2,2;0,-2}(s,t,u) = \frac{Q_+ t u \left(Q_+ \left(\beta _-^2-2 \beta _+ \beta _{\text{FP}}\right)-2 \beta _- Q_- \beta _{\text{FP}}\right)}{8 \sqrt{6} s \beta _{\text{FP}}^2 m_{\text{Pl}}^2}$ & $E^{2}$ \\

$\displaystyle \mathcal{A}^{MM\to MM}_{2,2;-1,-1}(s,t,u) = -\frac{\beta _-^2 Q_+^2 \left(s^2+s t+t^2\right)}{48 \beta _{\text{FP}}^2 m_{\text{FP}}^2 m_{\text{Pl}}^2}$ & $E^{4}$ \\

$\displaystyle \mathcal{A}^{MM\to MM}_{2,2;-1,-2}(s,t,u) = \frac{\beta _-^2 Q_+^2 (s+2 t) \sqrt{\frac{t u}{s}}}{24 \beta _{\text{FP}}^2 m_{\text{FP}} m_{\text{Pl}}^2}$ & $E^{3}$ \\

$\displaystyle \mathcal{A}^{MM\to MM}_{2,2;-2,-2}(s,t,u) = -\frac{\beta _-^2 Q_+^2 t u}{4 s \beta _{\text{FP}}^2 m_{\text{Pl}}^2}$ & $E^{2}$ \\

$\displaystyle 
\begin{aligned}
\mathcal{A}^{MM\to MM}_{2,1;2,1}(s,t,u) = \frac{1}{32 s t \beta _{\text{FP}}^2 m_{\text{Pl}}^2} \Bigg(4 &\beta _{\text{FP}}^2 \left(Q_+^2 \left(t^3-3 t^2 u+7 t u^2+2 u^3\right)+4 s^3\right) \\
-4 &\beta _- Q_- Q_+ \beta _{\text{FP}} (t-u) \left(t^2+u^2\right) \\
+4 &\beta _+ Q_+^2 t^2 u \beta _{\text{FP}}+\beta _-^2 Q_+^2 t \left(t^2-t u-u^2\right) \Bigg)
\end{aligned}
$ & $E^{2}$ \\

$\displaystyle \mathcal{A}^{MM\to MM}_{2,1;2,0}(s,t,u) = \frac{Q_+ u \left(2 \beta _- Q_- \beta _{\text{FP}} \left(2 s^2+5 s t+4 t^2\right)+12 Q_+ t u \beta _{\text{FP}}^2-\beta _-^2 Q_+ t^2\right)}{16 \sqrt{6} \beta _{\text{FP}}^2 m_{\text{FP}} m_{\text{Pl}}^2 \sqrt{s t u}}$ & $E^{3}$ \\

$\displaystyle \mathcal{A}^{MM\to MM}_{2,1;2,-1}(s,t,u) = \frac{\beta _- Q_+ u \left(12 Q_- s \beta _{\text{FP}} (2 s+t)-\beta _- Q_+ t (5 s+t)\right)}{96 s^2 \beta _{\text{FP}}^2 m_{\text{Pl}}^2}$ & $E^{2}$ \\

$\displaystyle \mathcal{A}^{MM\to MM}_{2,1;2,-2}(s,t,u) = \frac{\beta _- Q_+ u m_{\text{FP}} \sqrt{s t u} \left(18 Q_- s \beta _{\text{FP}}+\beta _- Q_+ (2 s+t)\right)}{48 s^3 \beta _{\text{FP}}^2 m_{\text{Pl}}^2}$ & $E^{1}$ \\

$\displaystyle \mathcal{A}^{MM\to MM}_{2,1;1,1}(s,t,u) = \frac{Q_+ t u (s+2 t) \left(Q_+ \left(6 \beta _{\text{FP}}-\beta _+\right)-2 \beta _- Q_-\right)}{16 \beta _{\text{FP}} m_{\text{FP}} m_{\text{Pl}}^2 \sqrt{s t u}}$ & $E^{3}$ \\

$\displaystyle \mathcal{A}^{MM\to MM}_{2,1;1,0}(s,t,u) = -\frac{Q_+ u \left(Q_+ \left(4 \beta _{\text{FP}}^2 (s+3 t)+4 \beta _+ t \beta _{\text{FP}}+\beta _-^2 (s-t)\right)-4 \beta _- Q_- \beta _{\text{FP}} (s+2 t)\right)}{32 \sqrt{6} \beta _{\text{FP}}^2 m_{\text{FP}}^2 m_{\text{Pl}}^2}$ & $E^{4}$ \\

$\displaystyle \mathcal{A}^{MM\to MM}_{2,1;1,-1}(s,t,u) = \frac{Q_+ u \sqrt{s t u} \left(\beta _-^2 Q_+ t-6 s \beta _{\text{FP}} \left(Q_+ \left(\beta _++2 \beta _{\text{FP}}\right)-2 \beta _- Q_-\right)\right)}{96 s^2 \beta _{\text{FP}}^2 m_{\text{FP}} m_{\text{Pl}}^2}$ & $E^{3}$ \\

$\displaystyle \mathcal{A}^{MM\to MM}_{2,1;1,-2}(s,t,u) = -\frac{\beta _-^2 Q_+^2 t u^2}{48 s^2 \beta _{\text{FP}}^2 m_{\text{Pl}}^2}$ & $E^{2}$ \\

$\displaystyle \mathcal{A}^{MM\to MM}_{2,1;0,0}(s,t,u) = \frac{Q_+ (s+2 t) \left(2 \beta _- Q_- \beta _{\text{FP}}-4 \beta _+ Q_+ \beta _{\text{FP}}+\beta _-^2 Q_+\right) \sqrt{s t u}}{96 \beta _{\text{FP}}^2 m_{\text{FP}}^3 m_{\text{Pl}}^2}$ & $E^{5}$ \\

$\displaystyle \mathcal{A}^{MM\to MM}_{2,1;0,-1}(s,t,u) = \frac{Q_+ u \left(4 \beta _- Q_- s t \beta _{\text{FP}}-4 \beta _+ Q_+ s \beta _{\text{FP}} (s+3 t)+\beta _-^2 Q_+ \left(2 u^2-t^2\right)\right)}{32 \sqrt{6} s \beta _{\text{FP}}^2 m_{\text{FP}}^2 m_{\text{Pl}}^2}$ & $E^{4}$ \\

$\displaystyle \mathcal{A}^{MM\to MM}_{2,1;0,-2}(s,t,u) = \frac{Q_+ u \sqrt{t u} \left(2 s \beta _{\text{FP}} \left(\beta _- Q_-+2 \beta _+ Q_+\right)-\beta _-^2 Q_+ (3 s+t)\right)}{16 \sqrt{6} s^{3/2} \beta _{\text{FP}}^2 m_{\text{FP}} m_{\text{Pl}}^2}$ & $E^{3}$ \\

$\displaystyle \mathcal{A}^{MM\to MM}_{2,1;-1,-1}(s,t,u) = \frac{\beta _-^2 Q_+^2 (s+2 t) \sqrt{s t u}}{96 \beta _{\text{FP}}^2 m_{\text{FP}}^3 m_{\text{Pl}}^2}$ & $E^{5}$ \\

$\displaystyle \mathcal{A}^{MM\to MM}_{2,1;-1,-2}(s,t,u) = -\frac{\beta _-^2 Q_+^2 u (s+2 t)}{48 \beta _{\text{FP}}^2 m_{\text{FP}}^2 m_{\text{Pl}}^2}$ & $E^{4}$ \\

$\displaystyle \mathcal{A}^{MM\to MM}_{2,0;2,0}(s,t,u) = \frac{Q_+ u \left(4 u \beta _{\text{FP}} \left(Q_+ \beta _{\text{FP}}-\beta _- Q_-\right)-\beta _-^2 Q_+ t\right)}{48 \beta _{\text{FP}}^2 m_{\text{FP}}^2 m_{\text{Pl}}^2}$ & $E^{4}$ \\

$\displaystyle \mathcal{A}^{MM\to MM}_{2,0;2,-1}(s,t,u) = \frac{\beta _- Q_+ \left(-\frac{u}{s}\right)^{3/2} \left(2 Q_- s \beta _{\text{FP}} (2 s+t)+\beta _- Q_+ t u\right)}{16 \sqrt{6} \sqrt{-t} \beta _{\text{FP}}^2 m_{\text{FP}} m_{\text{Pl}}^2}$ & $E^{3}$ \\

$\displaystyle \mathcal{A}^{MM\to MM}_{2,0;2,-2}(s,t,u) = \frac{\beta _- Q_+ u^2 \left(2 Q_- s \beta _{\text{FP}}+\beta _- Q_+ t\right)}{8 \sqrt{6} s^2 \beta _{\text{FP}}^2 m_{\text{Pl}}^2}$ & $E^{2}$ \\

$\displaystyle \mathcal{A}^{MM\to MM}_{2,0;1,1}(s,t,u) = \frac{Q_+ t u \left(Q_+ \left(\beta _-^2+8 \beta _{\text{FP}}^2-2 \beta _+ \beta _{\text{FP}}\right)-4 \beta _- Q_- \beta _{\text{FP}}\right)}{16 \sqrt{6} \beta _{\text{FP}}^2 m_{\text{FP}}^2 m_{\text{Pl}}^2}$ & $E^{4}$ \\

$\displaystyle \mathcal{A}^{MM\to MM}_{2,0;1,0}(s,t,u) = -\frac{Q_+ u \left(4 Q_+ \beta _{\text{FP}}^2-4 \beta _- Q_- \beta _{\text{FP}}+\beta _-^2 Q_+\right) \sqrt{s t u}}{96 \beta _{\text{FP}}^2 m_{\text{FP}}^3 m_{\text{Pl}}^2}$ & $E^{5}$ \\

$\displaystyle \mathcal{A}^{MM\to MM}_{2,0;1,-1}(s,t,u) = -\frac{Q_+ u^2 \left(4 Q_+ \beta _{\text{FP}}^2-4 \beta _- Q_- \beta _{\text{FP}}+\beta _-^2 Q_+\right)}{32 \sqrt{6} \beta _{\text{FP}}^2 m_{\text{FP}}^2 m_{\text{Pl}}^2}$ & $E^{4}$ \\

$\displaystyle \mathcal{A}^{MM\to MM}_{2,0;1,-2}(s,t,u) = -\frac{\beta _- Q_+ u^2 m_{\text{FP}} (s+3 t) \left(2 Q_- s \beta _{\text{FP}}+\beta _- Q_+ (t-s)\right)}{16 \sqrt{6} s^{5/2} \beta _{\text{FP}}^2 m_{\text{Pl}}^2 \sqrt{t u}}$ & $E^{1}$ \\

$\displaystyle \mathcal{A}^{MM\to MM}_{2,0;0,0}(s,t,u) = \frac{Q_+ s t u \left(\beta _- Q_--\beta _+ Q_+\right)}{24 \sqrt{6} \beta _{\text{FP}} m_{\text{FP}}^4 m_{\text{Pl}}^2}$ & $E^{6}$ \\

$\displaystyle \mathcal{A}^{MM\to MM}_{2,0;0,-1}(s,t,u) = \frac{Q_+ u \sqrt{s t u} \left(2 s \beta _{\text{FP}} \left(\beta _- Q_--2 \beta _+ Q_+\right)+\beta _-^2 Q_+ (2 s+t)\right)}{96 s \beta _{\text{FP}}^2 m_{\text{FP}}^3 m_{\text{Pl}}^2}$ & $E^{5}$ \\

$\displaystyle \mathcal{A}^{MM\to MM}_{2,0;0,-2}(s,t,u) = -\frac{Q_+^2 u^2 \left(\beta _-^2 (2 s+t)-2 \beta _+ s \beta _{\text{FP}}\right)}{48 s \beta _{\text{FP}}^2 m_{\text{FP}}^2 m_{\text{Pl}}^2}$ & $E^{4}$ \\

$\displaystyle \mathcal{A}^{MM\to MM}_{2,0;-1,-1}(s,t,u) = -\frac{Q_+^2 t u \left(\beta _-^2-2 \beta _+ \beta _{\text{FP}}\right)}{8 \sqrt{6} \beta _{\text{FP}}^2 m_{\text{FP}}^2 m_{\text{Pl}}^2}$ & $E^{4}$ \\

$\displaystyle \mathcal{A}^{MM\to MM}_{2,-1;2,-1}(s,t,u) = \frac{\beta _-^2 Q_+^2 t u^2}{48 s \beta _{\text{FP}}^2 m_{\text{FP}}^2 m_{\text{Pl}}^2}$ & $E^{4}$ \\

$\displaystyle \mathcal{A}^{MM\to MM}_{2,-1;2,-2}(s,t,u) = \frac{\beta _-^2 Q_+^2 t (-u)^{5/2}}{24 \beta _{\text{FP}}^2 m_{\text{FP}} m_{\text{Pl}}^2 \sqrt{-s^3 t}}$ & $E^{3}$ \\

$\displaystyle \mathcal{A}^{MM\to MM}_{2,-1;1,1}(s,t,u) = -\frac{\beta _-^2 Q_+^2 m_{\text{FP}} (s+2 t) \sqrt{s t u}}{24 s^2 \beta _{\text{FP}}^2 m_{\text{Pl}}^2}$ & $E^{1}$ \\

$\displaystyle \mathcal{A}^{MM\to MM}_{2,-1;1,0}(s,t,u) = \frac{\beta _-^2 Q_+^2 t u^2}{32 \sqrt{6} s \beta _{\text{FP}}^2 m_{\text{FP}}^2 m_{\text{Pl}}^2}$ & $E^{4}$ \\

$\displaystyle \mathcal{A}^{MM\to MM}_{2,-1;1,-1}(s,t,u) = \frac{\beta _-^2 Q_+^2 t (-u)^{3/2} u}{96 \beta _{\text{FP}}^2 m_{\text{FP}}^3 m_{\text{Pl}}^2 \sqrt{-s t}}$ & $E^{5}$ \\

$\displaystyle \mathcal{A}^{MM\to MM}_{2,-1;1,-2}(s,t,u) = -\frac{\beta _-^2 Q_+^2 u^3}{48 s \beta _{\text{FP}}^2 m_{\text{FP}}^2 m_{\text{Pl}}^2}$ & $E^{4}$ \\

$\displaystyle \mathcal{A}^{MM\to MM}_{2,-1;0,0}(s,t,u) = \frac{\beta _-^2 Q_+^2 t u (s+2 t)}{24 \beta _{\text{FP}}^2 m_{\text{FP}} m_{\text{Pl}}^2 \sqrt{s t u}}$ & $E^{3}$ \\

$\displaystyle \mathcal{A}^{MM\to MM}_{2,-1;0,-1}(s,t,u) = \frac{\beta _- Q_+ u \left(\beta _- Q_+ \left(s^2+2 s t-t^2\right)-Q_- s t \beta _{\text{FP}}\right)}{8 \sqrt{6} s^2 \beta _{\text{FP}}^2 m_{\text{Pl}}^2}$ & $E^{2}$ \\

$\displaystyle \mathcal{A}^{MM\to MM}_{2,-1;-1,-1}(s,t,u) = \frac{\beta _- Q_+ m_{\text{FP}} (s+2 t) \left(3 Q_- \beta _{\text{FP}}-2 \beta _- Q_+\right) \sqrt{s t u}}{24 s^2 \beta _{\text{FP}}^2 m_{\text{Pl}}^2}$ & $E^{1}$ \\

$\displaystyle \mathcal{A}^{MM\to MM}_{2,-2;2,-2}(s,t,u) = \frac{u^3 \left(6 \left(Q_+^2-3\right) s \beta _{\text{FP}}^2-\beta _-^2 Q_+^2 t\right)}{12 s^2 t \beta _{\text{FP}}^2 m_{\text{Pl}}^2}$ & $E^{2}$ \\

$\displaystyle \mathcal{A}^{MM\to MM}_{2,-2;1,1}(s,t,u) = \frac{\beta _- Q_+ t u \left(12 Q_- \beta _{\text{FP}}-\beta _- Q_+\right)}{48 s \beta _{\text{FP}}^2 m_{\text{Pl}}^2}$ & $E^{2}$ \\

$\displaystyle \mathcal{A}^{MM\to MM}_{2,-2;1,0}(s,t,u) = \frac{\beta _- Q_+ \sqrt{-t} \left(-\frac{u}{s}\right)^{3/2} \left(2 Q_- s \beta _{\text{FP}}+\beta _- Q_+ (t-s)\right)}{16 \sqrt{6} \beta _{\text{FP}}^2 m_{\text{FP}} m_{\text{Pl}}^2}$ & $E^{3}$ \\

$\displaystyle \mathcal{A}^{MM\to MM}_{2,-2;1,-1}(s,t,u) = -\frac{\beta _-^2 Q_+^2 t u^2}{48 s \beta _{\text{FP}}^2 m_{\text{FP}}^2 m_{\text{Pl}}^2}$ & $E^{4}$ \\

$\displaystyle \mathcal{A}^{MM\to MM}_{2,-2;0,0}(s,t,u) = \frac{\beta _-^2 Q_+^2 t u}{48 \beta _{\text{FP}}^2 m_{\text{FP}}^2 m_{\text{Pl}}^2}$ & $E^{4}$ \\

$\displaystyle \mathcal{A}^{MM\to MM}_{1,1;1,1}(s,t,u) = -\frac{\beta _-^2 Q_+^2 s^3}{192 \beta _{\text{FP}}^2 m_{\text{FP}}^4 m_{\text{Pl}}^2}$ & $E^{6}$ \\

$\displaystyle \mathcal{A}^{MM\to MM}_{1,1;1,0}(s,t,u) = -\frac{Q_+^2 s \left(2 \beta _{\text{FP}}-\beta _+\right) (s+2 t) \sqrt{\frac{t u}{s}}}{16 \sqrt{6} \beta _{\text{FP}} m_{\text{FP}}^3 m_{\text{Pl}}^2}$ & $E^{5}$ \\

$\displaystyle \mathcal{A}^{MM\to MM}_{1,1;1,-1}(s,t,u) = \frac{Q_+^2 t u \left(\beta _-^2-4 \beta _{\text{FP}}^2\right)}{32 \beta _{\text{FP}}^2 m_{\text{FP}}^2 m_{\text{Pl}}^2}$ & $E^{4}$ \\

$\displaystyle \mathcal{A}^{MM\to MM}_{1,1;0,0}(s,t,u) = \frac{Q_+^2 s \left(4 \beta _+ t u \beta _{\text{FP}}+\beta _-^2 \left(-s^2+2 s t+2 t^2\right)\right)}{192 \beta _{\text{FP}}^2 m_{\text{FP}}^4 m_{\text{Pl}}^2}$ & $E^{6}$ \\

$\displaystyle \mathcal{A}^{MM\to MM}_{1,1;0,-1}(s,t,u) = \frac{Q_+^2 \left(\beta _-^2-2 \beta _+ \beta _{\text{FP}}\right) (s+2 t) \sqrt{s t u}}{16 \sqrt{6} \beta _{\text{FP}}^2 m_{\text{FP}}^3 m_{\text{Pl}}^2}$ & $E^{5}$ \\

$\displaystyle \mathcal{A}^{MM\to MM}_{1,1;-1,-1}(s,t,u) = -\frac{\beta _-^2 Q_+^2 s t u}{64 \beta _{\text{FP}}^2 m_{\text{FP}}^4 m_{\text{Pl}}^2}$ & $E^{6}$ \\

$\displaystyle \mathcal{A}^{MM\to MM}_{1,0;1,0}(s,t,u) = -\frac{Q_+^2 s t u \left(\beta _-^2+4 \beta _{\text{FP}} \left(\beta _{\text{FP}}-\beta _+\right)\right)}{192 \beta _{\text{FP}}^2 m_{\text{FP}}^4 m_{\text{Pl}}^2}$ & $E^{6}$ \\

$\displaystyle \mathcal{A}^{MM\to MM}_{1,0;1,-1}(s,t,u) = \frac{Q_+^2 s u \left(\beta _+-2 \beta _{\text{FP}}\right) \sqrt{\frac{t u}{s}}}{16 \sqrt{6} \beta _{\text{FP}} m_{\text{FP}}^3 m_{\text{Pl}}^2}$ & $E^{5}$ \\

$\displaystyle \mathcal{A}^{MM\to MM}_{1,0;0,0}(s,t,u) = -\frac{Q_+^2 \left(-\beta _-^2+8 \beta _{\text{FP}}^2-4 \beta _+ \beta _{\text{FP}}\right) (s+2 t) \sqrt{s t u}}{48 \sqrt{6} \beta _{\text{FP}}^2 m_{\text{FP}}^3 m_{\text{Pl}}^2}$ & $E^{5}$ \\

$\displaystyle \mathcal{A}^{MM\to MM}_{1,0;0,-1}(s,t,u) = \frac{Q_+^2 u \left(\beta _-^2 \left(s^2+4 s t+t^2\right)-4 \beta _+ s t \beta _{\text{FP}}\right)}{192 \beta _{\text{FP}}^2 m_{\text{FP}}^4 m_{\text{Pl}}^2}$ & $E^{6}$ \\

$\displaystyle \mathcal{A}^{MM\to MM}_{1,-1;1,-1}(s,t,u) = -\frac{\beta _-^2 Q_+^2 u^3}{192 \beta _{\text{FP}}^2 m_{\text{FP}}^4 m_{\text{Pl}}^2}$ & $E^{6}$ \\

$\displaystyle \mathcal{A}^{MM\to MM}_{1,-1;0,0}(s,t,u) = \frac{Q_+^2 s t u \left(\beta _-^2+4 \beta _{\text{FP}} \left(\beta _{\text{FP}}-\beta _+\right)\right)}{192 \beta _{\text{FP}}^2 m_{\text{FP}}^4 m_{\text{Pl}}^2}$ & $E^{6}$ \\

$\displaystyle \mathcal{A}^{MM\to MM}_{0,0;0,0}(s,t,u) = -\frac{Q_+ s t u \left(2 \beta _- Q_- \beta _{\text{FP}}+Q_+ \left(3 \beta _-^2+3 \beta _{\text{FP}}^2-8 \beta _+ \beta _{\text{FP}}\right)\right)}{144 \beta _{\text{FP}}^2 m_{\text{FP}}^4 m_{\text{Pl}}^2}$ & $E^{6}$ \\
\bottomrule
\caption{Leading high-energy behaviour of the $MM\to MM$ helicity amplitudes.}
\label{tab:HighEnergyMMMM}
\end{longtable}
\endgroup

\printbibliography

\end{document}